\documentclass[aps,preprintnumbers,showpacs,twocolumn,superscriptaddress,floatfix,nofootinbib]{revtex4-1}

\usepackage{latexsym} \usepackage{amssymb} \usepackage{amsfonts}
\usepackage{amsmath} \usepackage{bm} \usepackage[dvips]{graphicx}
\usepackage{color}
\usepackage{subfigure} \usepackage{times} \usepackage{units}
\usepackage{hyperref}

\usepackage[utf8x]{inputenc} \usepackage{amssymb,amsmath}
\usepackage{graphicx} \usepackage[squaren]{SIunits} 

\newcommand{\dt}{(\partial_t - {\mathcal L}_\beta)\;}

\newcommand{\bea}{\begin{eqnarray}} \newcommand{\eea}{\end{eqnarray}}
\newcommand{\beq}{\begin{equation}} \newcommand{\eeq}{\end{equation}}
\newcommand{\Eref}[1]{Eq.~\eqref{#1}}

 \newcommand{\tg}{\tilde\gamma}
\newcommand{\Q}{Q_{\nu_i}} \newcommand{\R}{R_{\nu_i}}

\newcommand{\be}{\begin{equation}} \newcommand{\ee}{\end{equation}}
\newcommand{\tpd}[3]{\left.\frac{\partial #1}{\partial
      #2}\right|_{#3}} 

\newcommand{\eg}{e.g.,~} 
\newcommand{\ie}{i.e.,~}
\newcommand{\cf}{cf.,~}

\newcommand{\referee}[1]{{{#1}}}

\newcommand{\gij}{\gamma}        
\newcommand{\detgij}{\gamma}     

\newcommand{\Lapse}{\alpha}

\newcommand{\Rmd}{\rho} 
\newcommand{\Sed}{\epsilon}
\newcommand{\Efr}{Y_e}
\newcommand{\Lf}{W}

\newcommand{\Ced}{\tau} 
\newcommand{\Crmd}{D}
\newcommand{\Cmom}{S} 
\newcommand{\Cefr}{\hat{Y}_e} 

\newcommand{\nue}{\nu_e}
\newcommand{\nua}{\bar{\nu}_e} 
\newcommand{\nux}{\nu_{\tau,\mu}}
\newcommand{\nuax}{\bar{\nu}_{\tau,\mu}}

  \allowdisplaybreaks

\usepackage[]{mathrsfs}

\begin{document}
\title{Implementation of a simplified approach to radiative transfer in
  general relativity}
\author{Filippo \surname{Galeazzi}}\affiliation{Departamento de
  Astronom\'{\i}a y Astrof\'{\i}sica, Universitat de Val\`encia,
  Calle del Doctor Moliner 50, 46100, Burjassot (Val\`encia), Spain} \affiliation{
  Max-Planck-Institut f\"ur Gravitationsphysik, Albert-Einstein-Institut,
  Potsdam-Golm, Germany}

\author{Wolfgang \surname{Kastaun}} \affiliation{ Max-Planck-Institut
  f\"ur Gravitationsphysik, Albert-Einstein-Institut, Potsdam-Golm,
  Germany}
  
\author{Luciano \surname{Rezzolla}} \affiliation{ Max-Planck-Institut
  f\"ur Gravitationsphysik, Albert-Einstein-Institut, Potsdam-Golm,
  Germany} \affiliation{Institut f\"ur Theoretische Physik, 
  Max-von-Laue-Straße 1, 60438 Frankfurt, Germany}

\author{Jos\'e A.~\surname{Font}} \affiliation{Departamento de
  Astronom\'{\i}a y Astrof\'{\i}sica, Universitat de Val\`encia,
  Calle del Doctor Moliner 50, 46100, Burjassot (Val\`encia), Spain}

\begin{abstract}
  We describe in detail the implementation of a simplified approach to
  radiative transfer in general relativity by means of the well-known
  neutrino leakage scheme (NLS). In particular, we carry out an extensive
  investigation of the properties and limitations of the NLS for isolated
  relativistic stars to a level of detail that has not been discussed
  before in a general-relativistic context. Although the numerous tests
  considered here are rather idealized, they provide a well-controlled
  environment in which to understand the relationship between the matter
  dynamics and the neutrino emission, which is important in order to
  model the neutrino signals from more complicated scenarios, such as
  binary neutron-star mergers. \referee{When considering nonrotating hot 
  neutron stars we confirm earlier results of one-dimensional simulations, 
  but also present novel results about the equilibrium properties and 
  on how the cooling affects the stability of these configurations.} In our idealized
  but controlled setup, we can then show that deviations from the
  thermal and weak-interaction equilibrium affect the stability of these
  models to radial perturbations, leading models that are stable in the
  absence of radiative losses, to a gravitational collapse to a black
  hole when neutrinos are instead radiated.
\end{abstract}
\maketitle

\section{Introduction}
\label{section:intro}

Neutron stars (NS) are a unique astrophysical laboratory to study the
microphysics of superdense matter. From their births in a supernova
explosion until their deaths in a merger of compact objects or in a
gravitational collapse to a black hole (BH), NSs undergo the most extreme
physical conditions with respect to density, temperature and
composition. For example, the temperature can vary from \referee{a
  surface temperature of a few $\mathrm{eV}$}, during the inspiral of two
old NSs in a binary system (BNSs), up to $\sim 100~\textrm{MeV}$ at the
time of merger~\cite{Bauswein2012}. Similarly, a NS can reach up to
$10$--$20$ times nuclear saturation density $\rho_0 =
2.7\times10^{14}~{\rm g~cm}^{-3}$~\cite{Baiotti04}. The merger of two NSs
is considered one of the most promising astrophysical events in which to
test the models of nuclear matter in regimes that are unattainable in
laboratory experiments. In addition to the bulk properties of the NSs,
such as mass and radius, also the details of the NS equation of state
(EOS), are expected to be imprinted in the electromagnetic emission (EM)
and gravitational wave (GW) signal produced by these events.

In particular, the GW signal that accompanies the merger of BNSs is
expected to be the prime target for the advanced generation of kilometer
length ground-based detectors, such as LIGO \cite{Abbott:2009qj}, Virgo
\cite{Acernese:2008dd}, and KAGRA \cite{Kuroda2010}. Individual merger
signals can reveal the NS radii with an accuracy of ${\sim}1$~km together
with their masses \cite{Hinderer09}. These observations will strongly
constrain the EOS that describes the NS matter at zero
temperature. Moreover, GWs, together with their EM counterparts, will be
of great relevance to determine the properties of the engine behind short
gamma-ray bursts (GRBs) which are thought to arise from NS/NS and BH/NS
mergers~\cite{Eichler89,Rezzolla:2011}. The incorporation of
microphysical nuclear EOSs, general relativity, magnetic fields, neutrino
radiation, viscosity and other transport coefficients in the numerical
modeling of those events, is essential to increase the realism of the
simulations and help us explore the possible outcomes from a large set of
initial conditions and parameters.

A great deal of effort has been spent in the field of nuclear physics on
the development of an EOS appropriate to describe the behavior of matter
at both supranuclear and subnuclear densities
(see~\cite{Lattimer00,Haensel07} for reviews). Unfortunately, only few
constraints on the properties of the EOS are available from relativistic
heavy-ion collision experiments and astronomical observations (see
\cite{Lattimer2012rev} for a review on the topic). The recent discoveries
of two massive NSs in a binary system with a mass of $\sim 2~M_{\odot}$
\cite{Demorest2010,Antoniadis2013} seem to rule out some of the NS matter
models. Nevertheless, the lack of precise estimates of their radii did
not allow to significantly reduce the parameter space for the NS EOSs.
Combined measurements of mass and radius of NSs can only be obtained from
x-ray bursts and from the NS surface thermal emission, something unlikely
to happen in quiescent systems as the one described in
\cite{Demorest2010} and \cite{Antoniadis2013}. These measurements are
subject to many uncertainties, as the distance of the NS and its magnetic
field, that can spoil the accuracy of the estimates of the radius
\cite{Ozel:2010}. \referee{More importantly, the estimates of the physical 
parameters from the observations are sensitive to the models
adopted for their analysis ~\cite{Steiner2010}.}

As a result of the complexity involved in the theoretical models of NS
matter, only few EOSs cover temperatures up to $100~\textrm{MeV}$,
densities that reach several times nuclear saturation density and a range
of composition that goes from pure neutron matter to symmetric matter
(equal neutron and proton number, $N=Z$). Amongst the few EOSs that have
gained popularity for astrophysical simulations, particularly well
studied are the EOS of \citet{Lattimer91} (hereafter, LS-EOS) and the EOS
proposed by \citet{shen98,shen98b}. More recently, a new EOS has been
derived by \citet{ShenG2010,ShenG2011} (hereafter, SHT-EOS), in which the
meson mean fields and the Dirac wave function describing the nucleons are
solved self-consistently for each Wigner-Seitz cell. Despite the accuracy
reached in the theoretical models, these EOSs not always satisfy the
constraints imposed by both experiments (symmetry energy and
incompressibility) and astronomical observations (mass and radius)
\cite{Steiner2010,Hebeler2013}.

Over the last few years, a lot of progress has been made to understand
the dynamics of BNS systems in general relativity with respect to the
possible detection of GWs from such systems (for a review see
\cite{Faber2012:lrr}). The level of sophistication reached by present-day
general-relativistic simulations includes important microphysical aspects
\cite{Sekiguchi2011, Sekiguchi2011b, Bauswein2012} that in the past were
only accessible to Newtonian simulations
\cite{Ruffert96b,Ruffert97,Ruffert01,Rosswog:2003b}. Together with
nuclear EOSs to describe the NS matter, current approaches also include
weak-interaction processes, albeit modeled by means of approximate
neutrino treatments. Under the extreme thermodynamic conditions reached
during the merger, powerful neutrino bursts are produced from the hot and
shock heated NS matter~\cite{Sekiguchi2011}. When the nuclear matter is
strongly compressed and its temperature reach values of about
$1~\textrm{MeV}$, weak-interaction processes become increasingly
important, leading the material out of the original chemical equilibrium
with respect to $\beta$ processes while copious amounts of neutrinos are
emitted. Neutrinos are responsible for the luminosities of the order $\sim
10^{53}~\mathrm{erg~s}^{-1}$ attained in BNS mergers and they may play a
role in powering the relativistic jet needed for the beamed emission of a
GRB. The energy required for the outflow can be efficiently deposited
along the baryon-free axis of rotation of the BH by neutrino pair
annihilation \cite{Ruffert96b}.

The aim of this paper is to present the details of the implementation in
the new \texttt{Whisky} code (\texttt{WhiskyThermal})
\cite{Baiotti03a,Baiotti04b} of a ''neutrino leakage'' scheme (NLS) along
with state-of-the-art nuclear EOSs with density, temperature and
composition dependence. The NLS is a pragmatic approach to neutrino
radiation transfer that estimates the local changes in the lepton number
and the associated energy losses via neutrino emission. The relative
simplicity of the implementation along with the small computational costs
turn the NLS into an interesting approach to upgrade numerical relativity
codes that evolve the equations of general-relativistic hydrodynamics
(GRHD) and magneto-hydrodynamics together with the Einstein
equations for a dynamical spacetime. 

\referee{Neutrino interactions are influenced by density, temperature and
  composition of the hadronic matter and by the energy of the emitted
  neutrinos.  At rest-mass densities of few times $10^{12}
  ~\mathrm{g~cm}^{-3}$ and temperatures around $10
  \usk\mega\electronvolt$, the scattering of neutrinos onto baryons is so
  efficient that neutrinos quickly reach the thermal equilibrium with the
  NS matter and can be considered as trapped (\ie with mean free path
  $\lambda_{\nu} \simeq 50 \usk\meter$). When the density drops below
  $10^{11} ~\mathrm{g~cm}^{-3}$, neutrinos with energy below $10
  \usk\mega\electronvolt$ interact rarely with the nuclear matter and can
  therefore be considered as free-streaming (\ie with mean free path
  $\lambda_{\nu} \gtrsim 10 \usk\kilo\meter$).}  The NLS scheme is
particularly suited to study the evolution of a NS because of the sharp
density gradient at the surface where the density drops by several orders
of magnitude. More accurate schemes such as the direct Monte Carlo method
for solving the Boltzmann transport equation \cite{Abdikamalov12}, are
able to capture properly the dynamic in the semitransparent
regime. Unfortunately, the computational cost involved limits the use of
such schemes to one-dimensional problems. Three-dimensional simulations
are performed instead using approximate ''ray-by-ray'' \cite{Scheck2006}
multi-energy-group neutrino schemes [\eg multigroup flux-limited
  diffusion \cite{Mezzacappa1993} and isotropic diffusion source
  approximation \cite{Liebendoerfer2009}]. Even more advanced, but
still rudimentary, approaches to the solution of the radiative-transfer
problem have been presented recently in Ref.~\cite{Radice2013}.

Besides carrying out an extensive investigation of the properties and
limitations of the NLS to a level of detail that, to the best of our
knowledge, has not discussed before in a general-relativistic context, we
also present novel results regarding the equilibrium properties of
nonrotating hot NSs and how the cooling affects the stability of these
configurations. In particular, we use several temperature profiles close
to the one reached by protoneutron stars (PNSs) during their formation
in a supernova core collapse or by the hypermassive neutron star (HMNS)
produced in a BNS merger and investigate how the energy losses can
influence the stability of these objects. In our idealized but
well-controlled setup we can then show that deviations from the thermal
and weak-interaction equilibrium of these models will affect their
stability to radial perturbations and thus lead to a neutrino-induced
collapse of stable nonrotating NS models.

The organization of the paper is as follows: In
Sec.~\ref{section:hde_micro} we summarize our mathematical framework
regarding the gravitational field equations and the GRHD equations in the
presence of neutrino emission. Sec.~\ref{sec:leakage} describes the
neutrino leakage scheme while Sec.~\ref{sec:eos} discusses the nuclear
equations of state for NS matter used in this
work. Sec.~\ref{section:numframe} describes in detail the numerical
framework adopted for our simulations. Sec.~\ref{sec:tests}, presents
a series of tests and results obtained with our code. In particular, we
show simulations of nonrotating stable NS with nuclear EOS at finite
temperature and in $\beta$-equilibrium, comparing with predictions from
linear-perturbation theory. Finally, Sec.~\ref{sec:nic} reports the
main result of this paper, namely, the neutrino-induced collapse of a
stable nonrotating NS and the associated neutrino bursts when using
different thermal EOSs. Appendix~\ref{app_a} reports the details of the
free emission rates and opacity sources used in our code and that could
be of direct use for anyone wanting to reproduce our implementation,
while a detailed description of the convergence tests of our scheme are
presented in Appendix~\ref{sec::convergence}. A detailed description of a
novel and robust technique for the transformation from conserved to
primitive variables is finally presented
in Appendix~\ref{apdx:con2prim}. Unless stated otherwise, we use
geometrized units in which $h=c=G=1$. Latin indices run from 1 to 3 while
greek indices run from 0 to 3.


\section{Mathematical Framework}
\label{section:hde_micro}

Although much of our mathematical and numerical framework for the
solution of the Einstein-Euler equations has been presented in a
scattered manner in a number of papers, \ie \cite{Baiotti04,Baiotti08},
we include a short review here both for completeness and as an aid to
anyone wanting to implement our approach to the handling of the radiative
transfer.

\subsection{Einstein equations}
\label{sec:einstein_eq}

We evolve a conformal-traceless ''$3+1$'' formulation of the Einstein
equations~\cite{Nakamura87, Shibata95, Baumgarte99}, BSSNOK. The
spacetime is foliated into three-dimensional spacelike hypersurfaces,
with a three-metric $\gamma_{ij}$, extrinsic curvature $K_{ij}$, and the
gauge functions $\alpha$ (lapse) and $\beta^i$ (shift) to specify the
coordinate frame.
The three-metric $\gamma_{ij}$ is conformally transformed as
\begin{equation}
  \label{eq:def_g}
  \phi = \frac{1}{12}\ln (\det \gamma_{ij})\,, \qquad
  \tilde{\gamma}_{ij} = e^{-4\phi} \gamma_{ij}
\end{equation}
and the conformal factor $\phi$ is evolved as an independent variable,
whereas $\tilde{\gamma}_{ij}$ is subject to the constraint $\det
\tilde{\gamma}_{ij} = 1$.
The extrinsic curvature is subject to the same conformal transformation
and its trace $K \equiv K_{i}^{i}$ is evolved as an independent variable,
while in place of $K_{ij}$ we evolve its conformal-traceless equivalent, 
\ie
\begin{equation}
  \label{eq:def_K}
  \tilde{A}_{ij} = e^{-4\phi} (K_{ij} - \frac{1}{3}\gamma_{ij} K)\,,
\end{equation}
with $\tilde{A}_{i}^{i}=0$. In addition, new evolution variables defined
in terms of the Christoffel symbols $\tilde{\Gamma}^i_{jk}$ of the
conformal three-metric are introduced, \ie
\begin{equation}
  \label{eq:def_Gamma}
  \tilde{\Gamma}^i = \tilde{\gamma}^{jk}\tilde{\Gamma}^i_{jk}\,,
\end{equation}
and solved as independent evolution variables. As a result, the complete
set of the evolution equations for the Einstein equations is given by
\begin{align}
  \label{eq:evolution}
  &\dt \tilde{\gamma}_{ij} = -2 \alpha \tilde{A}_{ij}\,,  \\ \nonumber \\
  &\dt \phi = - \frac{1}{6} \alpha K\,, \\ \nonumber \\
  &\dt \tilde{A}_{ij} = e^{-4\phi} [ - D_i D_j \alpha
  + \alpha (R_{ij} - 8 \pi {S}_{ij}) ]^{\rm TF} \nonumber\\
  & \hskip 2.0cm + \alpha (K \tilde{A}_{ij} - 2 \tilde{A}_{ik} \tilde{A}^k_j)\,, \\
  &\dt K  = - D^i D_i \alpha \nonumber \\
  & \hskip 2.0cm + \alpha \Big [\tilde{A}_{ij} \tilde{A}^{ij} +
  \frac{1}{3} K^2 +
  4\pi ({E}+{S})\Big ], \\ \nonumber \\
  &
  \partial_t \tilde{\Gamma}^i = \tilde\gamma^{jk} \partial_j\partial_k
  \beta^i + \frac{1}{3} \tilde\gamma^{ij} \partial_j\partial_k\beta^k
  \nonumber \\
  & \hskip 1.0cm + \frac{2}{3} \tilde\Gamma^i
  \partial_j\beta^j
  - 2 \tilde{A}^{ij} \partial_j\alpha + 2 \alpha (
  \tilde{\Gamma}^i_{jk} \tilde{A}^{jk} + 6 \tilde{A}^{ij}
  \partial_j \phi \nonumber\\
  & \hskip 1.0cm - \frac{2}{3} \tg^{ij} \partial_j K - 8 \pi \tg^{ij}
  {S}_j)\,,
\end{align}
where $\mathcal L_{\beta}$ is the Lie derivative with respect to the
shift vector, $R_{ij}$ is the three-dimensional Ricci tensor, $D_i$
the covariant derivative associated with the three-metric
$\gamma_{ij}$, ''TF'' indicates the trace-free part of tensor objects
and $ {E}$, ${S}_j$, and ${S}_{ij}$ are the matter source terms defined as
\begin{align}
  {E}&\equiv n_\alpha n_\beta T^{\alpha\beta}\,,  \\
  {S}_i&\equiv -\gamma_{i\alpha}n_{\beta}T^{\alpha\beta}\,, \\
  {S}_{ij}&\equiv \gamma_{i\alpha}\gamma_{j\beta}T^{\alpha\beta}\,,
\end{align}
where $n_\alpha\equiv (-\alpha,0,0,0)$ is the future-pointing four-vector
orthonormal to the spacelike hypersurface, $T^{\alpha\beta}$ is the
energy-momentum tensor for a perfect fluid [\cf Eq.~\eqref{eq:str-ene}],
and ${S}\equiv {S}^{i}_{i}$. The Einstein equations also comprise a set of
physical constraint equations that are satisfied within each spacelike
slice,
\begin{align}
  \label{eq:einstein_ham_constraint}
  \mathcal{H} &\equiv R + K^2 - K_{ij} K^{ij} - 16\pi {E} = 0, \\
  \label{eq:einstein_mom_constraints}
  \mathcal{M}^i &\equiv D_j(K^{ij} - \gamma^{ij}K) - 8\pi {S}^i = 0,
\end{align}
which are usually referred to as Hamiltonian and momentum constraints.
Here $R=R_{ij} \gamma^{ij}$ is the Ricci scalar on a three-dimensional
time slice. Our
specific choice of evolution variables introduces two additional
constraints,
\begin{align}
  \det \tilde{\gamma}_{ij} & = 1\,,
  \label{eq:gamma_one}\\
  \tilde{A}_{i}^{i} & = 0\,,
  \label{eq:trace_free_A}
\end{align}
which are enforced on each spacelike hypersurface. The remaining
constraints, $\mathcal{H}$ and $\mathcal{M}^i$ are not actively enforced
and can be used to monitor the accuracy of our numerical solution.

Our gauge choices reflect the experience matured over the last decade and
are already discussed in detail in~\cite{Baiotti08}. In particular, we
evolve the lapse according to the ''$1+\log$'' slicing
condition~\cite{Bona94b}:
\begin{equation}
  \partial_t \alpha - \beta^i\partial_i\alpha 
  = -2 \alpha (K - K_0)\,,
  \label{eq:one_plus_log}
\end{equation}
where $K_0$ is the initial value of the trace of the extrinsic
curvature and equals zero for the maximally sliced initial data we
consider here. The shift is evolved using the hyperbolic
$\tilde{\Gamma}$-driver condition~\cite{Alcubierre02a},
\begin{eqnarray}
  \label{eq:shift_evol}
  \partial_t \beta^i - \beta^j \partial_j  \beta^i & = & \frac{3}{4} \alpha B^i\,,
  \\
  \partial_t B^i - \beta^j \partial_j B^i & = & \partial_t \tilde\Gamma^i 
  - \beta^j \partial_j \tilde\Gamma^i - \eta B^i\,,
\end{eqnarray}
where $\eta$ is a parameter which acts as a damping coefficient.
Following ~\cite{Alcubierre02a} and because we are not considering
binaries here, we set $\eta$ to be $\eta \approx 2/ M_{\rm tot}$, where
$M_{\rm tot}$ is the total gravitational mass of the system; a more
sophisticated prescription in case of unequal-mass binaries can be found
in~\cite{Alic:2010}.

All the equations discussed above are solved using the
\texttt{McLachlan} code, a three-dimensional finite-differencing code
based on the Cactus Computational Toolkit \cite{Loffler:2011ay}.

\subsection{Relativistic-hydrodynamic equations}
\label{sec:rhd_eq}

In the following, we describe the matter evolution equations implemented
in the new \texttt{Whisky} code. These are given by the
general-relativistic-hydrodynamic equations describing a perfect
compressible fluid governed by a generic EOS with temperature and
composition dependence. In particular, the changes in composition,
energy, and momentum due to neutrino radiation are modeled by
introducing additional source terms.

When written in covariant form, the equations for the conservation of
energy, momentum, baryon and lepton numbers are expressed as
\begin{align}
  \nabla_\alpha T^{\alpha\beta} &= \Psi^{\beta}\,, \label{hydro eqs1}\\
  \nabla_\alpha (n_b u^\alpha) &= 0\,, \label{hydro eqs2}\\
  \nabla_\alpha (n_e u^\alpha) &= N\,, \label{hydro eqs3}
\end{align}
where $n_b$ and $n_e$ are the baryon and electron number densities, 
respectively.
Note that the energy-momentum tensor $T^{\alpha\beta}$ accounts for the
ordinary matter as well as for the trapped neutrinos and photons, but it
does not include the nontrapped neutrinos. The corresponding
energy-momentum tensor is taken into account only in the form of the
source term $\Psi^{\beta}$ modeling the radiative losses of energy and
momentum. Furthermore, because we assume these nontrapped neutrinos to
behave essentially as a test fluid, the associated energy momentum tensor
is neglected when building the right-hand side of the Einstein equations.

As customary in a NLS, also in our implementation we do not evolve
directly the number density and the energy distribution of the
neutrinos. Instead, since the neutrinos are assumed to be in local
thermal equilibrium with the baryonic matter, we can obtain direct
estimates for the source terms $\Psi^{\beta}$ and $N$ by simply
considering the matter properties. Also, because we do not consider
lepton species other than electrons in our fluid, the only degree of
freedom in the composition is represented by the electron fraction
defined as $Y_e \equiv n_e / n_b$.
The electron fraction is changed only by the production of electron 
neutrinos and antineutrinos as described by the source term $N$.

The energy-momentum tensor used in \Eref{hydro eqs1} is that of a perfect
fluid, given by
\begin{equation}\label{eq:str-ene}
  T^{\alpha\beta} = \rho \left(1+\epsilon+\frac{p}{\rho}\right) u^{\alpha}
  u^{\beta} + p g^{\alpha\beta}\,,
\end{equation}
where $\Rmd$ is the baryon rest-mass density defined as $\Rmd=m_b n_b$,
where $m_b$ denotes the nucleon mass\footnote{
The formal nucleon mass is chosen differently for each EOS, 
such that $\epsilon>0$. In detail,
$m_b =
  922.316\,\textrm{MeV}=1.6442 \times 10^{-24}\,\textrm{g}$ for the
  LS-EOS, and by $m_b = 930.267\,\textrm{MeV}=1.6583 \times
  10^{-24}\,\textrm{g}$ for the SHT-EOS.}. The total isotropic pressure
$p$ and the total specific internal energy $\epsilon$ contain
contributions of baryons, electrons, photons, and trapped neutrinos
\begin{align}
  p        &= 
    p_{e} + p_{b} + p_{\gamma} + p_{\nue,\nua} \,, 
    \label{eq:press_contribs}\\
  \epsilon &= 
    \epsilon_{e} + \epsilon_{b} + \epsilon_{\gamma} 
    + \epsilon_{\nue,\nua}  \,,
    \label{eq:energy_contribs}
\end{align}
where the contribution from trapped electron-neutrinos $\nue$ and
antineutrinos $\nua$ in the dense baryonic component can be evaluated
from the thermodynamic state of the fluid and assuming that the neutrinos
are following a Fermi-Dirac distribution, \ie
\begin{eqnarray}
  p_{\nue,\nua}& \equiv & p_{\nue} + p_{\nua} = 
\frac{4\pi}{3} T^4
\left[F_3(\eta_{\nue})+F_3(\eta_{\nua})\right]\,,
\\
  \epsilon_{\nue,\nua}& \equiv & \epsilon_{\nue} + \epsilon_{\nua} 
= \frac{1}{3} \frac{p_{\nue,\nua}}{\rho}\,.
\end{eqnarray} 
Here, $\eta_{\nue} = \mu_{\nue}/ T$ and $\eta_{\nua} = \mu_{\nua}/ T$ are
the degeneracy parameters for the electron-neutrinos and antineutrinos,
$\mu_{\nue}, \mu_{\nua}$ the corresponding chemical potentials, $T$ is
the temperature, which is expressed in $\mega\electronvolt$ throughout
the paper, and $F_3(\eta_{\nu_i})$ is a Fermi integral [see
  \Eref{eq::fermi} and Appendix~\ref{app_a} for a more extended
  discussion]. Note also that in \Eref{eq:press_contribs} the electron
contributions to the total pressure and internal energy are computed
using the EOS for a semidegenerate gas of electrons as described in
Ref.~\cite{Timmes2000}.

At rest-mass densities $\rho \approx 10^{12} ~{\rm g~cm}^{-3}$ 
\referee{and temperatures $T \approx 10 ~ \mega\electronvolt$}, the
contributions $p_{\nue,\nua}$ can be significant and around $10\%$ of the
baryonic one. However, close to nuclear saturation density, it becomes
less than $1\%$ of the total pressure and thus smaller than the typical
uncertainties of the nuclear EOSs at such densities. Furthermore, the
addition of a neutrino \referee{isotropic} pressure to the fluid pressure is meaningful
only in the trapped regime and cannot be made when the neutrinos are free
streaming. For these reasons and to maintain a smooth transition between
the regimes of trapped and free-streaming neutrinos, we neglect
contributions of trapped neutrinos to pressure and internal energy,
setting $p_{\nue,\nua} = 0 = \epsilon_{\nue,\nua}$ in
Eqs.~(\ref{eq:press_contribs})--(\ref{eq:energy_contribs}). 

To compute the neutrino source term $N$ in \Eref{hydro eqs3}, which
describes the changes in electron fraction, we introduce the 
neutrino emission rates per baryon, 
$R_{\nue}$ and $R_{\nua}$, for
electron-neutrinos and -antineutrinos, respectively. As a result, in the
fluid rest frame we can express the change in electron fraction as
\begin{align}
  u^{\alpha} \nabla_{\alpha} (\Efr) 
    & = \mathcal{R} \equiv R_{\nua} - R_{\nue}  \,.
\end{align}
Similarly, in order to determine the source terms $\Psi^{\beta}$
describing the radiative losses of energy and momentum due to neutrinos
in \Eref{hydro eqs1}, we assume that the emission is isotropic in the
fluid rest frame, so that no net momentum change can take place in the
rest frame. We then obtain the following covariant equation that can be
most easily verified in the comoving frame
\begin{align}
\Psi^{\beta} & = -\Rmd \, m_b^{-1} \mathcal{Q} u^{\beta} \equiv
-\Rmd \,  m_b^{-1} \sum_I Q_{_I} u^{\beta} \nonumber \\
& = -\Rmd \, m_b^{-1}(Q_{\nue} + Q_{\nua} + Q_{\nu_{\tau,\mu}} + Q_{\bar{\nu}_{\tau,\mu}}) u^{\beta}
\,, 
\end{align} 
where we have introduced the neutrino emissivity $\mathcal{Q}$ as the
sum of the emissivities of the various neutrino species
$Q_{_I}$ with $I={\nue}, {\nua}, {\nu_{\tau,\mu}}, \bar{\nu}_{\tau,\mu}$,
which denote the radiated energy per unit time and baryon (\ie
it is expressed as ${\rm erg~s}^{-1}$ in cgs units). Note
that we collect the emissivity due to the $\tau$ and $\mu$ neutrinos 
into a single contribution $Q_{\nux}$. An important remark should
be made about \Eref{hydro eqs3}. Since the neutrino interaction time scale
depends largely on the microphysics of the process described, it can
differ significantly from the time scale over which the baryonic matter
evolves, resulting in a problem with a stiff source term. When this
happens, special numerical algorithms are needed to handle the evolution
(see, \eg \cite{Palenzuela:2008sf}).

For numerical evolution, the equations are cast into a flux conservative
formulation, based on the {Valencia formulation} originally developed
by~\cite{Marti91,Banyuls97}. In particular, we write Eqs.~(\ref{hydro
  eqs1})--(\ref{hydro eqs3}) as balance laws
\begin{align}
  \label{eq:consform1}
  \partial_t \left(\sqrt{\detgij} \boldsymbol{q}\right) + 
  \partial_i \left(\sqrt{\detgij} \boldsymbol{f}^{(i)} 
  (\boldsymbol{q})\right) &= 
  \boldsymbol{s} (\boldsymbol{q})\,,
\end{align}
where $\detgij$ is the determinant of the three-metric, while
$\boldsymbol{f}^{(i)} (\boldsymbol{q})$ and
$\boldsymbol{s}(\boldsymbol{q})$ are the flux vectors and source terms,
respectively (see~\cite{Font08} for details). The right-hand side
contains the geometrical terms and the neutrino reaction rates that
depend on the state of the fluid. The evolved {conserved variables}
\mbox{$\boldsymbol{q} \equiv (\Crmd, \Ced, \Cmom_i, \Cefr)$} are given in
terms of the {primitive variables} via the relations
\begin{align}
\label{eq:prim2con}
  \Crmd &\equiv 
       \rho W \,, \\
  \Ced &\equiv 
        \rho h W^2 - p - D  \,,  \\
  \Cmom_i &\equiv 
       \rho h W^2 v_i \,, \\
  \Cefr &\equiv 
      \Crmd \Efr \,.
\end{align}
In the equations above, $v^i$ is the three-velocity measured by an
observer orthogonal to the hypersurface of constant coordinate time (\ie
a Eulerian observer), $\Lf$ is the corresponding Lorentz factor, and $h
\equiv 1 + \epsilon + p/\rho$ is the specific enthalpy. When written out
explicitly, Eq. \eqref{eq:consform1} reads
\begin{align}
  \partial_t \bar{\Crmd} &= 
      -\partial_i \left( w^i \bar{\Crmd} \right)  \,,
  \label{rest_mass_evol}\\
  \begin{split}
    \partial_t \bar{\Ced} &= 
      -\partial_i \left( w^i \bar{\Ced} +
        \sqrt{\detgij} \alpha p v^i \right) 
      -\Lapse \mathcal{Q} \, m_b^{-1}\bar{\Crmd} \\ & \qquad
      +\alpha \bar{S}^{kl} K_{kl} 
      - \bar{S}^i \partial_i \alpha  \,,
  \end{split}
  \label{energy_density_evol}\\
  \begin{split}
    \partial_t \bar{\Cmom}_j &= 
      -\partial_i \left( w^i \bar{\Cmom}_j + \sqrt{\detgij} \alpha
        p \delta^i_j \right) 
      -\Lapse \mathcal{Q} \, m_b^{-1} \bar{\Crmd} v_i \\&\qquad 
      +\frac{\alpha}{2} \bar{S}^{kl} \partial_j \gij_{kl} 
      + \bar{\Cmom}_k \partial_j \beta^k 
      - (\bar{\tau} + \bar{D}) \partial_j \alpha  \,,
  \end{split}
  \label{momentum_evol} \\
  \partial_t \bar{Y}_e &= 
      -\partial_i \left( w^i \bar{Y}_e \right) 
      + \bar{\Crmd} \frac{\Lapse}{\Lf} \mathcal{R} \,,
  \label{ye_m_evol}
\end{align}
where $\bar{\Crmd} \equiv \sqrt{\gamma}{\Crmd}$, $\bar{\Ced} \equiv
\sqrt{\gamma}{\Ced}$, $\bar{S}_{i} \equiv \sqrt{\gamma}S_{i}$,
$\bar{S}_{ij} \equiv \sqrt{\gamma}S_{ij}$, $\bar{Y}_{e} \equiv
\sqrt{\gamma} D Y_e$, and $w^i \equiv \alpha v^i - \beta^i$ denotes the
advection speed with respect to the coordinates.

In order to close the system of equations, we need an EOS to relate the
pressure to the rest of the primitive quantities (see
Sec.~\ref{sec:eos}). The new \texttt{Whisky} code implements several
analytic EOSs, such as the polytropic EOS, the ideal-fluid ($\Gamma$-law)
EOS, as well as ''hybrid'' EOSs \cite{Janka93}. Finally, the code can use
microphysical finite-temperature EOSs in tabulated form to describe
nuclear matter and the neutrino interactions, as will be discussed in
Sec.~\ref{sec:eos}.

The following sections are devoted to a description of our treatment
of the radiative transport within the NLS and of the interactions
between neutrinos and ordinary matter when the latter is modeled with
the nuclear EOS described in Sec.~\ref{sec:eos}.

\section{Approximate radiative transfer: the neutrino leakage scheme}
\label{sec:leakage}

Neutrinos from hot, dense and neutron-rich nuclear material are produced
via nonequilibrium weak-interaction processes during the first minutes
(or months) of the life of a PNS and in the final stages of the merger of
BNSs, when a HMNS is produced. The cooling produced by these neutrinos
and the consequent influence on the stellar structure and equilibrium is
of course of great importance to model these processes accurately. In
order to study the possible influence of thermal effects on the onset of
dynamical instabilities in PNSs or HMNSs, it is necessary to evolve the
system for tens of dynamical time scales, which can be estimated to be
$\tau_{\rm dyn} \sim (M/R^3)^{1/2} \approx 1$ ms, where $M$ and $R$ are
the mass and size of the object. This time scale is short when compared
with the typical diffusion time scale of neutrinos in dense matter, which
is $\gtrsim 1$ s. As a result, it is reasonable to use as a first
approximation to the radiative transfer a simple leakage scheme that
estimates the instantaneous energy loss via neutrino emission and by the
total change in electron fraction of the nuclear matter. Clearly, this
represents only a first step towards more sophisticated and accurate
transport schemes such as those employed in stellar-core collapse
simulations~\cite{Scheck2006,Marek2009,Liebendoerfer2009,Abdikamalov12}.

The NLS was originally developed by van Riper \emph{et al.}
\cite{vanRiper1981}, to describe the neutrino cooling through weak
interactions during the core-collapse phase of a supernova. In Newtonian
gravity, it was first applied in BNS simulations by Ruffert \emph{et
al.}~\cite{Ruffert96b,Ruffert97,Ruffert99b,Ruffert01} and more recently by
Rosswog \emph{et al.}~\citep{Rosswog02,Rosswog:2003b,Rosswog:2003}. It
has undergone a revival in general-relativistic three-dimensional
supernova simulations with the work of \citet{Sekiguchi2010} and
\cite{OConnor10,Ott2012,Ott2012b} (see also \cite{Peres2013}). Finally,
the NLS was also used recently to describe the neutrino cooling in
general-relativistic BNS simulations \citep{Sekiguchi2011,Sekiguchi2011b}
and BH-NS mergers \cite{Deaton2013}. 
The details of the implementation of the NLS differ among these authors
and are sometimes fragmentary. For these reasons, and to help anyone
wishing to reproduce our implementation, we give below a detailed account
of our treatment.

In our scheme we account for three different neutrino species: electron
neutrinos $\nue$, electron-antineutrinos $\nua$ and heavy lepton
neutrinos, $\nux$ that are considered as a single component with a
statistical weight of 4. The creation of electron-neutrinos and
antineutrinos by $\beta$ processes involving neutrons ($n$) and protons
($p$) leads to changes in the electron number, \ie
\begin{align}
  e^+ + n \to p + \nua , \\
  \;\; e^- + p \to n + \nu_e \,.
\end{align}
that need to be taken into account. On the other hand, all three species
of neutrinos and antineutrinos are created by annihilation of
electron-positron pairs, \ie
\begin{align}
 & e^+ + e^-  \to \bar{\nu}_{e} + \nu_{e}\,, \\
 & e^+ + e^-  \to \bar{\nu}_{\tau,\mu} + \nu_{\tau,\mu}\,,
\end{align}
and, especially at high temperatures, by plasmon decay, \ie
\begin{align}
  &\gamma \to \nue + \nua\,,\\
  &\gamma \to \nux + \bar{\nu}_{\tau,\mu}\,.
\end{align}
In order to compute the emission coefficients, we assume that the
neutrinos are massless and in thermal equilibrium with the surrounding
matter. As a result, the 
energy spectrum of the neutrinos follows 
a Fermi-Dirac distribution for ultrarelativistic particles at the
temperature of the matter. A further assumption concerns the chemical
potentials of the electron-neutrinos and of the electron-antineutrinos,
which are assumed to be at $\beta$-equilibrium~\cite{Rosswog:2003b}, \ie
\begin{equation}
  \label{eq:beta-eq}
  \mu^{eq}_{\nue} = \mu_e - \mu_n - \mu_p = -\mu^{eq}_{\nua} \,,
\end{equation}
where $\mu_e$, $\mu_n$, $\mu_p$ are the relativistic chemical potentials
including the rest-mass of the particle. As a result, we can define
$\eta^{eq}_{_I} \equiv \mu_{_I}/T$, where now
$\eta^{eq}_{\nue}=-\eta^{eq}_{\nua}$ and we assume $\eta^{eq}_{\tau,
  \mu}= 0$ since the heavy neutrinos are only weakly interacting with
the matter. The detailed derivation of the number and energy rates for
the different emission processes is described in detail in Appendix
\ref{app_a}.

\referee{As mentioned in the Introduction, low-energy neutrinos (\ie with
  $\lesssim 5 \usk\mega\electronvolt$) interact very rarely at low
  densities, \ie for $\rho \lesssim 10^{10}\,{\rm g~cm}^{-3}$}. In this
regime, if $\lambda$ is the mean free path of a neutrino, the time
$t_{_I}$ needed by a neutrino of species $I$ to cross a slab of nuclear
matter of depth $\mathscr{L} \ll \lambda$, is simply given by
\begin{eqnarray}
  t_{_I} \sim \mathscr{L} \,,
\end{eqnarray}
where, we recall, $I={\nue}, {\nua}, {\nu_{\tau,\mu}}, \bar{\nu}_{\tau,\mu}$. 
In this limit, the amount of energy loss by neutrinos can be directly computed using the
free emission rates (see below and Appendix~\ref{app_a} for details). On
the other hand, at high densities, \ie for $\rho \gtrsim 10^{13}\,{\rm
  g~cm}^{-3}$, \referee{the scattering and absorption of neutrinos onto baryons is extremely
efficient already for neutrinos with $\simeq 10 \usk\mega\electronvolt$
} and a purely diffusive regime can be reached, where
${\lambda_{_I}}/{\mathscr{L}} \ll 1$. Under these conditions, neutrinos
cannot freely escape but can only diffuse via a random walk over a
time scale
\begin{align}\label{eq:tdiff_slab}
  t_{_I} &\sim 
  \mathscr{L}\left( 1+ \mathcal{D} \frac{\mathscr{L}}{\lambda_{_I}} \right)
  \underset{\lambda_{_I} \ll \mathscr{L}}{\longrightarrow}
  \mathcal{D} \frac{\mathscr{L}^2}{\lambda_{_I}}\,,
\end{align}
where $\mathcal{D}$ is a coefficient depending on the optical depth
evaluated along the direction of propagation~\cite{Mihalas84}. In order
to compute the total mean free path, in our NLS we take into account
scattering and absorption processes between neutrinos and the surrounding
matter. In particular, for the calculation of the opacities we include
the following processes:
\begin{itemize}
\item coherent neutrino scattering on heavy nuclei (with atomic mass
number $A$)
\begin{align}
    \nu_{e} + A &\to \nu_{e} + A\,, &
    \bar{\nu}_{e} + A &\to \bar{\nu}_{e} + A\,, \\
    \nu_{\tau, \mu} + A &\to \nu_{\tau, \mu} + A\,, &
    \bar{\nu}_{\tau, \mu} + A &\to \bar{\nu}_{\tau, \mu} + A\,, 
\end{align}
\item neutrino scattering on free nucleons
\begin{align}
    \nu_{e} + [n,p] &\to \nu_{e} + [n,p] \,, \\ 
    \bar{\nu}_{e} + [n,p] &\to \bar{\nu}_{e} + [n,p]\,, \\
    \nu_{\tau, \mu} + [n,p] &\to \nu_{\tau, \mu} + [n,p]\,, \\
    \bar{\nu}_{\tau, \mu} + [n,p] &\to \bar{\nu}_{\tau, \mu} + [n,p]\,,  
\end{align}
\item electron-flavor neutrinos absorption on free nucleons
  \begin{align}
    \nue + n &\to p + e^-\,, \qquad  \nua + p \to n + e^+
  \end{align}
\end{itemize}
Summing all the various contributions to the mean free paths
$\lambda_{_I}$, we define the optical depth simply as the integral of
the inverse of the mean free path along a line of sight which is 
specifically chosen for the geometry of the problem, \ie
\begin{equation}
  \label{eq:opt_depth}
  \tau_{_I}(E_{_I})=
  \int_{x_1}^{x_2} \frac{d s}{\lambda_{_I}(E_{_I})}\,,
\end{equation}
where $E_{_I}$ is the neutrino energy (see Appendix~\ref{app_a} for
details)\referee{, and $d s$ is the proper length.
Note we have to neglect the gravitational redshift of the neutrinos,
as well as special-relativistic corrections due to the velocity of 
the absorbing matter.}
The knowledge of the optical depth has two important
applications. First, it can be used to determine the 
\referee{extent of the optically thick regions,}
which are clearly different for different neutrino
species. 
Second, it can be used to estimate the time scale a neutrino needs to 
diffuse out of the dense matter region. A simple estimate
can be obtained from \Eref{eq:tdiff_slab} by replacing 
$\mathscr{L}/\lambda$ with the optical depth $\tau$, yielding
\begin{align}
  \label{eq:t_diff_janka}
  t_{_I} (E_{_I}) &= \mathcal{D}\lambda_{_I}(E_{_I}) \tau^2_{_I}(E_{_I})\,.
\end{align}
This expression is also derived in Ref.~\cite{Rosswog:2003b} from a 
one-dimensional stationary diffusion model, and a similar formula is 
used by Ref.~\cite{Ruffert97}. 
Results obtained from those prescriptions have been compared 
with a neutrino flux-limited diffusion transport scheme discussed in 
Ref.~\cite{Baumgarte1996}, providing also a calibration of the geometry 
parameter, which takes a value $\mathcal{D} = 3$.
Ref.~\cite{Rosswog:2003b} also proposes an alternative recipe 
given by
\begin{align}
  \label{ross_opcty}
  t_{_I} (E_{_I}) &= 
  \lambda_{_I}(E_{_I}) \left(\tau_{_I}(E_{_I}) + \frac{1}{2} \right)
  \left( 2\tau_{_I}(E_{_I}) +1\right)\,.
\end{align}
In practice, it is hard to assess which of the two prescriptions is the
most realistic one.
The first one has the advantage that the energy dependency can be factored
out, which will be crucial for the NLS described below.
Considering the approximate nature of the NLS and
the fact that the coefficient $\mathcal{D}$ is itself determined
empirically, we use hereafter only the
prescription~\eqref{eq:t_diff_janka}.

It should also be noted that the time scale over which our simulations are
performed is typically at least 1 order of magnitude smaller than the
typical diffusion time scale on which the neutrinos escape from the NS
dense core, \ie $\gtrsim 1$~s. As a result, the contribution of the
diffusive leakage of neutrinos is expected to be much smaller than the
abundant emission of neutrinos at densities below the ones of the neutrinosphere. 
For this reason, the NLS can be considered as a reasonable first
approximation to treat neutrino effects without resorting to more
sophisticated transport schemes.

As shown in Appendix~\ref{app_a}, we can separate algebraically the
neutrino energy dependence from the mean free path, introducing
an inverse mean free path $\zeta_{_I}$ which does not depend on the 
energy, \ie
\begin{equation}
  \zeta_{_I} \equiv \frac{1}{E_{_I}^2\lambda_{_I}}\,.
\end{equation}
Similarly, we can define an energy-independent optical depth $\chi_{_I}$
as
\begin{equation}
\label{eq:tau_I}
\tau_{_I}(E_{_I}) = E_{_I}^2 \chi_{_I} \equiv E_{_I}^2\int_{x_1}^{x_2}
\zeta_{_I}(x)~ds \,.
\end{equation}
Using \Eref{eq:t_diff_janka}, we can now also factor
out the energy dependence from the diffusion time scale.
With these definitions, we can finally compute the 
\emph{diffusive} number and energy emission rates per baryon as 
\begin{align}
  \label{eq:RD}
  R^{^D}_{_I}(\eta_{_I}^{eq})  
  &= \frac{\zeta_{_I}}{\mathcal{D} \chi_{_I}^2} 
     \int_0^\infty \frac{n_{_I}(E)}{E^2}\, dE 
  \,,\\
  \label{eq:QD}
  Q^{^D}_{_I}(\eta_{_I}^{eq})  
  &= \frac{\zeta_{_I}}{\mathcal{D} \chi_{_I}^2} \int_0^\infty \frac{n_{_I}(E)}{E}\, dE
  \,,
\end{align}
where, we recall, we assume the neutrino number densities $n_{_I}(E)$ to
be described by the Fermi-Dirac distribution for ultrarelativistic
particles. The integrals over energy can then be carried out analytically.
The diffusive emission rates~\eqref{eq:RD} and \eqref{eq:QD}
approximate the real emission rates only in the diffusive regime.
We will use them below in a prescription interpolating between optically 
thin and optically thick regimes.

The assumption of chemical equilibrium also impacts the opacities and the
radii of the neutrinospheres, which are in general slightly overestimated
in our approach. More specifically, as the region transparent to
neutrinos is approached, the chemical potentials tend to vanish as a
result of the decreased interaction of the nuclear matter with the
neutrino flow. Clearly, this behavior cannot be reproduced when adopting
the assumption of chemical equilibrium and in Ref.~\cite{Ruffert97} this
problem was overcome by using an iterative procedure that adjusted the
chemical potential of the neutrinos between the two regions. Here we
follow instead the approach suggested more recently in
Ref.~\cite{Rosswog:2003b} and use a simple interpolation of the emission
rates at chemical equilibrium between the purely diffusive regime and the
free-streaming one. This approach simplifies considerably the numerical
scheme, allowing us to tabulate all the emission rates and opacities, leaving
the computation of the optical depth as the only operation to be
performed during the evolution (see Appendix~\ref{app_a} for details).

As a result, using the \emph{free} 
neutrino emission rate per baryon, $R^{^F}_{_I}$, and neutrino luminosity
per baryon, $Q^{^F}_{_I}$, and the corresponding
quantities for the \emph{diffusive} emission $R^{^D}_{_I}$ and
$Q^{^D}_{_I}$, we can now compute the \emph{effective rates}, $R_{_I}$ and
$Q_{_I}$ via the following interpolation formulas
\begin{eqnarray}
  \label{eq:effective_rates_a}
  R_{_I}(\eta_{_I}) &=&  R^{^F}_{_I}(\eta_{_I}^{eq})
\left(1+\frac{R^{^F}_{_I}(\eta_{_I}^{eq})}{R^{^D}_{_I}(\eta_{_I}^{eq})}\right)^{-1}\,,\\
  \label{eq:effective_rates_b}
  Q_{_I}(\eta_{_I}) &=&  Q^{^F}_{_I}(\eta_{_I}^{eq})
\left(1+\frac{Q^{^F}_{_I}(\eta_{_I}^{eq})}{Q^{^D}_{_I}(\eta_{_I}^{eq})}\right)^{-1}\,.
\end{eqnarray}
Alternatively, we have also implemented a more elaborated
interpolation technique as suggested in~\cite{Sekiguchi2010}
\begin{eqnarray}
  R_{_I}(\eta_{_I}) &=&  R^{^F}_{_I}(0)e^{- \tau_{_I}}+R^{^D}_{_I}(\eta_{_I}^{eq})\ \left( 1 - e^{- \tau_{_I}}\right)\,,\\
  Q_{_I}(\eta_{_I}) &=&  Q^{^F}_{_I}(0)e^{- \tau_{_I}}+Q^{^D}_{_I}(\eta_{_I}^{eq})\ \left( 1 - e^{- \tau_{_I}}\right)\,.
\end{eqnarray}
This new prescription does not seem to lead to significant changes but we
reserve a more careful comparison of the two prescriptions to future
work.

After computing the effective emission rates, we can define a quantity
that will be very useful as a diagnostic tool to estimate where the
neutrino cooling is most effective. 
\referee{In particular, for each neutrino
species, we define as the \emph{neutrinosphere} the surface at which}
\begin{equation}
  \label{eq:last_int_surf}
  \frac{Q_{_I}}{Q^{^F}_{_I}}=\frac{2}{3}\,.
\end{equation}
Clearly, inside this region the neutrino cooling is strongly suppressed
and the time scale of the emission is dominated by the diffusion
processes.  \referee{Note that other definitions are possible and
  sometimes the neutrinosphere is defined as the surface at which the
  spectrally averaged optical depth $\tau$ is equal to $2/3$ (see, for
  instance, \cite{Rosswog:2003b}).}

Finally, the interpolated quantities can then be used to compute the
net emission rate per baryon appearing in Eq.~\eqref{ye_m_evol}
\begin{align}
& \mathcal{R} \equiv R_{\nua}(\eta_{\nua}) - R_{\nue}(\eta_{\nue}) = 
R_{pc}(\eta_{\nua}) - R_{ec}(\eta_{\nue}) \,,
\end{align}
and the luminosity per baryon $Q$ appearing in
Eqs.~\eqref{energy_density_evol}--\eqref{momentum_evol}
\begin{align}
\begin{split}
\label{eq:Q_full}
 \mathcal{Q} \equiv \sum_I  Q_{_I}(\eta_{_I})  &= 
Q_{pc}(\eta_{\nua}) + Q_{ec}(\eta_{\nue}) + \\
& \qquad \sum_I  \left[ Q_{e^+e^-}(\eta_{_I}) + 
Q_{\gamma}(\eta_{_I}) \right]\,,
\end{split}
\end{align}
where $R_{pc}(\eta_{\nua}), Q_{pc}(\eta_{\nua})$ are the emission rates
from proton capture ($pc$), $R_{ec}(\eta_{\nua}), Q_{ec}(\eta_{\nua})$
the emission rates from electron capture ($ec$), while
$Q_{e^+e^-}(\eta_{_I})$ and $Q_{\gamma}(\eta_{_I})$ are the emission
rates per unit mass from pair annihilation and plasmon decay,
respectively.  For the sum over these last two rates, here our notation
is compact but not perfect; indeed the term $\sum_I
Q_{e^+e^-}(\eta_{_I})$ in Eq. \eqref{eq:Q_full} should be read as
\begin{equation}
\sum_I Q_{e^+e^-}(\eta_{_I}) = 
Q_{e^+e^-}(\eta_{\nue},\eta_{\nua}) +
Q_{e^+e^-}(\eta_{{\nu}_{\tau,\mu}},\eta_{\bar{\nu}_{\tau,\mu}})\,, 
\end{equation}
and the same for the plasmon-decay rates.

While the assumption of chemical equilibrium can be considered natural in
the view of the many approximations already taken in a NLS, we should
also caution that this assumption could overestimate considerably the
emission rates, especially in the transition between the optically thin
and thick region.

\referee{The effective local neutrino source terms are also used to
  estimate the total luminosity as the neutrino energy that is emitted
  between the coordinate times $t$ and $t+dt$, where the neutrino energy
  emitted at each point is the one measured by a local 
  \emph{coordinate observer} with
  worldline tangent to the four-vector $t^\mu = \delta^\mu_0$. This
  observer is somewhat arbitrary, unless the spacetime is stationary with
  Killing vector $t^\mu$. In this case, and assuming no further
  interactions, the energy of a neutrino measured by such observer
  $E_\nu$ is related to the energy measured when the neutrino reaches
  infinity $E^\infty_\nu$ by the simple relation $E^\infty_\nu =
  \sqrt{-g_{00}} E_\nu$, regardless of the direction of the emission.  To
  compute the local luminosity, we use the effective emission rates,
  which are such that the contributions from the optically thick regions
  are strongly suppressed.  The emission rates are given in the fluid
  rest frame, where a 4-momentum $dp^\mu = \mathcal{Q} u^\mu d\tau$ is
  emitted isotropically per baryon and unit proper time $d\tau$.  Such a
  4-momentum $dp^\mu$ corresponds to an energy $dp^\mu t_\mu /
  \sqrt{-t^\nu t_\nu}$ which is measured by our coordinate observers. The
  relation between the fluid proper time and the coordinate time is $dt =
  d\tau u^0 = d\tau W / \alpha$, and the number of baryons per coordinate
  volume is $\bar{D} m_b^{-1}$. Using these expressions, we define for
  each neutrino species $\nu_{_I}$ the ''source luminosity'' $L_{_I}$, 
  which does not take into account the gravitational redshift, as
\begin{align}\label{eq:L_Is}
L_{_I} &= 
\int \tilde{f} Q_{_I} \bar{D} m_b^{-1} \alpha \,d^3x\,,
\end{align}
where the integral is taken over the whole computational domain, and
$\tilde{f} \equiv (\alpha - \beta^i v_i) / \sqrt{g_{00}}$.  In addition,
we can compute the luminosity observed at infinity, $L^\infty_{_I}$, by
replacing $\tilde{f}$ in Eq.~(\ref{eq:L_Is}) with $\tilde{f}
\sqrt{g_{00}}$, so as to account for the gravitational redshift.  The
expression for $L^\infty_{_I}$ is strictly valid only for a stationary
spacetime with Killing vector $t^\mu$. For our nonrotating stationary
stellar models, $\beta^i=0$ and thus $\tilde{f}=1$. We will adopt this
expression also for oscillating NSs, in which case the dominant error is
caused by the fact that a significant fraction of the emitted neutrinos,
even those emitted in the free-streaming regime, will hit the NS core and
never reach infinity. This could easily reduce the luminosity by a factor
around two.  }

\section{Nuclear equations of state for neutron-star matter}
\label{sec:eos}

The importance that the EOS has in our understanding of the properties of
NSs is so large that its study requires no justification. Besides
determining the composition and structure of NSs, the EOS also plays a
fundamental role in defining the properties of the GW signal that is
expected to be produced by NSs, either when isolated or in a binary
system. A number of studies about the inspiral and merger of BNS have
made this point very clearly, both when the EOS was idealized and
analytic, \eg \cite{Baiotti07, Baiotti08, Rezzolla:2010}, or when more
realistic ones were considered, \eg \cite{Rosswog99, Shibata05c,
  Bauswein:2010dn}. Unfortunately, however, our knowledge of matter at
nuclear densities is still plagued by too many uncertainties, making it
impossible to derive, from robust first principles, what is the most
realistic EOS for NS matter. These uncertainties leave ample room for a
large variety of possible models for the nuclear matter at zero
temperature, which is only mildly constrained by astronomical
observations.

\begin{figure*}
  \begin{center}
    \includegraphics[width=0.45\textwidth]{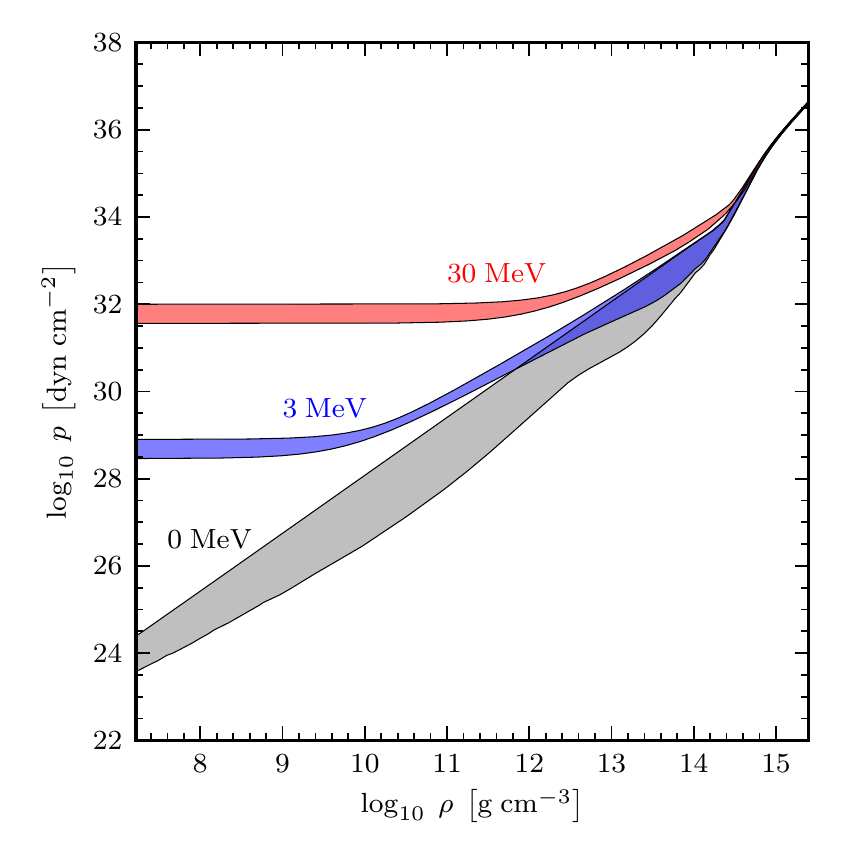}
    \hskip 1.0cm
    \includegraphics[width=0.45\textwidth]{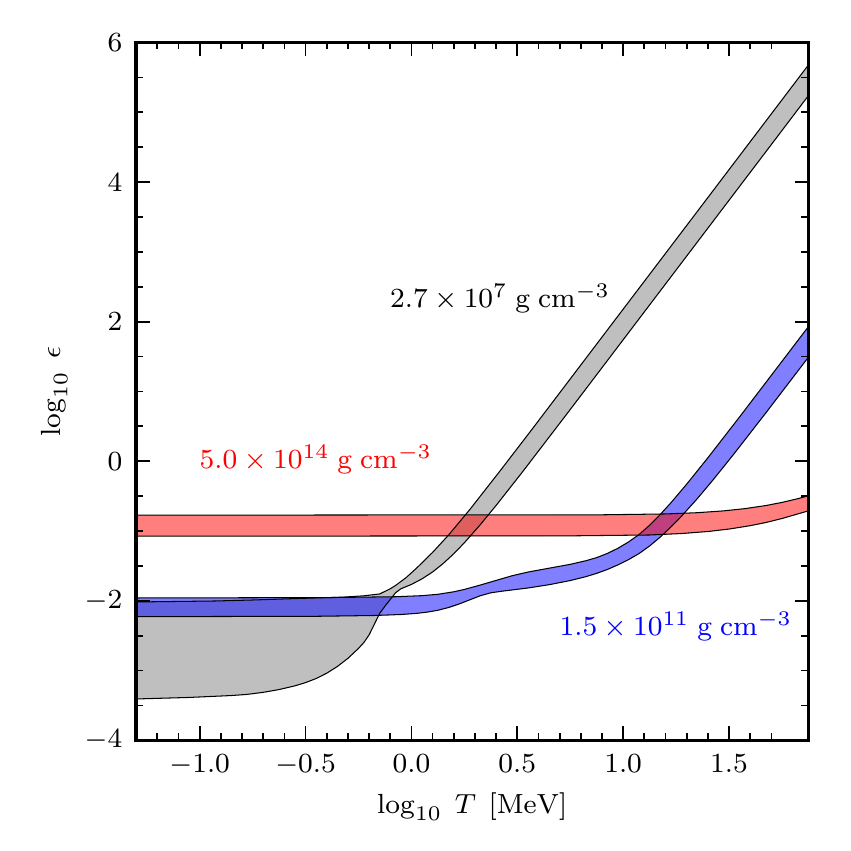}
    \caption{Left panel: Dependence of the pressure on the
      rest-mass density for the SHT-EOS. Right panel: Dependence
      of the internal energy on the temperature for the same EOS. The
      different colors refer to the different cuts of the table at
      constant temperature (left panel) or at constant rest-mass density
      (right panel). In both panels the shaded colored regions show the
      span of the variation in the pressure/energy density for different
      values of the composition.}
    \label{fig:Composition} 
  \end{center}
\end{figure*}

In practice, however, only a few nuclear EOSs are able to encompass the
wide range of temperatures, compositions and rest-mass densities
necessary to describe matter in supernova explosions and BNS
mergers. Here, we will make use of two nuclear EOSs for hot dense matter:
the Lattimer-Swesty (LS) EOS \cite{Lattimer91}, that adopts a
compressible liquid drop model with Skyrme interaction
\cite{Tondeur1984}, and the Shen-Horowitz-Teige (SHT) EOS, that adopts a
relativistic mean-field model for uniform matter with a modified NL3 set of
interaction parameters (see Refs. \cite{ShenG2010,ShenG2011}). 
These two EOSs do not include contributions from exotic matter like quarks, 
hyperons, and kaon or pion condensates. Furthermore, they assume charge neutrality.

In these publicly available EOSs, the pressure $p$ and the specific
internal energy $\epsilon$ are given as a function of three independent
variables: the baryon number density $n_b$, the electron/proton fraction
$\Efr$, and the temperature $T$. The SHT-EOS is given in a tabulated
form, while the LS-EOS is provided in form of computer routines, which
provide several choices for the incompressibility modulus, namely $K =
180, 220, 375\usk\mega\electronvolt$. For our simulations we concentrate
on a value $K = 220 \usk\mega\electronvolt$ and cast also the LS-EOS into
a tabular form using the routines described in Ref.~\cite{OConnor10}.
Note that the original table for the SHT-EOS has a fairly low resolution
and we compensate this by computing a larger table via a
thermodynamically consistent interpolation of the Helmholtz free energy
\cite{Swesty1996,Timmes2000}. \referee{In a separate table the SHT-EOS
  provides also the $T=0.0 \usk\mega\electronvolt$ slice. Since we store
  the logarithm of the temperature, we cannot use directly this slice, but
  we use it to extend the original temperature range via linear
  interpolation.}  In Table \ref{TAB:eoss}, we summarize the ranges
covered by the two EOS tables used in our simulations and their
resolution.

When it comes to the range in composition, the SHT-EOS can describe the
properties of pure neutron matter, that is $\Efr = 0$, but this case is
provided as a separate table. To compute a regularly spaced table
covering the full range, we fill the gap into the original table using a
third-order polynomial interpolation. For the LS-EOS, on the other hand,
the smallest available value for the electron fraction is $\Efr = 0.035$
and is taken from the table described in Ref.~\cite{OConnor10}, and as
produced by the routines provided by the LS-EOS. The regime of low
electron fraction becomes relevant for NSs at low temperature and
$\beta$-equilibrium, which show a pronounced dip in the electron fraction
at the rest-mass densities typical of the inner crust \ie $\rho \approx
10^{12}$--$10^{14} ~{\rm g~cm}^{-3}$.

\begin{table}
\begin{ruledtabular}
  \begin{tabular}{l|cc} 
    {EOS} & {SHT-NL3} & {LS-220}\\
    \hline
    $\rho_{\mathrm{min}}~
    [{\rm g~cm}^{-3}]~~~$ &  $1.65\times{10^{5}}$& $1.56\times{10^{5}}$\\
    $\rho_{\mathrm{max}}~
    [{\rm g~cm}^{-3}]~~~$&  $2.48\times{10^{15}}$& $2.43\times{10^{15}}$\\
    Number of points in $\log_{10}(\rho)$ & $328$ & $223$\\
    $T_{\mathrm{min}}~
    \usk[\mega\electronvolt]$ & $0.05$ & $0.01$\\
    $T_{\mathrm{max}}~
    \usk[\mega\electronvolt]$ & $75.0$ & $250.0$\\
    Number of points in $\log_{10}(T)$ & $128$ & $136$\\
    $Y_{e,\mathrm{min}}$ &$0.0$ &
    $0.035$\\
    $Y_{e,\mathrm{max}}$ &$0.56$ &
    $0.55$\\
    Number of points in $\Efr$ & $57$ & $50$
  \end{tabular}
\end{ruledtabular}
  \caption{Temperature, density and electron fraction ranges for the two
    EOS tables used in our simulations, as well as the number of entries
    for each quantity.}
  \label{TAB:eoss} 
\end{table}

We note that the original table for the SHT-EOS contains some problematic
regions where derived quantities, such as the speed of sound $c_s$, are
extremely noisy or even unphysical. Since our numerical evolution scheme
requires a well-behaved sound speed, an additional smoothing step was
necessary to remove these irregularities. In particular, we compute the
sound speed at constant composition as
\begin{align}
  \begin{split}
    c_s^2  =  \frac{1}{h} &\left[ 
      \tpd{p}{\rho}{T, \Efr} 
        + \frac{p}{\rho^2} \tpd{p}{T}{\rho,\Efr} \left(
        \tpd{\epsilon}{T}{\rho,\Efr} \right)^{-1}\right.
      \\ &\quad \left.
      - \tpd{p}{T}{\rho,\Efr} \tpd{\epsilon}{\rho}{T,\Efr}
        \left( \tpd{\epsilon}{T}{\rho,\Efr} \right)^{-1} \right]\,,
  \end{split}
\end{align}
where the required derivatives are computed using the limiting procedure
described in \cite{Steffen1990}. On the other hand, the sound speed for
the LS-EOS can be computed analytically from the derivatives of the
Helmholtz free energy, yielding smoother results.
  
\begin{table*}
  \begin{ruledtabular}
    \begin{tabular}{lccccccccccccccc}
      Model & 
      EOS & 
      $T$ &
      $s$ &  
      $\rho_{c}$ & 
      $M_\text{b}$ & 
      $M$ &
      $R$ & 
      \multicolumn{2}{c}{$F$} & 
      \multicolumn{2}{c}{$H_1$} & 
      \multicolumn{2}{c}{$H_2$} &  
      Spacetime &
      $E_{\nu}$ \\
      &     
      & 
      $[\mega\electronvolt]$ &
      $[k_{_{B}}]$ &
      $[{\rm g~cm}^{-3}]$ & 
      $[M_{\odot}]$ & 
      $[M_{\odot}]$ &
      $[\kilo\meter$] & 
      \multicolumn{2}{c}{$[\kilo\hertz]$} & 
      \multicolumn{2}{c}{$[\kilo\hertz]$} & 
      \multicolumn{2}{c}{$[\kilo\hertz]$} & 
      &
      $[{\rm erg}]$ \\
      \hline
      \texttt{sTOV-CW1} &
      LS-220-$\beta$ & 
      $0.01$  & -- &
      $0.930 \times{10^{15}}$ & 
      $1.898$ & 
      $1.690$ & 
      $12.40$ & 
      $3.95$ & $3.89$ & 
      $6.85$ & $6.81$ & 
      $9.78$ & $9.72$ & 
      fixed &
      --    \\
      \texttt{sTOV-FW1} &
      LS-220-$\beta$ & 
      $0.01$  & -- &
      $0.930 \times{10^{15}}$ & 
      $1.898$ & 
      $1.690$ & 
      $12.40$ & 
      $2.39$ & --  & 
      $6.09$ & --  & 
      -- & --  & 
      evolved &
      -- \\
      \texttt{sTOV-CW2} &
      SHT-$\beta$ & 
      $0.05$ & -- &
      $0.930 \times{10^{15}}$ & 
      $3.268$ & 
      $2.732$ & 
      $13.85$ & 
      $3.55$ & $3.49$ & 
      $5.91$ & $5.86$ & 
      $8.34$ & $8.24$ & 
      fixed &
      --    \\
      \hline
      \texttt{sTOV-SHT}  &
      SHT-$\beta$    & 
      --  & $1$ & 
      $0.930 \times{10^{15}}$ & 
      $3.260$ & 
      $2.741$ & 
      $14.00$ & 
      $3.53$ & $3.42$ & 
      --     & $5.73$ & 
      --     & $8.01$ & 
      fixed & 
      $1.68 \times{10^{51}}$\\
      \texttt{uTOV-LS} &
      LS-220-$\beta$ & 
      -- & $1$ & 
      $2.200 \times 10^{15}$ & 
      $2.937$ & $2.218$ & 
      $11.73$ & --
      & --    & --
      & --    & --
      & -- & 
      evolved & 
      $5.34 \times{10^{49}}$\\
      \texttt{uTOV-SHT}  &
      SHT-$\beta$    & 
      -- & $1$ & 
      $1.300 \times{10^{15}}$ & 
      $3.259$ & $2.743$ & 
      $12.97$ & --
      & --    & --
      & --    & --
      & -- & 
      evolved & 
      $4.49 \times{10^{49}}$\\
      \texttt{mTOV-SHT}  &
      SHT-$\beta$    & 
      -- & $1$ & 
      $2.200 \times 10^{15}$ & 
      $2.832$ & $2.485$ & 
      $11.56$ & --
      & --    & --
      & --    & --
      & -- & 
      evolved & 
      $5.61 \times{10^{52}}$\\
      \texttt{uTOVh-SHT}  &
      SHT-$\beta$    & 
      $30.0$ & -- & 
      $1.800 \times 10^{15}$ & 
      $3.018$ & $2.616$ & 
      $14.90$ & --
      & --    & --
      & --    & --
      & -- & 
      evolved & 
      $3.30 \times{10^{51}}$\\
    \end{tabular}
     \end{ruledtabular}
   \caption {Parameters of the TOV models evolved in the numerical tests
     discussed in Sec.~\ref{sec:tests}. For the stable models we also
     show the frequencies of the fundamental radial mode, first and
     second overtones ($F$, $H_1$ and $H_2$ respectively), extracted from
     the evolution of the central rest-mass density (first value), and
     computed with a linear-perturbation code (second value). $E_{\nu}$
     is the total energy radiated by neutrinos during the
     simulations.}
     \label{tab:freq} 
\end{table*}

Both the LS-EOS and the SHT-EOS tables provide the necessary
microphysical quantities to compute the neutrino interactions such as the
nucleon and electron chemical potentials, the abundances of heavy nuclei
and alpha particles, the proton to nucleon ratio $Z/A$, and the atomic
mass number $A$ for the ''representative'' heavy nuclei.

Fig.~\ref{fig:Composition} illustrates the dependence of the pressure and
specific energy as a function of the three independent variables
$(\rho,\Efr,T)$ for the SHT-EOS (the LS-EOS shows a similar behavior).
The left panel shows the pressure as a function of the rest-mass density
for three different temperatures (distinguished by different colors),
while the right panel shows the specific internal energy as a function of
the temperature for three different densities. Both panels report the
influence of the electron fraction as shaded regions that represent the
variation over the whole range of $\Efr$. As one can see, the influence
of the composition on the pressure is quite large at low temperatures and
densities below the crust-core transition, \ie at $\rho_{0} \lesssim 2.5
\times 10^{14}~{\rm g~cm}^{-3}$. In the core region, on the other hand,
the pressure is relatively unaffected by changes in composition and
temperature. Note also that the specific internal energy becomes almost
temperature independent at densities larger than nuclear saturation
density, in stark contrast with simplified analytic EOSs such as the
$\Gamma$-law EOS.

We recall that the numerical scheme we use to solve the
relativistic-hydrodynamic equations provides us at each time step with
the rest-mass density $\rho$, the electron fraction $\Efr$, and internal
energy $\epsilon$, but not the temperature $T$. The EOS tables, on the
other hand, use $\rho, \Efr$ and $T$ as independent variables. As a
result, a conversion is needed between $\epsilon$ and $T$ for any given
couple of $(\rho, \Efr)$. This is done during the evolution by a discrete
bisection algorithm that finds the nearest tabulated temperatures and
then performs a linear interpolation.

At regions with high density but low temperature, the conversion becomes
very inaccurate because of the very weak dependence of $\epsilon$ on $T$
(\cf right panel of Fig.~\ref{fig:Composition}), with the consequence
that small errors in $\epsilon$ can be strongly amplified. Fortunately,
this does not affect the hydrodynamic evolution since the pressure shows
the same weak dependence on temperature. On the other hand, the local
neutrino emission rates do depend strongly on the temperature, \ie as
$T^6$, so that the free and the effective emission rates can be computed
with a large relative error. Luckily, also in this case the errors do
not have a serious impact because under these high-density conditions the
neutrinos are trapped and the resulting effective luminosity is small. As
a result, the influence of inaccurate emission rates on the total
luminosity is small as long as the latter is dominated by the emission
from hot matter outside the optically thick region.

\section{Numerical Framework}
\label{section:numframe}

Much of our numerical infrastructure has been discussed in other papers,
\eg \cite{Baiotti08,Rezzolla:2010}, to which we refer the interested
reader for details. Below, we only give a brief overview on the numerical
methods used. For the time integration of the coupled set of the
hydrodynamic and Einstein equations we use the method of lines (MOL) in
conjunction with an explicit fourth-order Runge-Kutta method. In all our
simulations we prescribe a Courant-Friedrichs-Lewy (CFL) factor of $0.3$
to compute the time step. The Einstein equations are spatially
discretized using a fourth-order finite-difference operator, implemented
by the publicly available \texttt{McLachlan} code. The hydrodynamic
equations, on the other hand, are discretized in space using a
finite-volume scheme based on a piecewise parabolic reconstruction
(PPM)~\cite{Colella84} and the Harten-Lax-van Leer-Einfeldt
(HLLE)~\cite{Harten83} approximate Riemann solver. Differently from the
original PPM implementation reported in Ref.~\cite{Baiotti07}, we
reconstruct here the quantities $W v^i$ instead of the three-velocities
$v^i$. This guarantees that the velocities reconstructed at the cell
boundaries remain subluminal even under extreme conditions like those
encountered near the central region of a BH. As is customary when solving
the relativistic-hydrodynamic equations, the vacuum regions are treated
by enforcing an artificial atmosphere with a rest-mass density and
temperature close to the lowest values covered by the EOS considered (see
Table~\ref{TAB:eoss}) and with a constant electron fraction of $\Efr =
0.4$.

A discussion on the global convergence order of the simulations reported
here is presented in Appendix~\ref{sec::convergence} and is generally of
order $\sim 1.7$--$1.8$, in agreement with what was found also in previous
calculations using the same code with ideal-fluid
EOS~\cite{Baiotti:2009gk}.

\subsection{Adaptive-mesh refinement and singularity handling}
\label{sec:NumericalSpecifications}

The use of mesh-refinement techniques is of fundamental importance in our
simulations. For this reason, we use the \texttt{Carpet} driver that
implements a vertex-centered adaptive-mesh-refinement scheme adopting
nested grids~\cite{Schnetter-etal-03b}. For the evolution of isolated NSs
presented in this work, it is sufficient to use a fixed hierarchy of
nested boxes centered around the origin, with a 2:1 refinement factor
between successive grid levels.

In some of the simulations presented here, the final state of the
evolution is a BH. We then use the isolated horizon
framework~\cite{Thornburg2003:AH-finding_nourl} to measure its properties
every few time steps. We do not make use of the excision
technique~\cite{Baiotti04b} in our simulations, neither for the spacetime
variables nor for the fluid. As described in
Refs.~\cite{Baiotti06,Baiotti07}, in order to extend the simulations well
past the BH formation, we add a small amount of dissipation to the
evolution equations for the metric and gauge variables and rely on the
singularity-avoiding gauge~\eqref{eq:one_plus_log} (note that no
dissipation is added to the evolution of matter variables). More
specifically, we use an artificial dissipation of the Kreiss-Oliger-type~\cite{Kreiss73}, 
\ie we augment the right-hand side of the evolution
equation of a given quantity $u$ with a term ${\cal L}_\text{diss} u =
-\varepsilon \Delta^3 \partial^4_i u$, where $\Delta$ is the grid
spacing. The dissipation is active on the whole computational domain
with constant strength $\varepsilon=0.1$. The use of a numerical
viscosity is necessary because all the field variables develop very steep
gradients in the region near the BH center. Under these conditions, small
high-frequency oscillations can lead to instabilities if not dissipated.

\subsection{optical depth grid}\label{sec:optical_depth}

Different prescriptions are possible in order to estimate the optical
depth that is needed in the NLS. A simple choice would be to perform the
integrals \eqref{eq:tau_I} along the Cartesian coordinate lines and then
take the minimum value over all directions. A better choice is to
calculate the integrals along directions that are better adapted to the
underlying geometry of the matter distribution. In this respect, since we
are mainly interested in objects with an overall spherical geometry, a
spherical polar coordinate system is better suited than a Cartesian one.
Hence, we introduce a spherical grid and compute along a number of a
radial coordinate directions the integrals of the energy-independent
optical depths $\chi_{_I}$, \ie
\begin{align}
  \chi_{_I}(r,\theta,\phi) &= 
  \int^{r_\text{max}}_{r} \zeta_{_I} (r',\theta,\phi)\sqrt{\gamma_{rr}} dr' \,.
  \label{eq:radial_opt_depth}
\end{align}
where $\zeta_{_I}$ is the energy-independent part of the inverse mean
free path (given in Appendix~\ref{app_a}). The spherical grid is uniform
and we interpolate the three-metric and the opacities $\zeta_{_I}$ from
the Cartesian grid hierarchy onto the spherical grid. Next, we integrate
from an outer radius $r_\text{max}$, which is large enough to ensure that
the region further out is always optically thin, down to $r=0$. As we
march inwards, we store the result of the integrals on the various radial
grid points as the intermediate values can be useful to determine, for
example, the location of the neutrinospheres. Thanks to this reversed
integration direction, it is not necessary to compute the integrals
(\ref{eq:radial_opt_depth}) from each point to $r=r_\text{max}$
separately, but only once from $r=r_\text{max}$ to $r=0$ for each angular
coordinate on the grid. Once the energy-independent optical depths are
known for each point on the spherical grid, they are linearly
interpolated back onto the Cartesian grids, where they are used to
compute the effective emission rates from the local emission rates
according to the NLS prescription (see the discussion in the previous
Section).

\begin{figure*}[t]
  \begin{center}
    \includegraphics[width=0.45\textwidth]{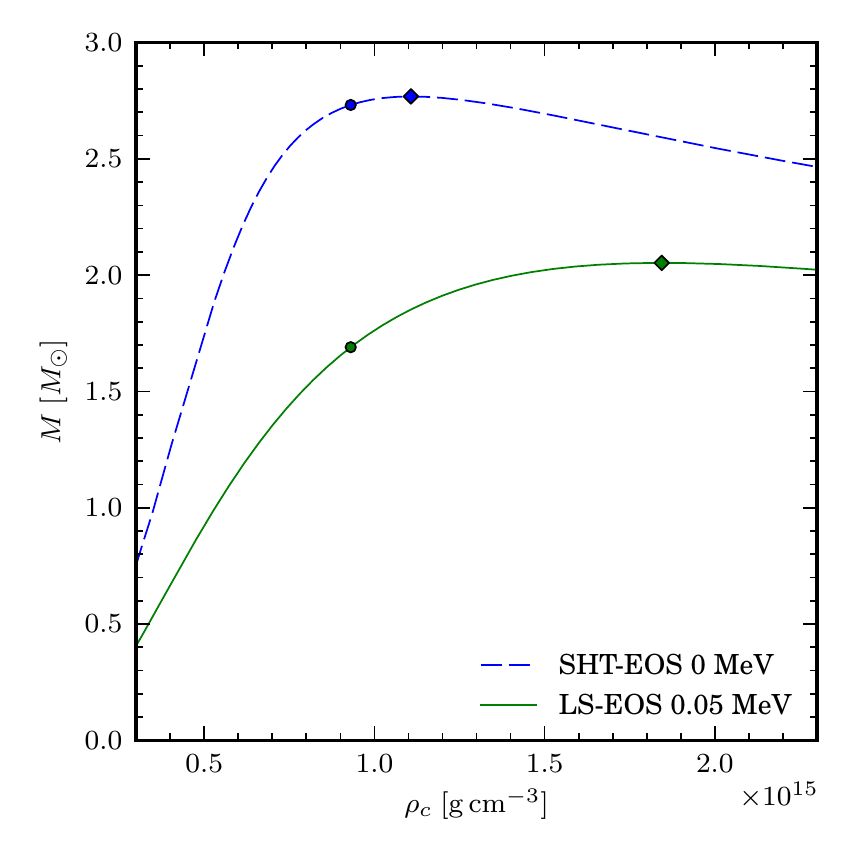}
    \hskip 1.0cm
    \includegraphics[width=0.45\textwidth]{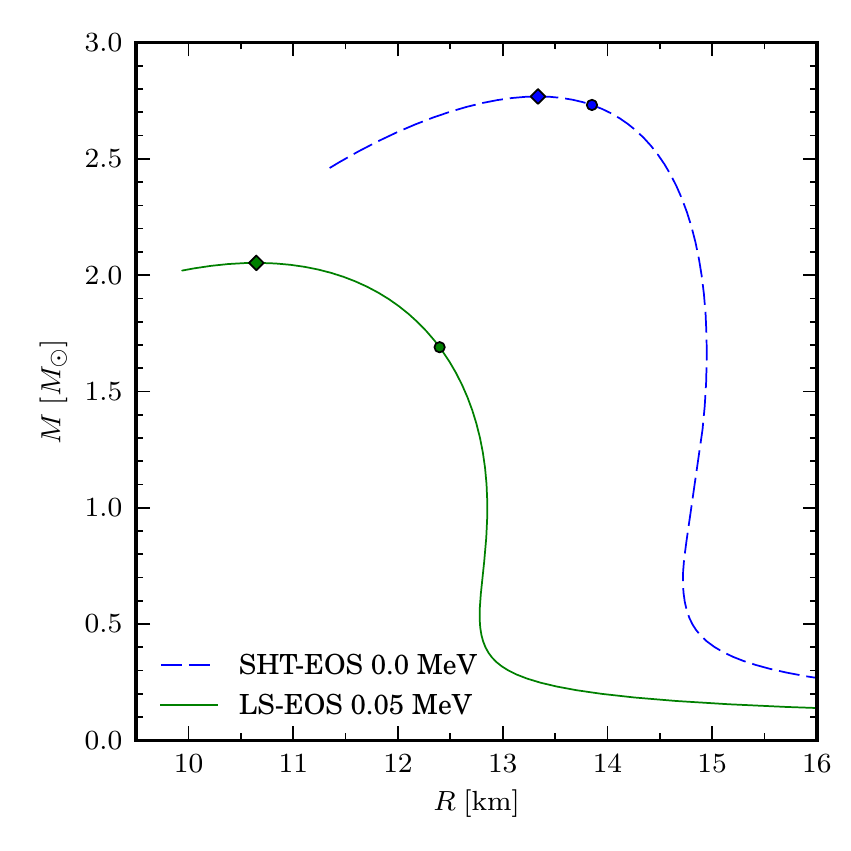}
    \caption{Left panel: Gravitational mass versus central
      density of TOV star sequences computed using the SHT-EOS at
      $T{=}0 \usk\mega\electronvolt$ (blue line) and the LS-EOS at
      $T{=}0.01 \usk\mega\electronvolt$ (green line) imposing the
      $\beta$-equilibrium condition. The models with the maximum
      gravitational mass are indicated with a diamond while dots mark the
      models described in Sec.~\ref{sec:cold_stars}. Right
        panel: Mass versus circumferential radius of the same
      configurations.}
    \label{fig:TOVMeV} 
  \end{center}
\end{figure*}

To reduce computational costs and simplify the implementation, the
procedure described above is performed only at each evolution step of the
coarsest refinement level. In Appendix~\ref{sec::convergence} we show
that the error caused by the assumption of slowly varying optical depth
is negligible for the case of a collapsing NS. In our implementation, the
radial spacing of the spherical grid roughly coincides with the spacing
on the finest Cartesian grid, and the spherical grid is large enough to
cover entirely the finest Cartesian grid. We note that although it would
be easy to use a nonregular grid spacing which is much finer near the
radiating regions, the accuracy would still be limited by the resolution
of the Cartesian grids. Furthermore, the resolution of the spherical
grid is essentially limited by two factors. First, by the scalability to
a large number of CPUs, since only the interpolation back to the
Cartesian grids is implemented in a fully parallel way, while the
interpolation to the spherical grid is not. Second, by the available
memory on each compute node, since the whole spherical grid needs to be 
kept in the memory of each parallel process.


\section{Tests of the NLS: stable and unstable stars}
\label{sec:tests}

In the following we present the results of several tests to assess the
stability and accuracy of our code when implementing the NLS. In these
simulations we cover several astrophysical scenarios from quasiradial
oscillations of nonrotating relativistic stars in fixed spacetimes and in
full general relativity (Sec..~\ref{sec:cold_stars} and
\ref{sec:hot_stars}), to the gravitational collapse of hot NSs to a BH
(Sec.~\ref{sec:hot_collapse}).

\subsection{Cold stars: Radial oscillations}
\label{sec:cold_stars}

We start our battery of tests by considering the radial oscillations of
nonrotating \emph{cold} NSs. 
The cold models are not expected to radiate and are evolved without NLS.
First, we test only the hydrodynamic part by using the so-called
Cowling approximation, that is, the spacetime is not evolved,
but set to the solutions of the Tolmann-Oppenheimer-Volkoff
(TOV) equations~\cite{MTW1973} describing a stationary NS.

\begin{figure*}[t]
  \begin{center}
    \includegraphics[width=0.45\textwidth]{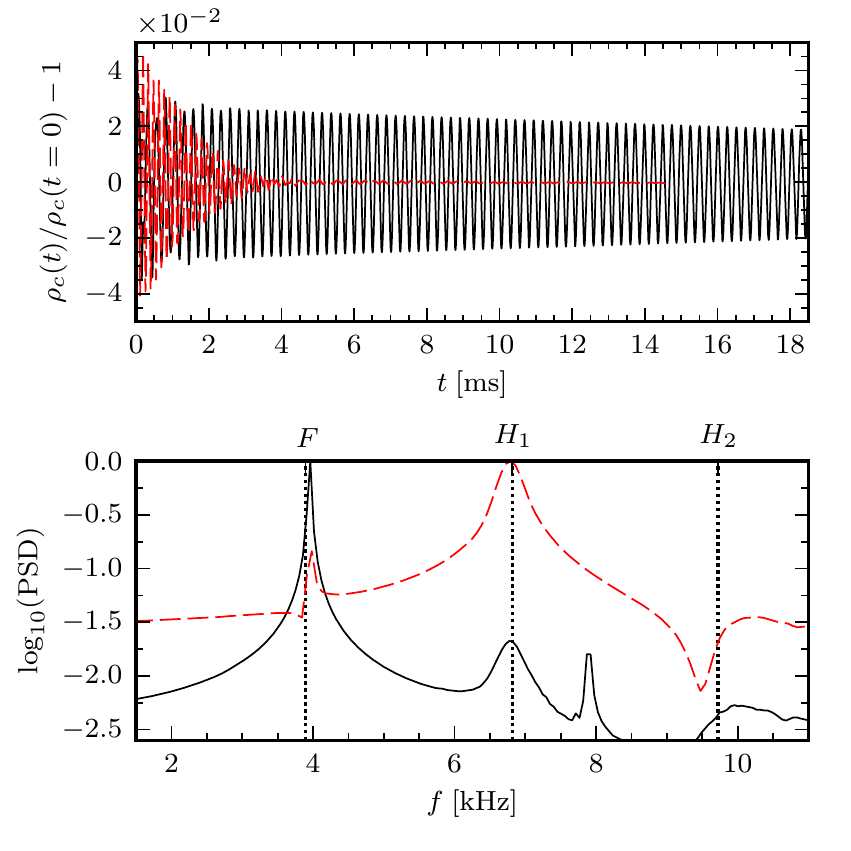}
    \hskip 1.0cm
    \includegraphics[width=0.45\textwidth]{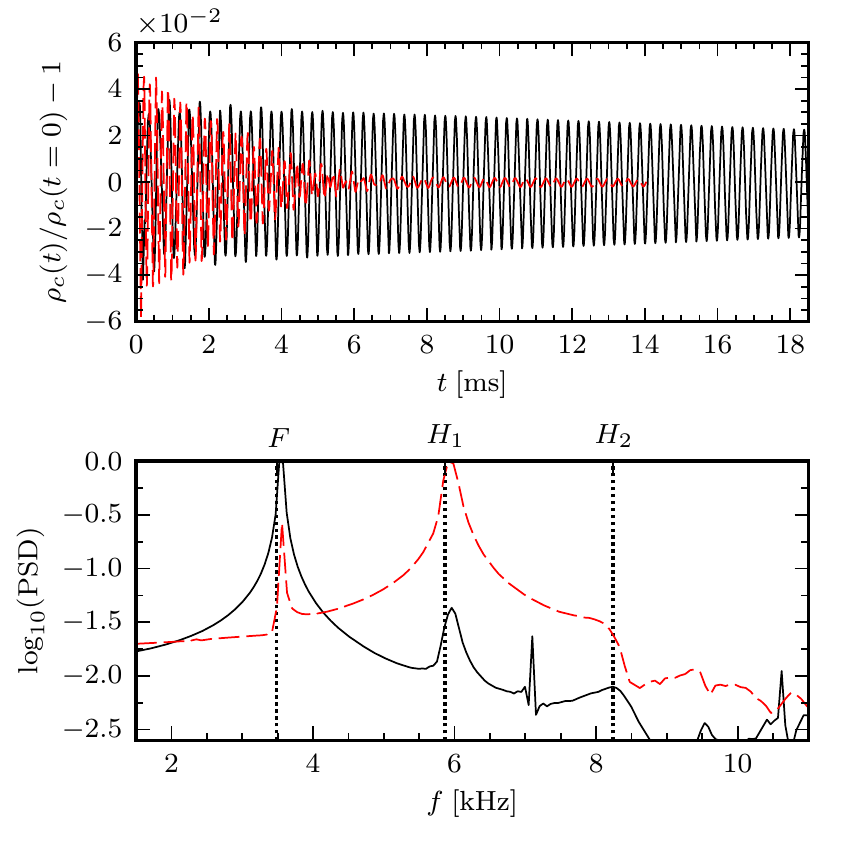}
    \caption{Evolution of stable TOV stars, described in
      Table~\ref{tab:freq}, using the Cowling approximation. Top
        panels: Time evolution of the maximum rest-mass density
      normalized to the initial value $\rho_c$. The left column shows the
      results for the LS-EOS (model \texttt{sTOV-CW1} in
      Table~\ref{tab:freq}), and the right column for the SHT-EOS (model
      \texttt{sTOV-CW2} in Table~\ref{tab:freq}). Bottom panels
      The logarithm of the power spectral density (PSD) normalized to the
      amplitude of the mode used in the initial perturbation. The black
      lines correspond to models initially perturbed with the $F$-mode,
      the red lines to a perturbation with the $H_{1}$-mode.
      The vertical lines show the frequencies expected from linear 
      perturbation theory in the Cowling approximation.}
    \label{fig:fft_rhoc} 
  \end{center}
\end{figure*}

We should remark that the evolution of a NS in hydrostatic equilibrium is
frequently used as a test for general-relativistic-hydrodynamics codes
\citep{Shibata99c,Font02c,Duez:2002bn} and general-relativistic MHD codes
\citep{Giacomazzo:2007ti,Cerda07a}. Despite the relative simplicity and
the well-developed theory of linear-perturbations describing such
configurations, this test in the presence of radiative emissions is
particularly delicate for a nonlinear evolution code like
\texttt{Whisky}, not specifically suited to study equilibrium
configurations as those considered in~\cite{kastaun_2006_hrs}.

We start by analyzing the radial oscillations of NSs initially in
$\beta$-equilibrium with the lowest temperature available in our EOS
tables ($T=0.01 \usk\mega\electronvolt$ for the LS-EOS and $T=0.05
\usk\mega\electronvolt$ for the SHT-EOS ). Integrating the TOV equations
for several central densities, we obtain the relationships between the
mass and the radius, and between the mass and the central rest-mass
density as plotted in Fig.~\ref{fig:TOVMeV}. Note that probably because
of the difference in the symmetry energy between the two EOSs, the
SHT-EOS yields models that are systematically less compact. For example,
for a $1.4 \usk M_{\odot}$ NS model the SHT-EOS gives a radius that is
$15\%$ larger than that of the LS-EOS. Also, because the SHT-EOS is a
particularly ''stiff'' EOS, the corresponding maximum mass is
significantly larger than for the LS-EOS. Interestingly, even for a NS
with $2.8\, M_{\odot}$, the central rest-mass density is only a few times
that of the nuclear-matter density $\rho_{\mathrm{nuc} }= 4 \times
10^{14} ~{\rm g~cm}^{-3}$.

For our tests we choose two TOV models, one for each EOS, situated on the
stable branch with identical central rest-mass density of $\rho_c=9.304
\times 10^{14} ~{\rm g~cm}^{-3}$. Their properties are reported in
Table~\ref{tab:freq} as models \texttt{sTOV-CW1} and
\texttt{sTOV-CW2}. The tests are performed using a grid with two levels
of refinement and a spatial resolution on the finest grid of $\Delta
=0.148 \usk\kilo\meter$, \referee{while the size of
the computational domain is $R_{\rm out} \simeq 26.5 \usk\kilo\meter$}.
In order to cover the star completely with the
finest refinement level, we use 100 points for the model
\texttt{sTOV-CW1} and 120 points for model \texttt{sTOV-CW2}, with the
coarser levels containing the same number of points. 

The initial equilibrium configurations are perturbed through the
injection of the fundamental quasiradial $F$-mode in one case, and of
the first radial overtone $H_1$ in the other. The necessary
eigenfunctions are computed under the assumption that the EOS used to
construct the background model is also valid during the oscillation,
which in our case implies the approximation that the $\beta$-equilibrium
holds in the perturbed model. The amplitude of the initial velocity
perturbation is $\Delta v = 0.01$ and thus still in the linear regime.

In the top panels of Fig.~\ref{fig:fft_rhoc} we show the evolution of the
central rest-mass density, for the two EOS considered, LS-EOS (left
panels) and SHT-EOS (right panels). We observe the presence of a
significant damping in the oscillations, especially for the model
perturbed with the $H_1$-mode (red solid lines in
Fig.~\ref{fig:fft_rhoc}) for which the eigenfunction has a node closer to
the surface of the star when compared to the $F$-mode (black solid lines
in Fig.~\ref{fig:fft_rhoc}). This is a purely numerical effect and can be
mainly attributed to the treatment of the stellar
surface~\cite{kastaun_2006_hrs}. The amplitude of this numerical damping
is comparable with the one found by other codes using similar numerical
methods for the evolution of the hydrodynamic
equations~\cite{Baiotti04,Dimmelmeier06}. The use of an artificial
atmosphere to approximate the vacuum region outside the NS, in fact,
leads to small spurious standing shocks which increase the internal
energy at the surface. This artificially heated region hardly affects the
global dynamics of the star, but it has a sizeable effect in terms of the
total neutrino luminosity.

This is caused by the steep increase of the local neutrino emission rates
with temperature, which we recall are proportional to $T^6$. The surface
heating therefore results in a large and unphysical neutrino emission
from the surface of the NS, which is not suppressed because the affected
regions are optically thin. This problem is somewhat inevitable within a
NLS because of the intrinsic assumption of thermal equilibrium between
nuclear matter and neutrino radiation. In order to estimate the amount
of this emission, we also evolve model \texttt{sTOV-CW1} with activated
NLS. During the evolution, the
maximum neutrino luminosity reaches $10^{48}\, {\rm erg~s}^{-1}$, which
should therefore be considered as the \emph{minimum} luminosity that can
be modeled using our scheme.

After evolving for $20 \usk\milli\second$, we compute the Fourier
transform of the maximum rest-mass density timeseries (lower panels in
Fig.~\ref{fig:fft_rhoc}) to obtain the frequencies of the stellar
pulsations. For each model we also solve the linearized perturbation
equations in the Cowling approximation as an eigenvalue problem in order
to compute the full spectrum of eigenfunctions. In Table~\ref{tab:freq}
we compare the frequencies extracted from the dynamical evolution of the
models with the prediction of the linear-perturbation theory.
Remarkably, the difference for the fundamental mode and the first
overtone of both EOS models stays within $1\%$. The accuracy is similar
to the results in \cite{Baiotti04} for the evolution of a polytropic EOS,
giving us confidence in the correctness of our
implementation. Interestingly, we note also the presence of other
frequencies not predicted by the linear theory. This is the case for the
peaks in the PSD that are located at exactly twice the frequencies of the
respective $F$-mode, \ie around $7 \usk\kilo\hertz$ for the SHT-EOS and
$8 \usk\kilo\hertz$ for the LS-EOS. We conjecture that they are caused by
nonlinearities in the oscillations that are captured by the evolution
code, as it is typical of physical systems governed by nonlinear
equations in the limit of small
oscillations~\cite{Landau-Lifshitz1,Zanotti05}.

For the model with the LS-EOS we also perform a run in full general relativity, \ie model
\texttt{sTOV-FW1} in Table~\ref{tab:freq}. As expected (see 
\cite{Dimmelmeier06}), the frequency of
the $F$-mode differs significantly from the one obtained for a fixed
spacetime, which is larger by a factor $\sim  1.7$.
For this simulation we needed to move the outer boundaries further in the
weak-field regime by adding an extra refinement level with boundary at
$R_{\rm out} \simeq 53 \usk\kilo\meter$.

\subsection{Hot stars: Radial oscillations}
\label{sec:hot_stars}

Next, we analyze the evolution of TOV models with nonzero temperature to
study the effects of neutrino cooling on the radial pulsations of the
star. To this scope we build a TOV model using a slice of the SHT-EOS
which is isentropic and in $\beta$-equilibrium. In particular, we choose
an entropy per baryon of $s= 1\, k_{_{B}}$, corresponding to a
temperature $T \simeq 35 \usk\mega\electronvolt$ at densities larger than
nuclear saturation density (here $k_{_{B}}$ is the Boltzmann
constant). The initial TOV model, indicated as \texttt{sTOV-SHT} in
Table~\ref{tab:freq}, is evolved using the Cowling approximation, both
when the neutrino cooling is activated and when it is switched off; this
allows for a simple and direct comparison between the two cases. Also in
this case, the equilibrium model is perturbed with the fundamental radial
mode computed under the assumption that the perturbation satisfies the
same EOS as the background model.

\begin{figure}
  \begin{center}
    \includegraphics[width=0.45\textwidth]{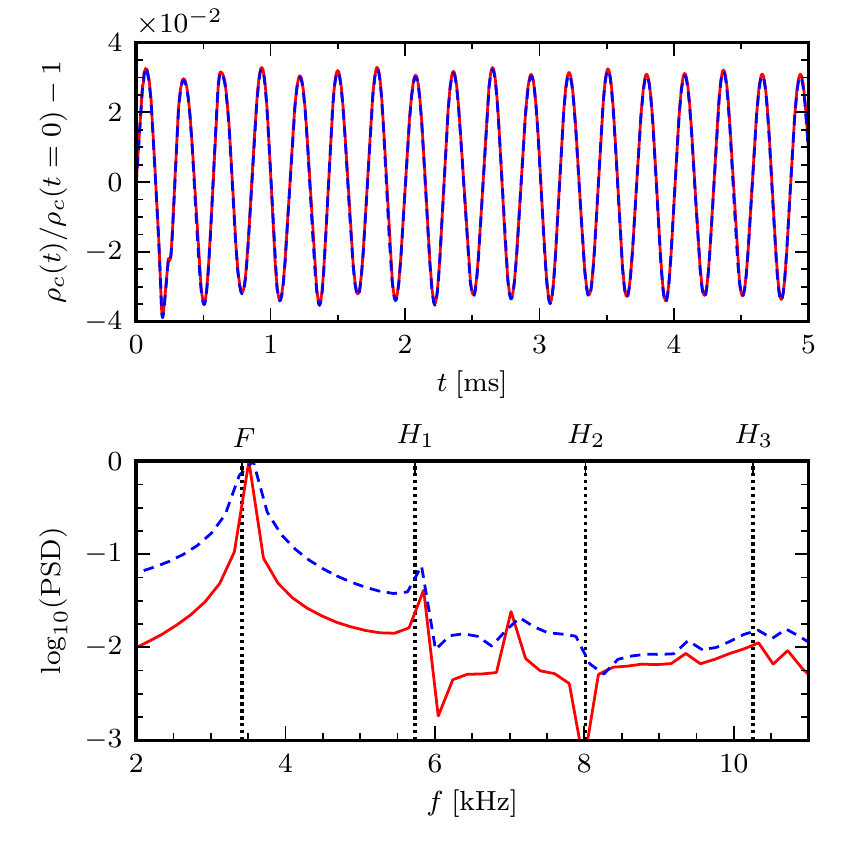}
    \caption{\label{fig:fft_rhoc_hot} Evolution of equilibrium
      configurations of hot NSs in the Cowling approximation (fixed
      spacetime). Upper panel: Evolution of the maximum rest-mass
      density $\rho_c$, normalized to the initial value, for the model
      \texttt{sTOV-SHT} described in Table~\ref{tab:freq}.  \referee{$t$
        is the coordinate time.}  Lower panel: corresponding power
      spectral density (normalized to the maximum amplitude),
      \referee{where $f$ is the frequency observed at infinity.} The red
      solid curve and the blue dashed curve represent the evolution with
      and without neutrino cooling, respectively.}
  \end{center}
\end{figure}

Note that the star is essentially opaque to neutrinos as the
neutrinosphere is located very close to the star surface. Furthermore,
because the diffusive neutrino emission is expected to take place on a
diffusion time scale, $t_\text{diff} \gtrsim 1$ s, that is much longer
than the dynamical time scale, $t_\text{dyn} \approx 1$ ms, we expect that
the star will cool only very slowly. Most of the emitted neutrinos, in
fact, will be reabsorbed in the hot central region. As a result, the net
luminosity will predominantly be produced by the outer stellar layers,
where the neutrinosphere is located, and will be modulated following the
smooth harmonic oscillations introduced by the perturbation. In general
the different neutrinospheres are located in a shell of $\sim 3$ km near
the stellar surface, with the electron-antineutrinos neutrinosphere being
located deeper inside the star when compared to the other neutrinos
species (\cf Fig.~\ref{fig:profile}).

\begin{figure}
  \begin{center}
    \includegraphics[width=0.45\textwidth]{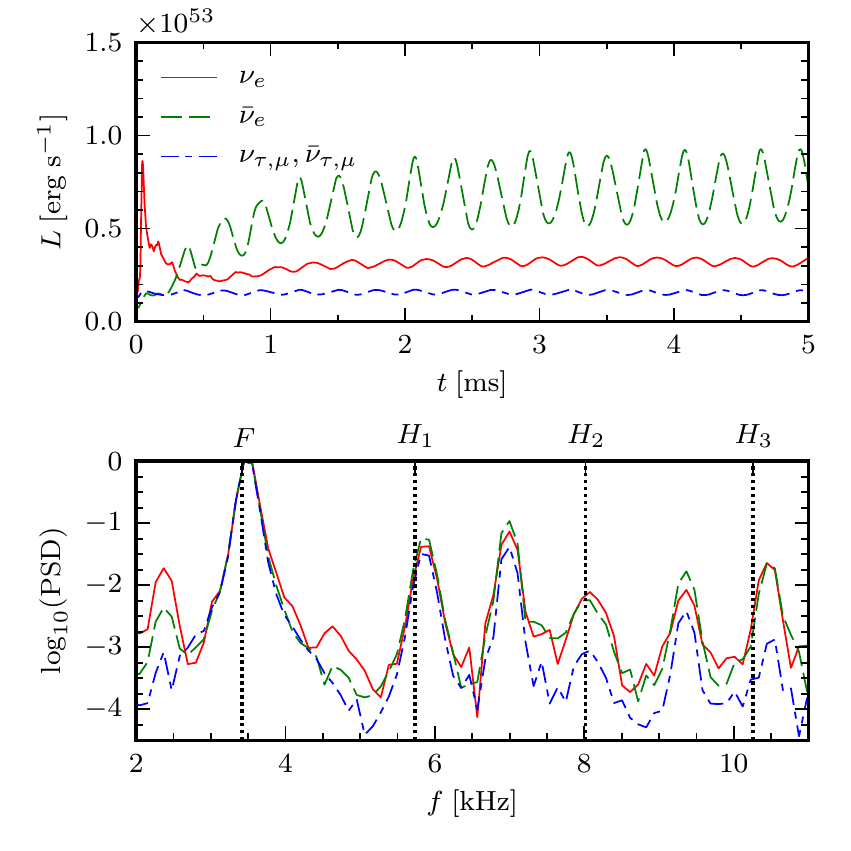}
    \caption{ \label{fig:hottov_nu} Top panel: :Luminosities of
      different neutrino species (in blue electron-neutrinos, in green
      electron-antineutrinos and in red $\mu$ and $\tau$ neutrinos) for
      the model \texttt{sTOV-SHT} described in
      Table~\ref{tab:freq}. Lower panel: The correspondent power
      spectral density (obtained after applying a Hanning window to 
      suppress the initial transient) compared
      with frequencies from linear-perturbation theory. }
  \end{center}
\end{figure}

In the top panel of Fig.~\ref{fig:fft_rhoc_hot}, we plot the evolution of
the normalized central rest-mass density for the model evolved without
neutrino emission (blue dashed line) and with the NLS (red solid
line). Note that the neutrino emission has a negligible effect on the
evolution of the bulk motion of the star. Only at late times, the two
evolutions start to show very small differences that can be hardly
noticed in the figure unless it is magnified. The lower panel of
Fig.~\ref{fig:fft_rhoc_hot} shows the PSDs of the central rest-mass
density for the two evolutions and compares them with the predictions
(vertical black dashed lines) from a linear-perturbation code for the
same models. Overall, the agreement between the linear and nonlinear
codes is rather good, with a difference of $2.4\%$ in the frequency of
the $F$-mode. As for the case of the radial pulsations of cold stars in
Sec.~\ref{sec:cold_stars}, we note the presence of modes that are not
predicted by the linear theory. Once again, we associate the peak around
$7 \usk\kilo\hertz$ to nonlinear couplings as it is located at exactly
twice the frequency of the $F$-mode. Finally, we note that the
frequencies for hot stellar models differ only slightly from the
corresponding values obtained for zero-temperature models with identical
baryon mass. In this case, the difference is only $0.6\%$ in the $F$-mode
frequency and increases to $1.3\%$ for the $H_3$-mode.

As anticipated, the harmonic time dependence of the fluid variables is
reflected also in the neutrino luminosities, which we compute
\referee{according to Eq.~(\ref{eq:L_Is}) with
  $\tilde{f}=1$.}
\begin{figure*}
  \begin{center}
    \includegraphics[width=0.33\textwidth]{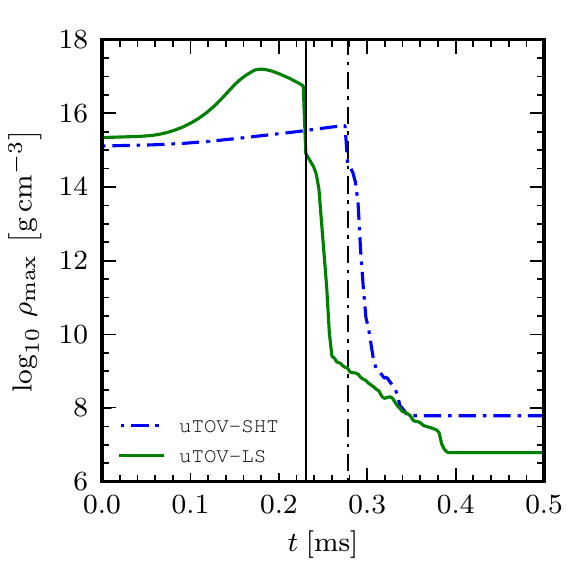}
    \includegraphics[width=0.33\textwidth]{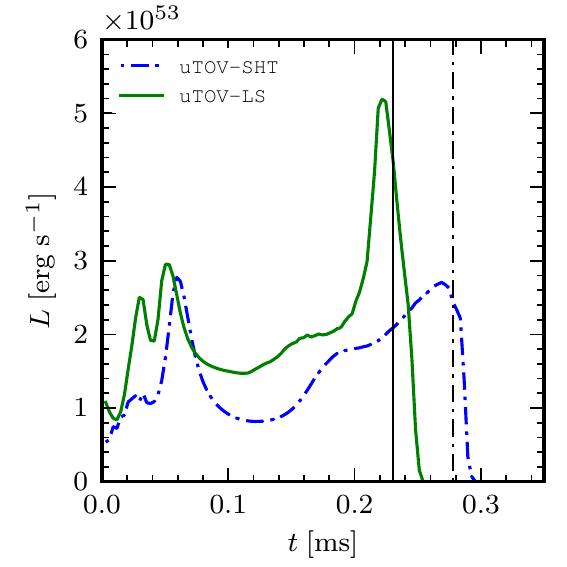}
    \includegraphics[width=0.33\textwidth]{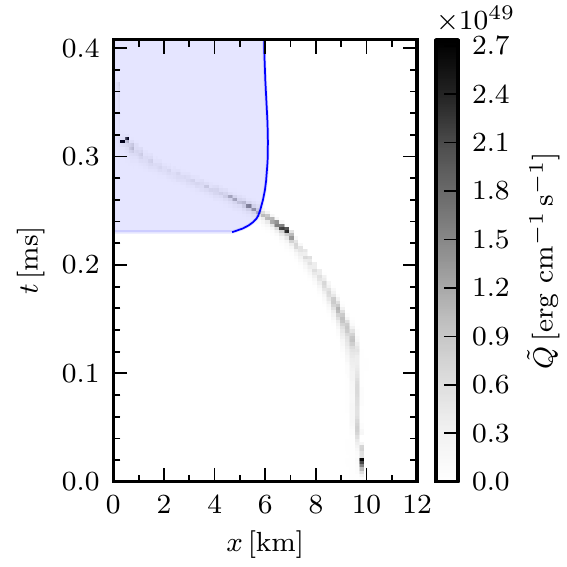}    
    \caption{ \label{fig:Coll_1S} Spherical collapse for the models
      \texttt{uTOV-LS} (blue) and \texttt{uTOV-SHT} (green) described in
      Table~\ref{tab:freq}. Left panel: Time evolution of the
      logarithm of \referee{the maximum over the rest-mass density,
      excluding the interior of the apparent horizon. Middle
      panel: Total neutrino luminosity. The vertical lines indicate the 
      time of the apparent horizon formation. Right panel: 
      Spacetime diagram of the local emissivity in terms of
      $\tilde{Q} \equiv 4\pi r^2 Q \bar{D} m_b^{-1} \alpha$ for model
      \texttt{uTOV-LS}. The blue solid line marks the apparent horizon, 
      the interior (shaded region) is excluded when computing the total
      luminosity in the middle panel.}
    }
  \end{center}
\end{figure*}
The evolution of the luminosity is shown in the top panel of
Fig.~\ref{fig:hottov_nu}, with different lines referring to different
neutrino species. Overall, it does not come as a surprise that the
modulation of the neutrino luminosities for the different species
reflects the oscillation frequencies of the star as shown in the PSDs
reported in the lower panel of Fig.~\ref{fig:hottov_nu}. Note however
that electron-neutrinos and heavy neutrinos show smaller-amplitude
oscillations in the luminosity than the electron-antineutrinos. This is
probably related to the relative position of the corresponding
neutrinospheres with respect the node of the eigenfunction
of the fundamental mode. In particular the electron-antineutrinos 
neutrinosphere is located deeper inside the star compared to the other 
species.                                                                 

\referee{We note that the local luminosity $L$ in
  Fig.~\ref{fig:hottov_nu} is effectively different from the one that
  would be measured by a distant observer. First, the modulation would be
  smaller than that in Fig.~\ref{fig:hottov_nu}; this is because $L$ sums
  up all the neutrinos emitted at a given coordinate time, which however
  do not reach the observer at the same time. As a result, the signal
  would resemble a running average of the luminosity $L$ over a time
  interval given by the light-crossing time, which is of the order
  $0.1\usk\milli\second$ for a NS. Second, the luminosity would also be
  reduced by a factor $\sim 1.5$ because of the gravitational
  redshift. Finally, because part of the neutrinos emitted inward would
  be absorbed by the NS core, the luminosity would be further reduced by
  a factor up to $\sim 2$.}

Even though the total evolution time is not sufficiently long to clearly
see the effects of the neutrino cooling on the lower order modes, there
are small differences between the eigenfrequencies \referee{when
  comparing to a cold NS with identical baryon mass.  For a star cooling
  down over longer time scales, we expect an impact mainly on the
  higher-order modes.} This can be
explained simply in terms of the nodal points of the corresponding
eigenfunctions, which are located closer to the surface in the case of
higher-order modes. As a result, they are affected more strongly by the
changes in composition taking place near the surface. 

\subsection{Hot stars: Migration or collapse to black hole}
\label{sec:hot_collapse}

Another important test for the correct implementation of the NLS is
represented by a classical benchmark in relativistic-hydrodynamics codes:
the dynamics of a nonrotating relativistic star on the unstable branch of
the configurations of equilibrium. This test, which was first considered
in three dimensions in Refs.~\cite{Baiotti04b,Font02c}, involves the
evolution of equilibrium models that are unstable to the fundamental
quasiradial mode and which, upon the introduction of a perturbation,
will either increase their central rest-mass densities and collapse to a
BH, or decrease their central rest-mass densities and hence migrate to a
new configuration on the stable branch of equilibria. We note that this
process can also exhibit a critical
behavior~\cite{liebling_2010_emr,Radice:10}, which however we will not
consider here. Both evolutionary paths are equally probable and in both
cases the stellar models are expected to maintain the same baryonic
mass. The degeneracy between the two possible evolutions is therefore
broken by the properties of the initial perturbation. In view of this we
introduce a well-defined velocity perturbation instead of leaving the
growth of the dynamical instability to the cumulative effect of the
numerical truncation error, since this might depend on the
grid resolution, reconstruction technique, symmetries and other details
of the evolution scheme. The perturbation consists of a simple radial
velocity profile with positive or negative sign in order to trigger
either the migration to the stable branch or the collapse to BH.

\referee{We consider three initial TOV models for this test, where the
  first two are constructed assuming $\beta$-equilibrium and a constant
  entropy distribution of $1\,k_{_{B}}$ for both the LS-EOS and SHT-EOS
  (\ie \texttt{uTOV-LS} and \texttt{uTOV-SHT}). The third model uses the
  SHT-EOS and has a constant temperature of $T=30 \usk\mega\electronvolt$
  (\ie \texttt{uTOVh-SHT}). The last three rows of Table~\ref{tab:freq}
  report the properties of the different models along with the maximum
  mass models for the two EOSs. Note that all the models have a central
  rest-mass density that is $\sim 20-30\%$ larger than the central
  rest-mass density of the model with the maximum mass. The temperature
  profile for these stars reaches $38 \usk\mega\electronvolt$ at the
  center for the LS-EOS and $52 \usk\mega\electronvolt$ for the SHT-EOS,
  while for both EOSs the surface temperature is around $5
  \usk\mega\electronvolt$. Hereafter we will concentrate only on models,
  \texttt{uTOV-LS} and \texttt{uTOV-SHT}, while model \texttt{uTOVh-SHT}
  will be used to show our convergence properties in
  Appendix~\ref{sec::convergence} .}

\referee{During the collapse to a BH, the central rest-mass density will
  increases, inevitably exceeding the upper limit of the EOS tables. For
  the purpose of this code test, we extended the EOSs to higher densities
  by the following prescription. First, we extend the EOS at zero
  temperature to higher densities by assuming that the pressure stays
  constant. Thermodynamic consistency then yields the specific energy
  $\bar{\epsilon}(\rho, Y_e)$ at zero temperature.  For higher
  temperatures, we use a zero-order extrapolation of the pressure along
  lines of constant $\epsilon - \bar{\epsilon}$. Of course, the
  prescription at higher densities will have an influence on the speed of
  the collapse as soon as those densities are reached.  However, the
  purpose of the extension is not physical realism, but testing the
  ability of the code to follow the collapse and BH formation.}

When driven by a negative radial velocity perturbation, all models evolve
towards a gravitational collapse. We follow the evolution of the
collapsing matter until the formation of a BH when an apparent horizon is
detected. The results of these collapsing simulations are shown in
Fig.~\ref{fig:Coll_1S}, where the left panel displays the evolution of
the maximum rest-mass density for both the extended SHT-EOS (blue solid
line) and the extended LS-EOS (green solid line), normalized to the
initial value. Note the central density exceeds the range of the EOS
tables during collapse (for the LS-EOS the overshooting is of more than
one decade).  We also note that after $\approx 0.1\usk\milli\second$, the
density profile develops a pronounced peak near the center, which is
numerically under-resolved.  We assume that this affects mainly the
central region and not the outer layers, which are still falling
in. Because of these two reasons, the decrease in central rest-mass
density observed for the LS-EOS at $t\gtrsim 0.175$ ms is likely an
artifact. The luminosity, however, should not be influenced by this since
it is mostly produced in the outer layers of the collapsing star. For
both models, the time of the collapse is of the order of a few tenths of
a millisecond before an apparent horizon is detected (marked with a
vertical solid or dashed line).

\referee{We also computed the neutrino luminosity $L$ according to
  Eq.~(\ref{eq:L_Is}), using $\tilde{f}=1$ and excluding
  the interior of the apparent horizon from the integral. The result is
  shown in the middle panel of Fig.~\ref{fig:Coll_1S}. For the models
  \texttt{uTOV-LS}, and \texttt{uTOV-SHT}, we find that the emitting
  region is a very thin shell close to the surface, resolved by only a
  few grid points. This is shown in the right panel, which represents a
  spacetime diagram of the local emissivity for model \texttt{uTOV-LS};
  note that most of the contribution to the luminosity takes place from a
  thin shell away from the apparent horizon (blue solid line). Such
  models represent an extremely challenging test for the NLS and the
  corresponding luminosities should therefore be taken with care. For the
  constant-temperature model \texttt{uTOVh-SHT} however, the emitting
  region is much larger and well resolved (see
  Appendix~\ref{sec::convergence}).}

\referee{Soon after an apparent horizon is formed, it covers most of the
  emitting material and indeed the neutrino signal terminates abruptly.
  As noted in Sec.~\ref{sec:hot_stars}, we should again stress that the
  luminosity we compute here is not necessarily a good approximation for
  the luminosity seen at infinity once the emitting shell comes close to
  the horizon. Already for the simpler case of a stationary BH, our
  assumption that all neutrinos reach infinity is less accurate than for
  a NS. This is because the neutrino luminosity seen by our coordinate
  observer is divided unevenly between highly redshifted outward-moving
  neutrinos and blueshifted inward-moving neutrinos which fall into the
  BH. Thus, Eq.~(\ref{eq:L_Is}) does not reproduce the very high redshift
  seen by a distant observer of the infalling matter close to the
  horizon. Another effect not taken into account is that neutrinos
  emitted closer to the horizon also need more time to reach a distant
  observer, while the luminosity we compute is a function of the
  coordinate time of the emission. Finally, estimating the signal that
  would observed at infinity is made even more difficult by the
  nonstationarity of the spacetime. For these reasons, we only reported
  the luminosity at the source, using $\tilde{f}=1$. We
  have checked, however, that most of the energy is always emitted at
  reasonable distances from the horizon (see again the right panel of
  Fig.~\ref{fig:Coll_1S}).}

\begin{figure}
  \begin{center}
    \includegraphics[width=0.45\textwidth]{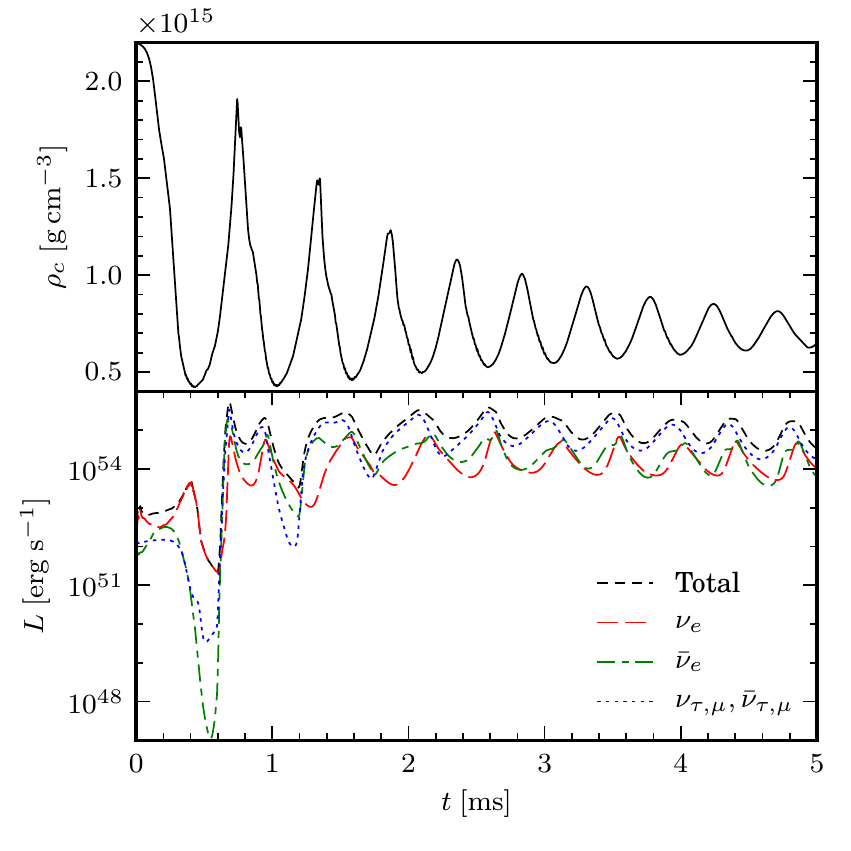}
    \caption{Migration to the stable branch of the
      unstable TOV model \texttt{mTOV-SHT} described in
      Table~\ref{tab:freq}. Top panel: oscillations of the maximum
      rest-mass density. Bottom panel: Neutrino luminosities for
      the different neutrino species.}\label{fig:mig_1s}
  \end{center}
\end{figure}

We now turn to discussing the dynamics of model \texttt{mTOV-SHT}, which
is perturbed with a positive radial velocity in order to trigger a
migration to the stable branch. The top panel of Fig.~\ref{fig:mig_1s}
displays the evolution of the central rest-mass density, while the bottom
panel shows the evolution of the three different luminosities. As
expected, the introduction of the perturbation triggers a very rapid
expansion of the star, followed by a series of oscillations as the star
tries to attain a new equilibrium on the stable branch of
configurations. Because the new equilibrium configuration has the same
baryonic mass but a smaller gravitational mass, the energy difference is
simply transformed into kinetic energy and dissipated in part by
shock heating at the stellar surface, which in turn leads to neutrino
emissions. Note that the evolution of the oscillation amplitude shows a
strong damping due to shock formation \cite{Radice2011,Kastaun2011}.
This behavior resembles closely the one found for an ideal-gas EOS as
discussed, for instance, in Refs. \cite{Font02c,Cordero2009}.

\begin{figure*}
  \begin{center}
    \includegraphics[width=0.49\textwidth]{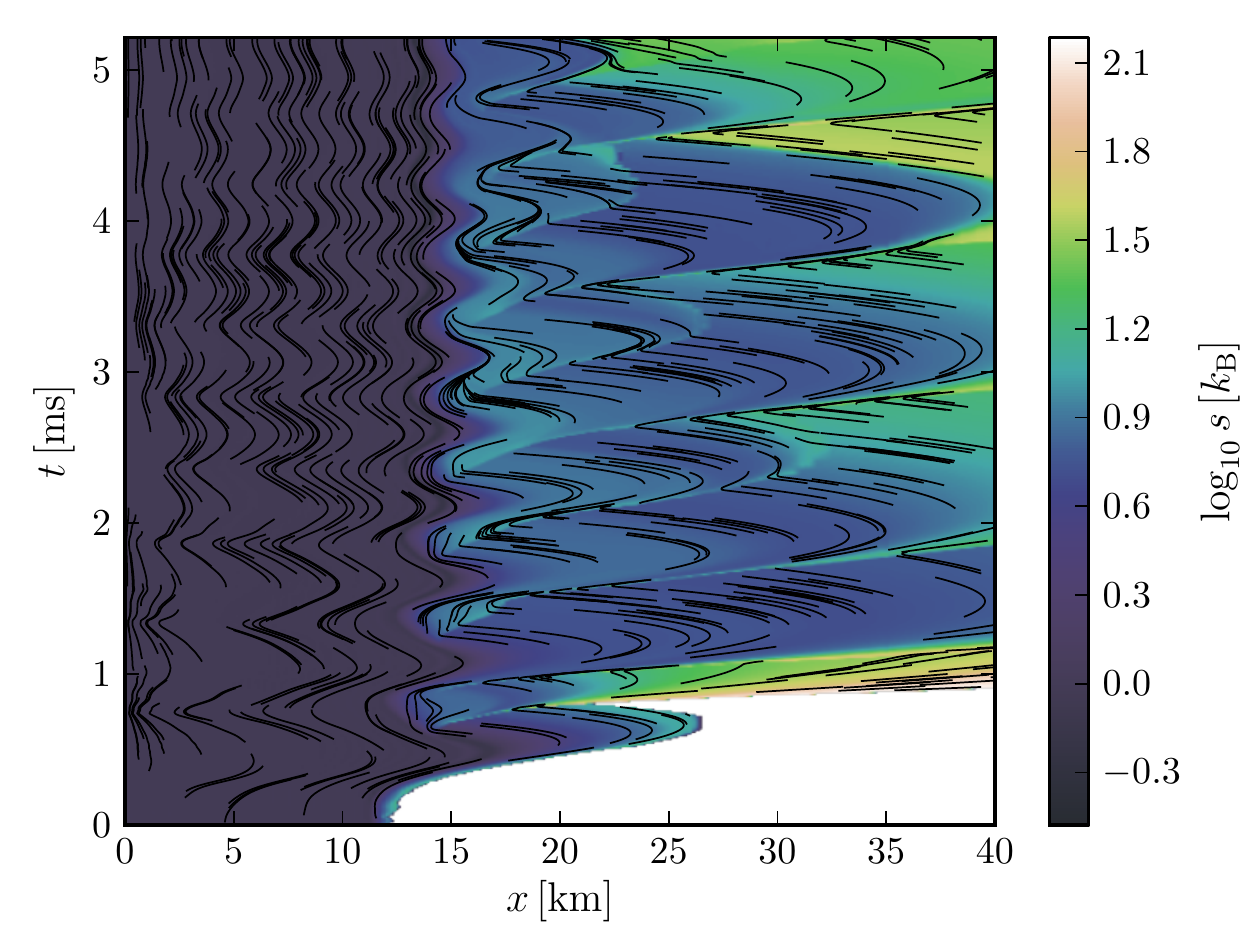}  
    \includegraphics[width=0.49\textwidth]{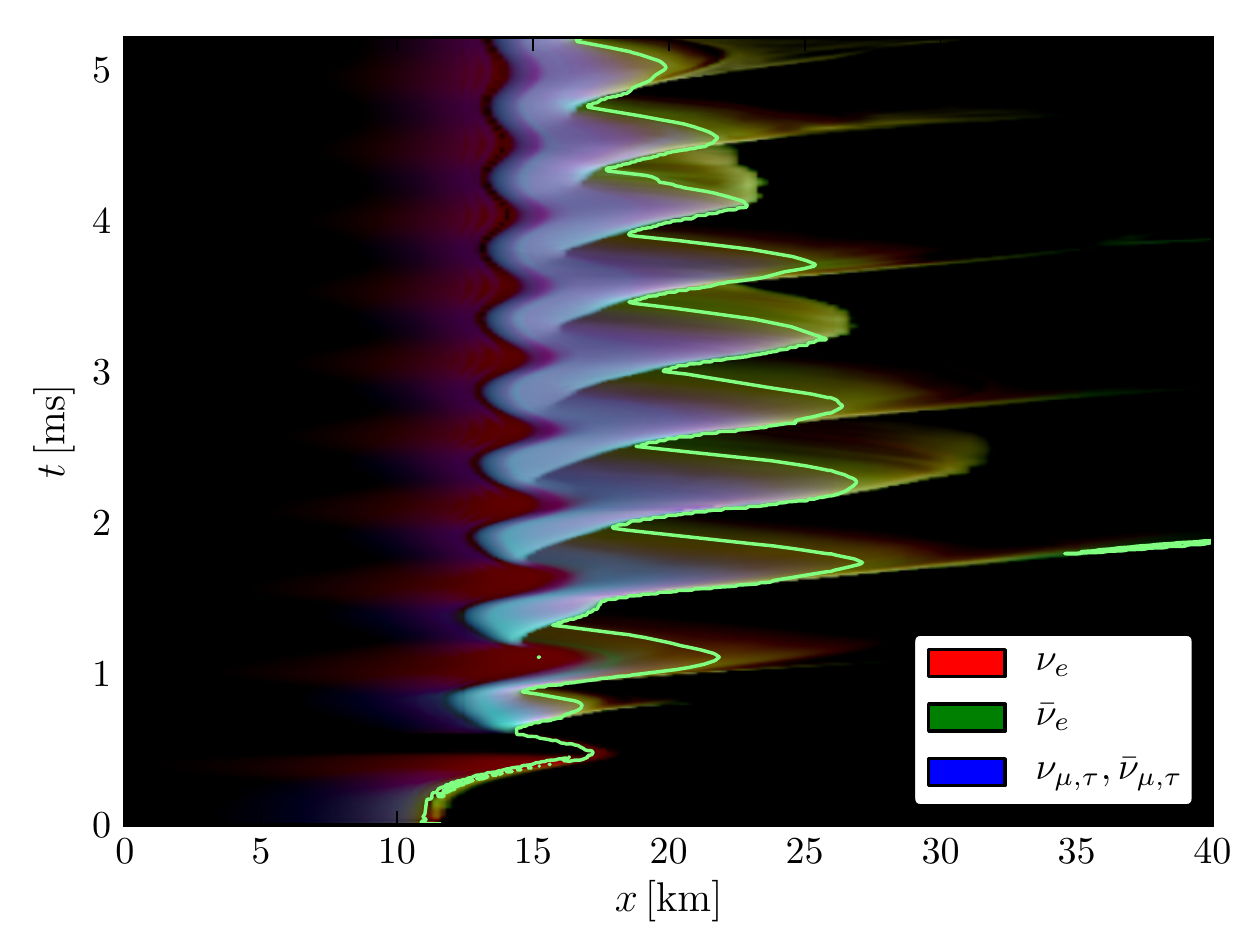}
    \caption{Evolution of specific entropy and neutrino emission along
    the $x$-axis for model \texttt{mTOV-SHT}. Left panel: 
    Specific entropy (color coded) 
    and trajectories of randomly placed test particles comoving with 
    the fluid (solid lines). The artificial atmosphere is drawn in 
    white. Right panel: Effective neutrino luminosities for
    the different neutrino species as a composite color plot. The
    red channel corresponds to the electron-neutrinos, green to the
    electron-antineutrinos, and blue to the heavy neutrinos. The 
    brightness corresponds to the logarithm of the luminosity $Q$.
    The location of the $\bar{\nu}_e$ neutrinosphere is marked by the 
    bright green solid line.
    }\label{fig:mig_xt}
  \end{center}
\end{figure*}

Particularly interesting is the evolution of the neutrino luminosity that
accompanies the various phases of expansion and contraction. We studied
the time evolution of entropy, neutrino emissivities, temperature, and
opacities along the $x$-axis. The first two are shown in
Fig.~\ref{fig:mig_xt}, from which we find that the total luminosity is
the result of several different effects.  We can divide the emitting
region into an inner one, at radii between $10$--$15 \usk\kilo\meter$,
where the star oscillates nonlinearly, but without significant shock
formation, and an outer region, extending to ${\sim}30 \usk\kilo\meter$,
where low density matter is heated up to $50 \usk\mega\electronvolt$ by
shock formation. Those shocks are caused by material from the outer layer
that is expanding very far and then falls back, colliding with denser
regions which are already expanding again. Those shocks are present at
each oscillation.  The neutrino luminosity from the inner region is
determined by the balance between two effects. First, the temperature
decreases when the star expands, lowering the local emission
rates. Second, the optical depth decreases when the star expands, thus
enhancing the effective emission rates. For our setup, the latter effect
is stronger, such that the luminosity is largest when the star is
expanded. The behavior of the inner region is similar to what we expect
for a linear oscillation of a stable star, and explains the modulation of
luminosity in Fig.~\ref{fig:hottov_nu}.  The luminosity in the outer
layers on the other hand follows essentially the temperature evolution
due to the shock heating, while the optical depth is irrelevant since the
densities are low. At radii between roughly $12$--$22 \usk \kilo\meter$
both heavy neutrinos and electron-antineutrinos are emitted, while at
radii larger than ${\sim}22\usk\kilo\meter$ mostly electron-antineutrinos
are emitted.  \referee{Note the intermediate regime between free emission
  and pure diffusion is more extended than for a stationary NS, and the
  position of the $\bar{\nu}_e$ neutrinosphere, shown in
  Fig.~\ref{fig:mig_xt} with a bright green solid line, just serves to
  approximately distinguish the two regimes (there can be multiple
  locations for the neutrinosphere at the same time, referring to
  different emitting regions).} The initial dip in the luminosities is
due to a pronounced peak in the opacity at roughly $15\usk\kilo\meter$,
which vanishes later when the material is heated. The nature of that peak
has yet to be identified. We also note that during the first 0.5 ms, when
the surface expands rapidly into the artificial atmosphere, the
luminosity may be influenced by treatment of the atmosphere.

The tests discussed in this section, although very idealized, provide us
with a well-controlled environment in which to understand the
relationship between the matter dynamics and the neutrino emission. The
understanding of the matter-radiation interaction within the NLS gained
in such simplified settings may help to understand the neutrino signals
from more complicated scenarios, such as BNS mergers~\cite{Sekiguchi2011,
  Sekiguchi2011b}.

\section{Applications of the NLS: neutrino-induced collapse}
\label{sec:nic}
After having discussed the ability of our code to properly account for
the neutrino emission \referee{from isolated NSs within the NLS}, 
we next consider the application of the code to explore novel aspects 
of the dynamics of these objects and in particular the conditions under 
which neutrino radiative losses can induce the collapse to a BH. 
More specifically, we will present the results of simulations of 
hot nonrotating NS models close to the maximum mass and show 
that the combined effects of neutrino cooling and deleptonization 
can drive a stable model to collapse gravitationally to a BH. 
Conversely, when these effects are neglected, the evolution leads to
a stable star oscillating in its eigenmodes. Although this is an
interesting scenario that has not been investigated before in consistent
three-dimensional simulations, we should also remark that for NSs with
temperatures close to the ones reached by the PNS in numerical
simulations of stellar-core collapse or by the HMNS in the case of BNS
mergers, the instability discussed here can be triggered only in an
extremely narrow range of masses near the maximum one.
Similar results are also reported in the study of equilibrium models 
of rotating hot NS \cite{Hashimoto1994} and more recently for HMNS 
\cite{Kaplan2013}. 

\subsection{Considerations from equilibrium models}

Before discussing in detail the results of the dynamical simulations, it
is instructive to make a number of considerations based on the properties
of equilibrium models and on quasistationary transformations along
sequences of equilibrium. We will then realize that the process of
cooling of a stable relativistic star which conserves baryon mass is far
less obvious than may appear at first sight and that it is actually very
difficult to make any reliable prediction about what is the outcome of
this process. 

Suppose here that we wish to represent the evolution of a cooling NS
through a sequence of quasistationary nonrotating equilibrium
configurations. To restrict the range of possibilities we need to make
the following assumptions: \textit{(i)} the star model moves through a
series of equilibria during the cooling process, keeping its baryonic
mass constant; \textit{(ii)} the cooling process is sufficiently slow
compared to the time scale of the $\beta$ processes, so that
$\beta$-equilibrium can be considered to be maintained at all times;
\textit{(iii)} the temperature profile in the star is either {isentropic}
(\ie the entropy per baryon is constant inside the star), or isothermal
(\ie the temperature is constant inside the star); \textit{(iv)} the
gravitational mass is reduced by the neutrino losses.

\begin{figure}
  \centering
    \includegraphics[width=0.45\textwidth]{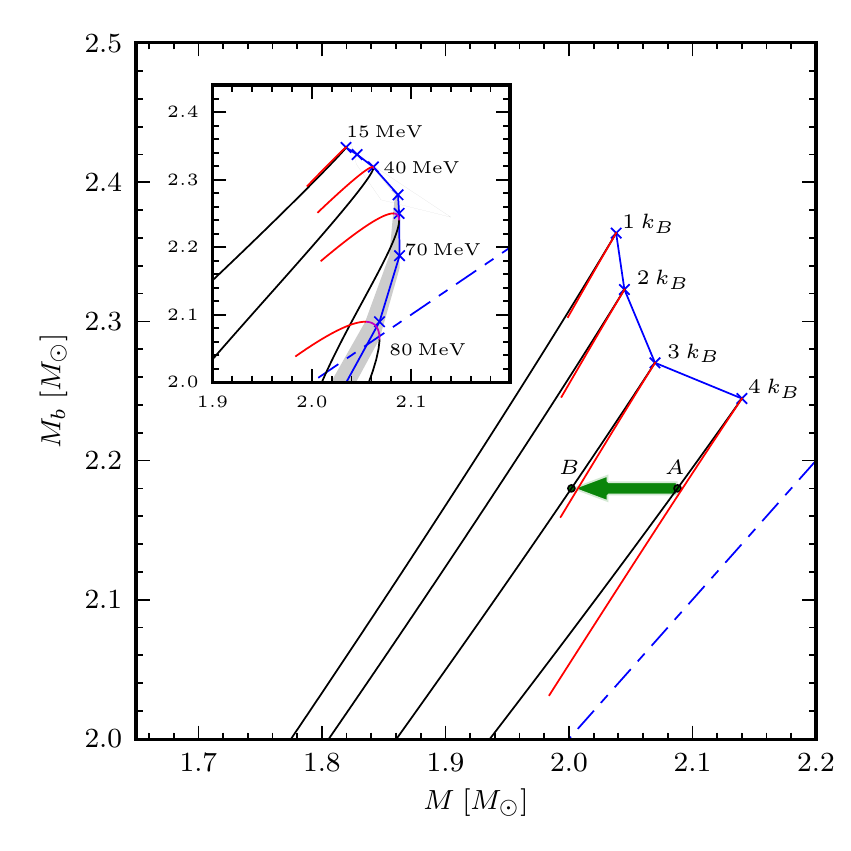}
    \caption{Baryonic versus gravitational mass
      for \emph{isentropic} star models in $\beta$-equilibrium, 
      constructed using the LS-EOS. The continuous black and red lines
      indicate, respectively, the stable and unstable configurations
      according to the ''turning-point'' criterion. The crosses mark
      the maximum gravitational and baryonic masses for each sequence.
      The inset shows \emph{isothermal} 
      sequences for the same EOS. The bounds of the 
      shaded region are given by the models with maximum baryon mass and
      those with maximum gravitational mass.
      The dashed blue line represents $M =
      M_b$. }\label{fig:TOV_Temp_Isen}
\end{figure}

Having made these assumptions, we can study the outcome of the cooling
process by comparing the gravitational and baryonic masses of TOV
sequences in $\beta$-equilibrium with different temperatures (this idea
was already explored in Ref.~\cite{Cabanell1981} in a more simplified
EOS). To this scope we have constructed isothermal or isentropic
sequences of nonrotating TOV solutions, shown in
Fig.~\ref{fig:TOV_Temp_Isen}. We first concentrate on the isentropic
case, to come back to the isothermal one later on. 

For an isentropic sequence with a given specific entropy, there is one
stellar model which maximizes both the gravitational mass $M$ and the
baryon mass $M_b$. We recall that the criterion of stability against
radial perturbations for a spherical relativistic star states that a
configuration with a central rest-mass density larger than the one of the
model with the maximum gravitational mass $M_{\mathrm{max}}$ is unstable
with respect to radial oscillations~\cite{Harrison65, Ipser1970}. This
criterion, which is also known as the ''turning-point''
criterion~\cite{Friedman88}, has been verified in a number of
simulations, for instance~\cite{Gondek1997, Goussard1998, Baiotti04,
  Baiotti06}, and represents a reasonable first approximation also in the
case of rotating stars~\cite{Takami:2011}\footnote{We also remark that
  for rotating stars, the turning-point criterion is only a sufficient
  condition for secular instability and not a necessary condition for
  secular and dynamical instability. Hence, a stellar model which is
  stable according to the turning-point criterion, can be nevertheless
  dynamical unstable, as shown by~\cite{Takami:2011} through dynamical
  simulations.}. Hence a NS model with $M=M_{\mathrm{max}}$ divides the
sequence into a stable branch and an unstable branch; models on the
unstable branch are unstable to radial perturbations, which cause the
stars to either collapse to a BH or to expand and oscillate around a
distinct and stable equilibrium configuration, as discussed in the
previous Section.

The question is now whether it is possible to move continuously along
equilibrium configurations belonging to sequences with decreasing
specific entropy, while decreasing the gravitational mass and conserving
the baryon mass. To answer this, we plot a number of sequences in a $(M,
M_b)$ diagram, as shown in Fig.~\ref{fig:TOV_Temp_Isen} for the specific
example of the LS-EOS. The green arrow indicates a cooling process
satisfying our conditions and connecting two putative starting and ending
states $A$ and $B$, both on the stable branches of equilibrium
configurations.

Note that in general the maximum baryonic mass decreases with increasing
entropy and, as one would expect for a star losing energy via radiation, 
that the gravitational mass decreases along sequences with
decreasing entropy. For example, a model with a baryonic mass of $2.1 \usk
M_\odot$ will radiate about $\sim 9\%$ of its gravitational mass when
going from an isentropic configuration with $s=4\,k_{\mathrm{B}}$ to one
with zero temperature. 
We thus find that the idealized cooling process through stable equilibrium 
models is not prohibited by energy or mass conservation.

However, a careful look at the figure will reveal that the trajectory
from $A$ to $B$ also intersects the unstable branches.  This raises the
question if it is possible for a model on the stable branch to change
into a model which has identical baryon mass and gravitational mass, but
lies on the unstable branch of another sequence.  When only considering
energy and mass conservation, there is nothing that would prevent such a
transition.  It would however require a finite, discontinuous reduction
of the stellar radius.  Since we allow only for transformations that
reduce the gravitational mass, also the specific entropy has to be
reduced by a finite amount when transitioning to an unstable branch, \eg
as a result of a phase transition~\cite{Abdikamalov2009b}, as can be
seen from Fig.~\ref{fig:TOV_Temp_Isen}.  Assuming that the star does not
change its radius discontinuously, a transition to the unstable branch
cannot happen through a sequence of isentropic equilibrium models. It
would however be difficult to prove conclusively that no such transition
is possible through any conceivable dynamical process or through
nonisentropic equilibrium models, e.g. when cooling first the outer
layers and then the core. This question is not to be confused with the
usual stability analysis, which describes adiabatic perturbations.

\begin{figure}
  \centering
    \includegraphics[width=0.45\textwidth]{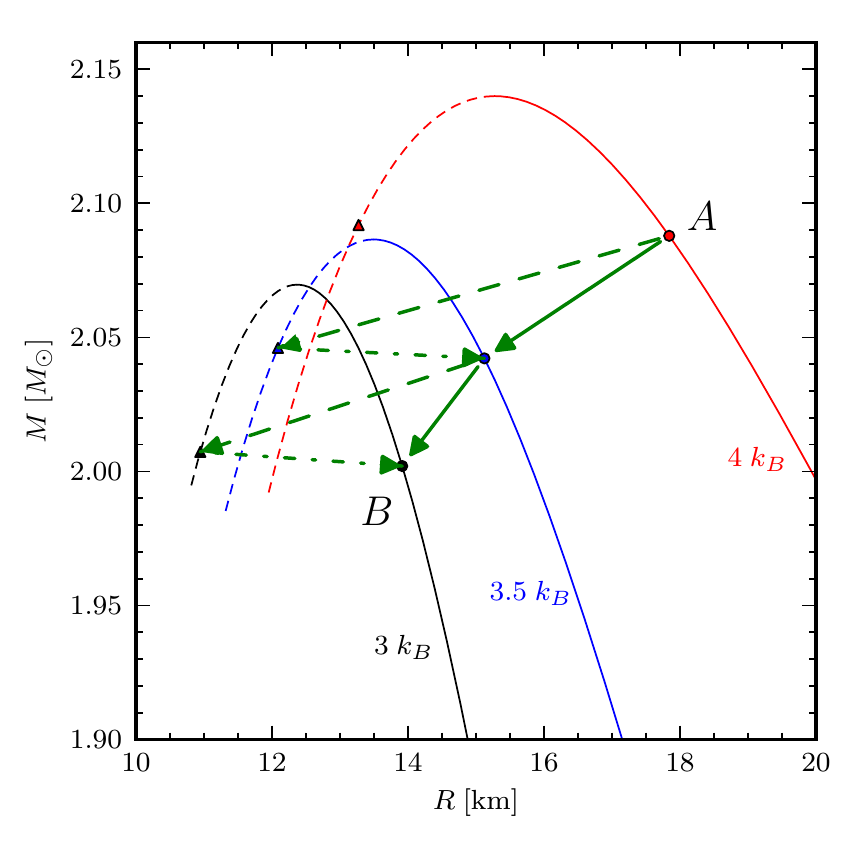}
    \caption{Trajectories in the mass-radius 
    diagram of a cooling star. The solid green arrow represents a 
    smooth transition along stable equilibrium models. The green dashed
    arrows represent an alternative path when allowing discontinuous 
    transitions from stable to unstable branches, the green dash-dotted
    arrows represent migration back to the stable branch via dynamical 
    instability.}
    \label{fig:Temp_trajects}
\end{figure}

In any case, whether or not the above crossings to unstable branches could 
happen in nature cannot be decided from the idealized cooling model, but 
requires a fully dynamical description. 
This is particularly relevant for models approaching the maximum mass,
where the jumps in radius and central density between stable and unstable models
become arbitrary small and can be more easily overcome by finite perturbations. 
Let us therefore consider possible trajectories in parameter space when allowing the above 
transitions between stable and unstable models, but still restricting ourselves
to isentropic models for simplicity.
This is illustrated in Fig.~\ref{fig:Temp_trajects}, which shows the
gravitational masses and radii along three isentropic sequences with
$s=3.0, 3.5$ and $4.0\,k_{\mathrm{B}}$. Also here we have marked the two
states $A$ and $B$ and shown with a  solid green arrow the simplest of
the possible trajectories, namely, one leading from $A$ to $B$ across
models on stable branches. 
Alternatively, the star could cool down along stable equilibrium 
sequences, then cross over to an unstable branch via some hypothetical 
process, reducing the radius and the specific entropy, but conserving 
baryonic and gravitational mass.
It could then continue to cool down along unstable equilibrium sequences.
Such a process is summarized by the green dashed arrow, connecting
model $A$ to a model on the unstable branch with $s=3.5\,k_{\mathrm{B}}$.
Once on the unstable branch, the star can either collapse to a BH (moving further
to the left) or ''migrate'' back to the stable branch (green dotted arrow),
following the dynamic instability discussed in the previous Section. In
doing so it will expand, going from a radius of, say, $\sim 11$ km to 
$\sim 16$ km and further cool (note that the gravitational mass
decreases). Once on the stable branch, it can further cool and go to
state $B$ or move again to
the unstable branch of the sequence with $s=3.0\,k_{\mathrm{B}}$, where
the same considerations made above can be repeated.

As anticipated above and as shown rather clearly in
Fig.~\ref{fig:Temp_trajects}, considering equilibrium sequences
and energy arguments does not allow us to reach any firm conclusion 
on the fate of a relativistic star
that cools via radiative losses. It may seem reasonable to expect that
the transition involving only states on the stable branches is favored
as in this case the star undergoes the minimal changes in radius and
density, without abrupt contractions and expansions. However, no
energetic argument can really be used in support of this conjecture.

We now return to the question of what happens under the assumption of
isothermal cooling. The corresponding equilibrium sequences are depicted
in the inset of Fig.~\ref{fig:TOV_Temp_Isen}. Also here, the maximum
baryonic mass is decreasing with increasing temperature. Note however
that in this case the maximum gravitational mass is not anymore
monotonically increasing with temperature. In fact, at temperatures above
$50 \usk\mega\electronvolt$ the maximum mass decreases with increasing
temperatures as larger and larger portions of the star become dominated
by the pressure support of photons and electrons. Also note that for the
highest temperature models we even find that $M > M_b$, \ie one would
gain energy by dispersing the stellar matter to infinity. Another
intriguing property of the isothermal solutions is that maximum
gravitational and maximum baryonic mass are not reached at the same
central rest-mass density. As shown by
Refs. \cite{Gondek1997,Gourgoulhon1993}, the turning-point criterion
needs to be extended under these conditions, although we still regard it
as a good estimate. Although in Fig.~\ref{fig:TOV_Temp_Isen} we have
employed the LS-EOS, very similar qualitative results are obtained
when using the SHT-EOS. Finally, it is not yet clear whether the region
with $M > M_b$ can really be considered to represent realistic models and
is not instead just an artifact of the EOSs for very large
temperatures. What is clear, however, is that the relation between
temperature, entropy and gravitational mass becomes more complicated when
thermal effects are considered as these can change significantly the
rest-mass density profile.

In summary, we have shown that no conclusive predictions can be made on
the evolution of a cooling relativistic star on the basis of equilibrium
considerations. This represents a perfect motivation to make use of fully
numerical simulations which can shed light on this process and account,
for instance, for the fact that cooling processes are far from being
homogeneous throughout the star. The outer layers of the hot NS would cool
down much faster than the core, creating an entropy or a temperature
gradient that might lead to the formation of convective cells or the
contraction of the outer stellar layers. This is indeed the focus of the
following Sec..

\subsection{Dynamical simulations of neutrino-induced collapse}

Having constructed equilibrium models of hot NSs, we now address the
question of how the stability of a nonrotating relativistic star against
radial perturbations is affected by changes in temperature and
composition. It is important to remark that the turning-point criterion
was obtained under the simplified assumptions that the radial
perturbations conserve entropy and that the chemical equilibrium is
reached essentially instantaneously (or on time scales much smaller than
those of the radial oscillation), such that the composition is always the
one at $\beta$-equilibrium. If this last condition is not satisfied, and
instead the time scale of the weak-interaction process is longer than (or
comparable to) the oscillation time scale, then the composition of the
initial configuration is simply advected along with the fluid
elements~\cite{Chanmugam1977,Gourgoulhon1993}. As a result, for central
rest-mass densities near that of the maximum-mass model, \ie near
$\rho_{c,\mathrm{max}}$, where the oscillation period of the fundamental
radial mode tends to diverge and where infinitesimal perturbation might
destabilize the configuration, the time scale of the changes in
composition could have an important dynamical effect (this was pointed
out already by Ref.~\cite{Gourgoulhon1995}). Clearly, the only way to
assess the stability properties under conditions in which the effects of
weak processes are taken into account and can be influential, is by
performing the fully nonlinear numerical simulations that we report
below.

\begin{figure*}[t]
  \begin{center}
    \includegraphics[width=0.45\textwidth]{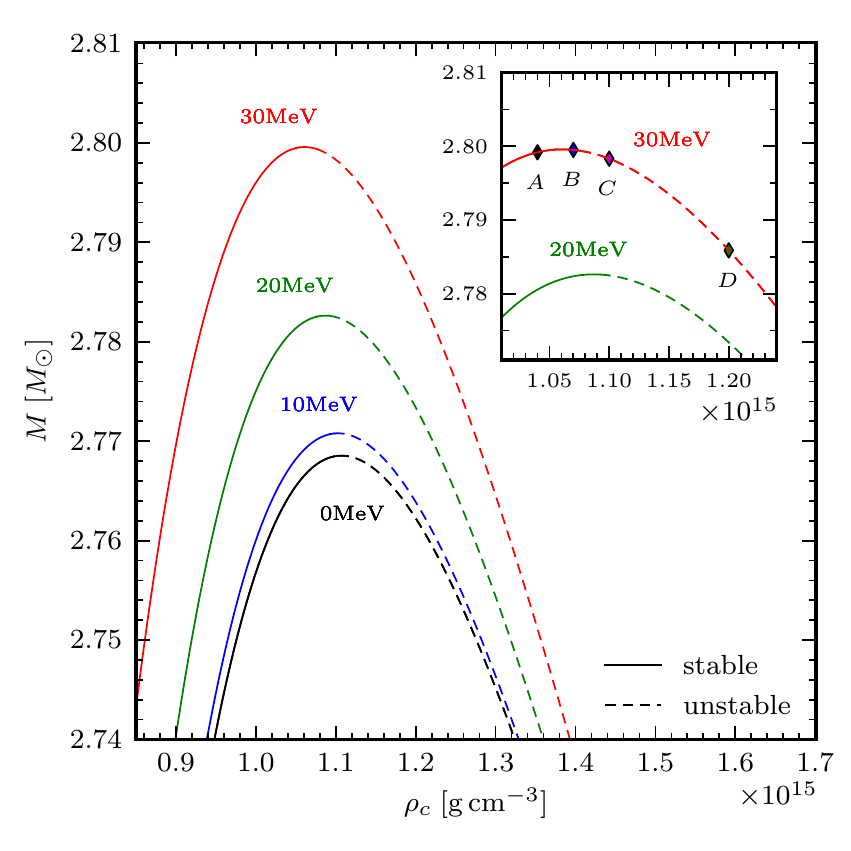}
    \hskip 1.0cm
    \includegraphics[width=0.45\textwidth]{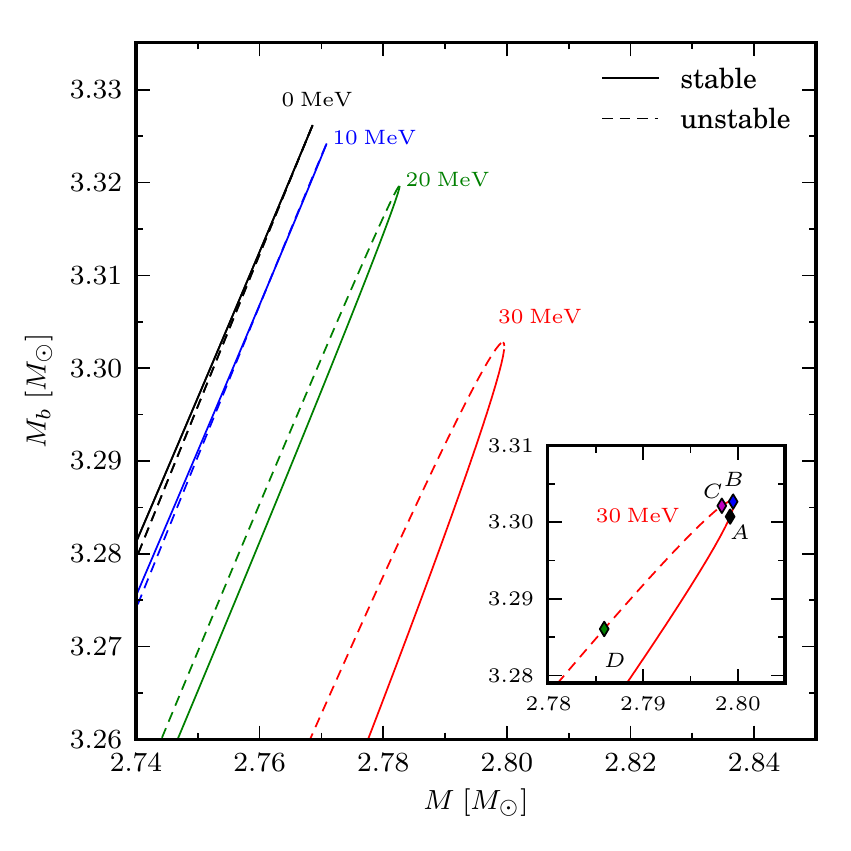}
    \caption{ \label{fig:models_nic} Left panel: Mass versus
      central rest-mass density for sequences of isothermal,
      $\beta$-equilibrated TOV solutions with temperatures ranging from
      $0$ (black) to $30 \usk\mega\electronvolt$
      (red), all obeying the SHT-EOS. Right panel: Gravitational
      mass versus baryonic mass for the same models. The insets show the
      models close to the maximum mass, the points $A$--$D$ indicate the
      models described in Table~\ref{tab:nic_models}. }
  \end{center}
\end{figure*}

\begin{figure*}[t]
  \begin{center}
    \includegraphics[width= .49\textwidth]{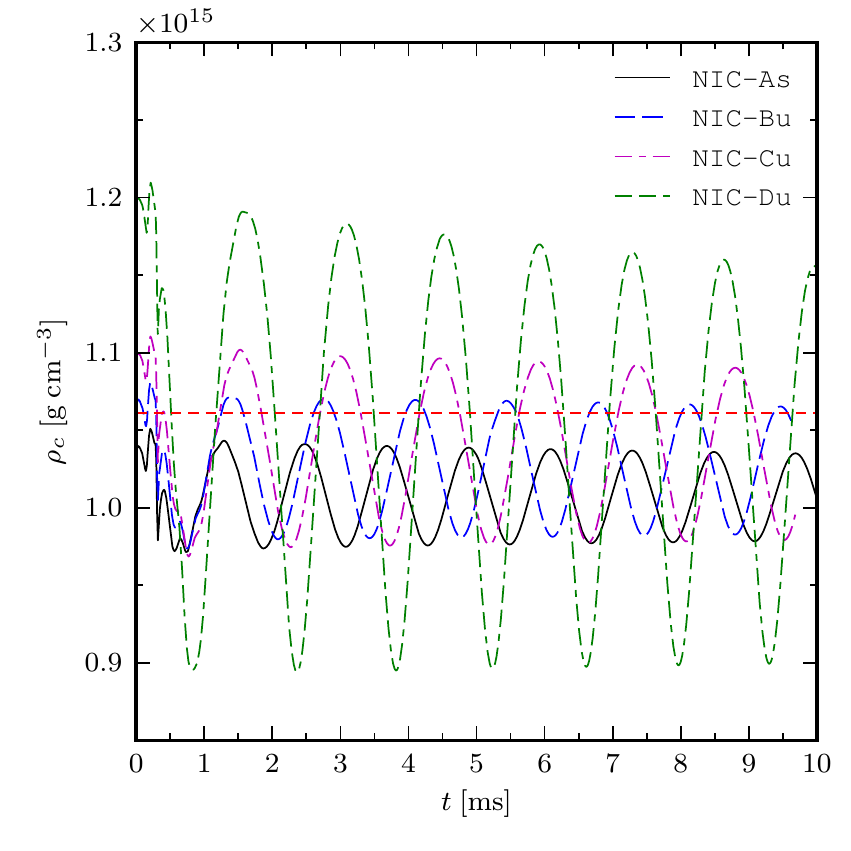}
    \includegraphics[width= .49\textwidth]{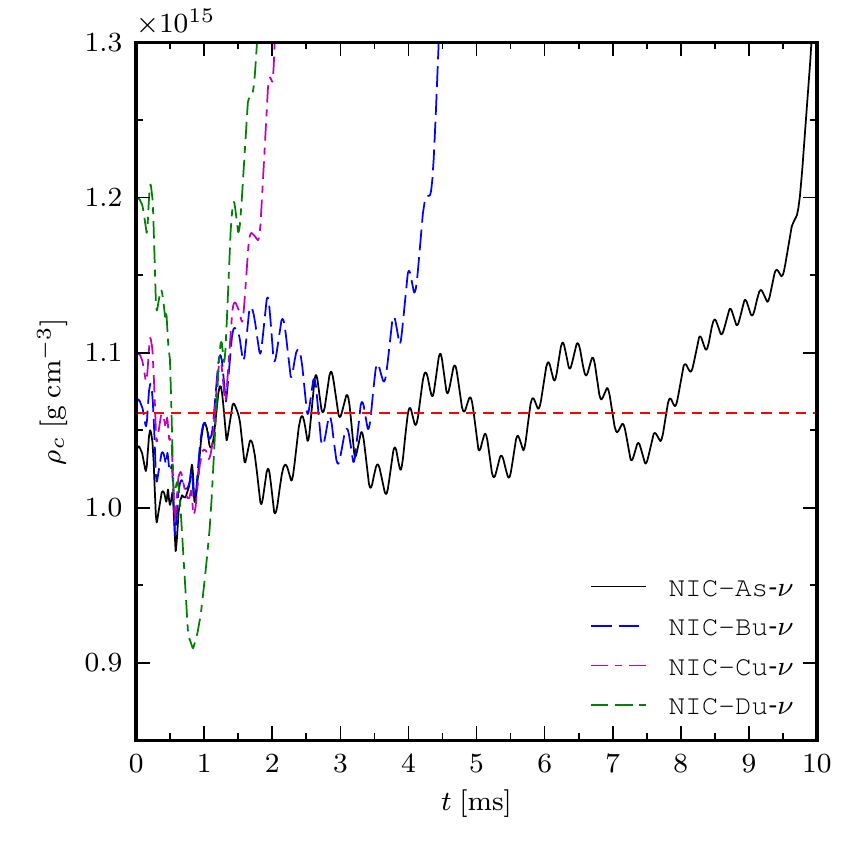}
    \caption{Evolution of the central rest-mass density for the different
      models described in Table~\ref{tab:nic_models}. Left panel:
      evolution of the central rest-mass density when the neutrino
      radiative losses \emph{are not taken into account}. Right
        panel: same models but evolved modeling the neutrino losses
      with the NLS. The horizontal dashed line indicates the central
      density $\rho_{c,{\rm max}}$ of the maximum-mass model at
      $30\usk\mega\electronvolt$. }
    \label{fig:nic_rhoc_ev}
  \end{center}
\end{figure*}

We have therefore selected a sequence of NS models with $T=30
\usk\mega\electronvolt$ in $\beta$-equilibrium close to the maximum
gravitational mass allowed by the SHT-EOS. The reason behind this choice
is that the results of the linear stability analysis of hot PNS models
performed by Ref. \cite{Gondek1997} have shown that the impact of
temperature and composition on the radial eigenfunctions is stronger for
isothermal configurations.  The properties of our models, named
\texttt{NIC-As} to \texttt{NIC-Du}, are listed in
Table~\ref{tab:nic_models}.

These models are also shown on a $(M,\rho_c)$ diagram in the left panel
of Fig.~\ref{fig:models_nic} as models $A$--$D$. For each sequence we
distinguish with solid and dashed lines models that are on the stable and
unstable branches, respectively. The right panel of
Fig.~\ref{fig:models_nic} shows the same sequences in a $(M_b,M)$
diagram. Note that the influence that the temperature has on the maximum
mass $M_{\rm max}$ is only very small (as mentioned already, the pressure
shows only a very weak dependence on temperature near nuclear densities;
\cf left panel of Fig.~\ref{fig:Composition}). The maximum mass of the
cold model (\ie $0 \usk\mega\electronvolt$) is only $\sim 1\%$ smaller
than the one of the hottest model at $30 \usk\mega\electronvolt$ (\cf
right panel of Fig.~\ref{fig:models_nic}). A closer comparison between
these two models is offered in Table~\ref{tab:nic_models}, where they are
denoted as \texttt{MAX-0} and \texttt{MAX-30}, respectively.

\begin{table}[h]
  \begin{tabular}{cccccccc}
    \hline
    \hline
    Model  & $T$& $\rho_{c}/{10^{15}}$ &$M_b$ & $M$ & $R$& $E/{10^{52}}$ &  $E/M$ \\
          & $[\mathrm{MeV}]$& $[{\rm g~cm}^{-3}]$  & $[M_{\odot}]$& $[M_{\odot}]$&$[\mathrm{km}]$& $[{\rm erg}]$ &  $[\%]$ \\
    \hline
    \texttt{MAX-0}   &$\phantom{3}0.0$ & $1.106$ & $3.3282$ & $2.7689$ &13.31& $-     $ & $-    $ \\
    \texttt{MAX-30}  &$30.0$& $1.061$ & $3.3024$ & $2.7996$ &17.40& $-     $ & $-    $ \\
    \texttt{NIC-As}  &$30.0$& $1.040$ & $3.3007$ & $2.7992$ &17.66& $2.0541$ & $0.410$ \\
    \texttt{NIC-Bu}  &$30.0$& $1.070$ & $3.3027$ & $2.7995$ &17.45& $1.3324$ & $0.266$ \\
    \texttt{NIC-Cu}  &$30.0$& $1.100$ & $3.3021$ & $2.7983$ &17.26& $1.0878$ & $0.217$ \\
    \texttt{NIC-Du}  &$30.0$& $1.200$ & $3.2861$ & $2.7859$ &16.72& $0.8917$ & $0.179$ \\
    \hline
    \hline
  \end{tabular}
  \caption{
    Parameters of the neutrino-induced collapse equilibrium models 
    discussed in Sec.~\ref{sec:nic}. $E$ is the 
     total energy radiated by neutrinos during the simulations.}
  \label{tab:nic_models}
\end{table}

Resolving the minuscule differences between these models in numerical
simulations is quite challenging, also in terms of the computational
costs. Our simulations make use of three fixed nested refinement grids
centered on the NS, with a resolution of the finest grid of
$\Delta{=}0.1846 \usk\kilo\meter$ (\ie $\sim 95$ points across the star).
The extent of the finest grid is large enough to include the entire star
and the outermost boundary is located at a distance of $118
\usk\kilo\meter$. The spherical grid used to compute the optical depth
has a radial resolution of $\Delta r {=} 0.00886 \usk\kilo\meter$, 100
points in the $\phi$-direction, and 50 along the $\theta$-direction. Also
in this case we set the density of the artificial atmosphere to the
lowest value available in the SHT-EOS table (\ie $1.65 \times 10^{7}
~{\rm g~cm}^{-3}$).

The results of these simulations are collected in
Fig.~\ref{fig:nic_rhoc_ev}. The left panel depicts the evolution of the
central rest-mass density when the neutrino radiative losses \emph{are
  not} taken into account. The different lines refer to the four
different models. As a reference, the horizontal dashed line marks the
central rest-mass density of the maximum-mass model, $\rho_{c,{\rm
    max}}$. Note that no perturbation was introduced in the initial data
and the oscillations are triggered simply by the truncation error.

As expected in the absence of neutrino losses, model \texttt{NIC-As},
which is on the stable branch, simply oscillates in its fundamental mode
with an amplitude that is larger when compared to the simulations using
Cowling approximation studied in Sec.~\ref{sec:cold_stars}. The
corresponding frequency is of the order of $0.5 \usk\kilo\hertz$, in good
agreement with what was found in Ref.~\cite{Takami:2011} when computing
the neutral-stability line in relativistic stars. Interestingly, models
\texttt{NIC-Bu}, \texttt{NIC-Cu} and \texttt{NIC-Du}, which are instead
on the unstable branch, migrate on a dynamical time scale to the stable
branch, expanding and reducing their central rest-mass density while
conserving baryonic mass. Once on the stable branch they simply oscillate
with amplitudes which are large but with the central rest-mass density
remaining below the initial one [just compare $\rho_c(t)$ with
  $\rho_c(t=0)$ for models \texttt{NIC-Bu}--\texttt{NIC-Du}], but with
excursions that go above above $\rho_{c,{\rm max}}$ (only model
\texttt{NIC-As} has oscillations that remain below $\rho_{c,{\rm max}}$).

In all of these simulations, there is no significant mass loss neither
from the outer boundary or from the treatment of the atmosphere
surrounding the star. The conclusion to be drawn from the left panel of
Fig.~\ref{fig:nic_rhoc_ev} is therefore that, in the absence of neutrino
losses, none of the models considered collapses to a BH. Furthermore,
even models that are initially unstable, simply migrate to the stable
branch. We note however that the perturbations required as a trigger 
are given by discretization errors, and a collapse to a BH seems equally 
possible.

\begin{figure}
  \begin{center}
    \includegraphics[width=0.45\textwidth]{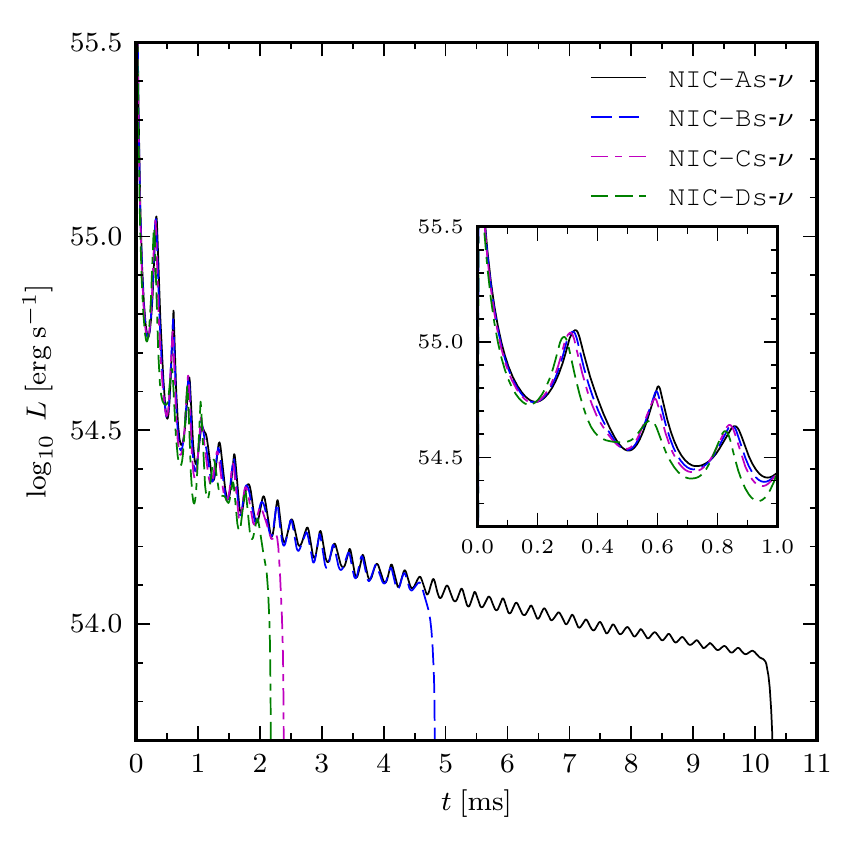}
    \caption{Evolution of the total neutrino luminosity for the models
      described in Table~\ref{tab:nic_models}. The inset shows the
      initial neutrino burst due to the rapid cooling of the outer
      envelope.}
    \label{fig:nic_lum_evol} 
  \end{center}
\end{figure}

The dynamics of these same models is very different if the neutrino
losses \emph{are taken into account}. This is shown in the right panel of
Fig.~\ref{fig:nic_rhoc_ev} and it is straightforward to realize that all
the models are now dynamically unstable over time scales of $1$--$10
\usk\milli\second$. Let us first analyze the unstable models \texttt{NIC-Bu}, 
\texttt{NIC-Cu} and \texttt{NIC-Du}. Overall, the cooling produced by the 
neutrino losses, which is already active after about $0.1 \usk\milli\second$ 
(\cf Fig.~\ref{fig:nic_lum_evol}) induces a perturbation that leads
to the final gravitational collapse. However, before collapsing to a BH,
all of the unstable models actually \referee{expand until the central 
density is well below the one of the maximum-mass model,}
just as they did in the absence of neutrino losses 
(compare the same lines in the
left and right panels of Fig.~\ref{fig:nic_rhoc_ev}). The continuous
radiative losses experienced by these models and the corresponding
pressure depletion, however, are such that these models
progressively collapse to a BH. Not surprisingly, the
first model to collapse is the one that was originally further away from
the maximum central density, \ie \texttt{NIC-Du}, and that is also
subject to the largest oscillations. Similarly, the duration until collapse 
of the other unstable models is progressively longer, the closer they are 
to the maximum-mass model. Before collapsing, each model loses a significant
fraction of the total thermal energy stored in the initial configuration.
The radiated energy can be as high as 0.4\% of the initial mass (see 
Table~\ref{tab:nic_models} for details).

\begin{figure*}
  \includegraphics[width=0.45\textwidth]{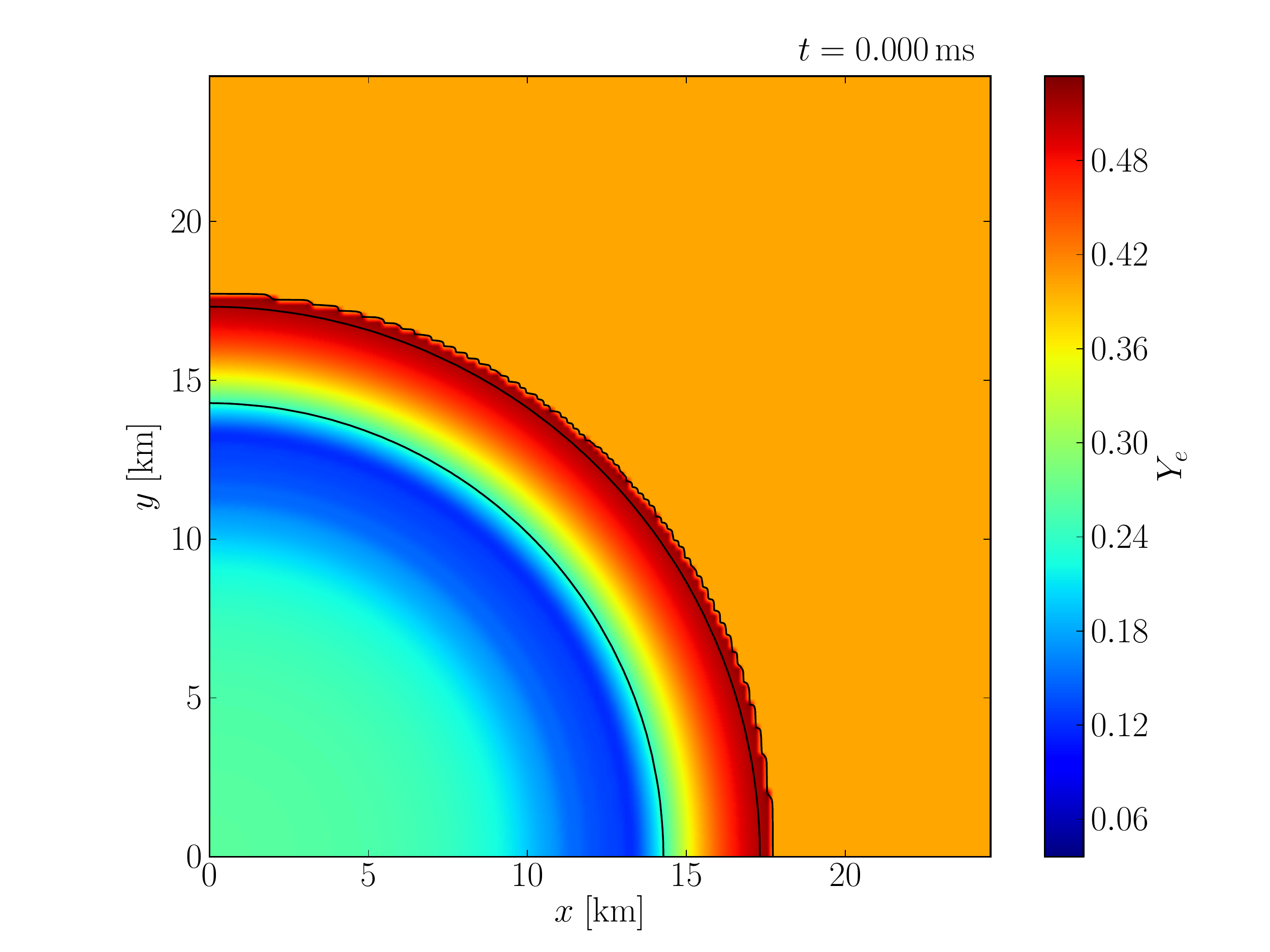}
    \hskip 1.0cm
  \includegraphics[width=0.45\textwidth]{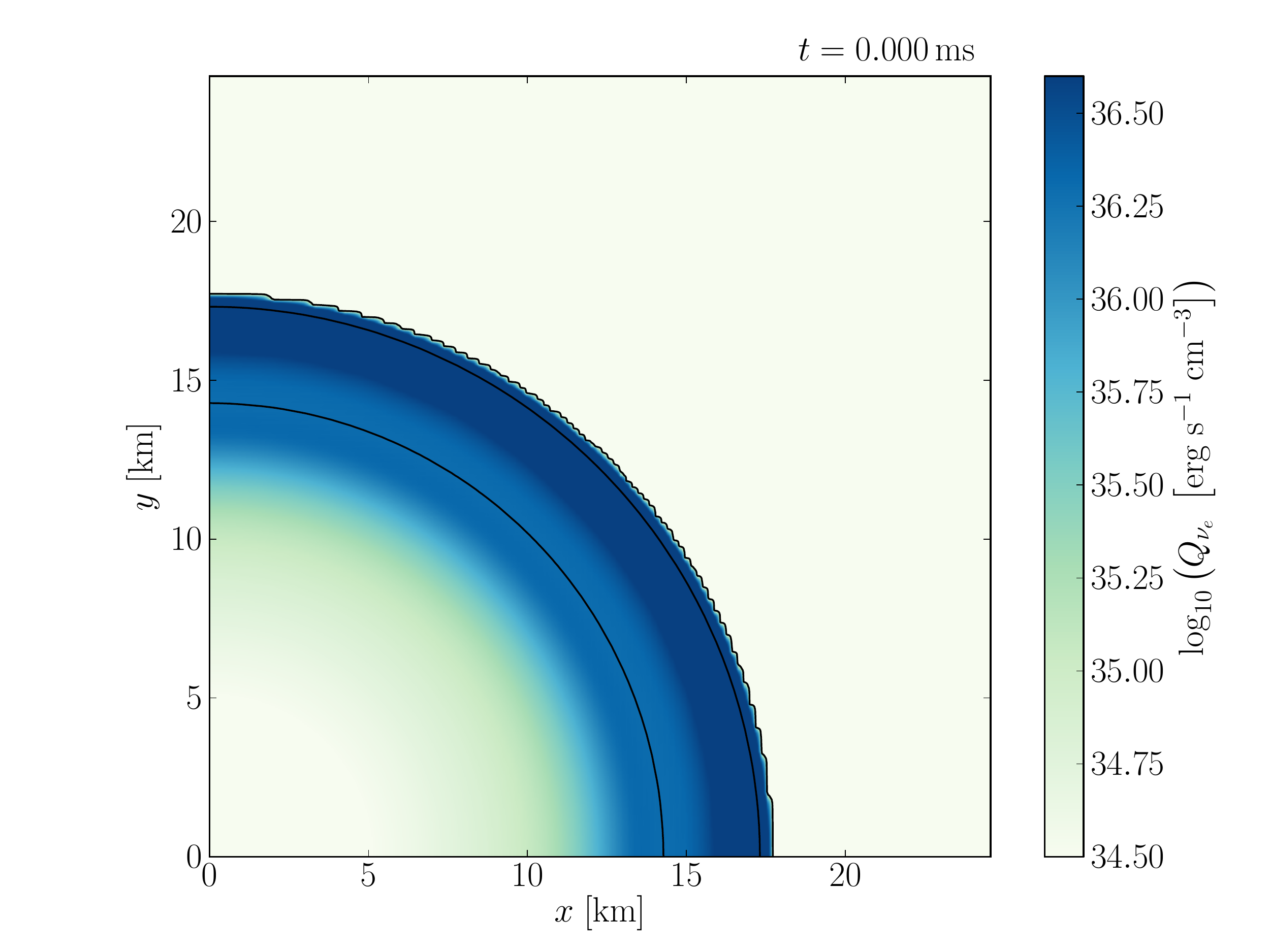}
  \includegraphics[width=0.45\textwidth]{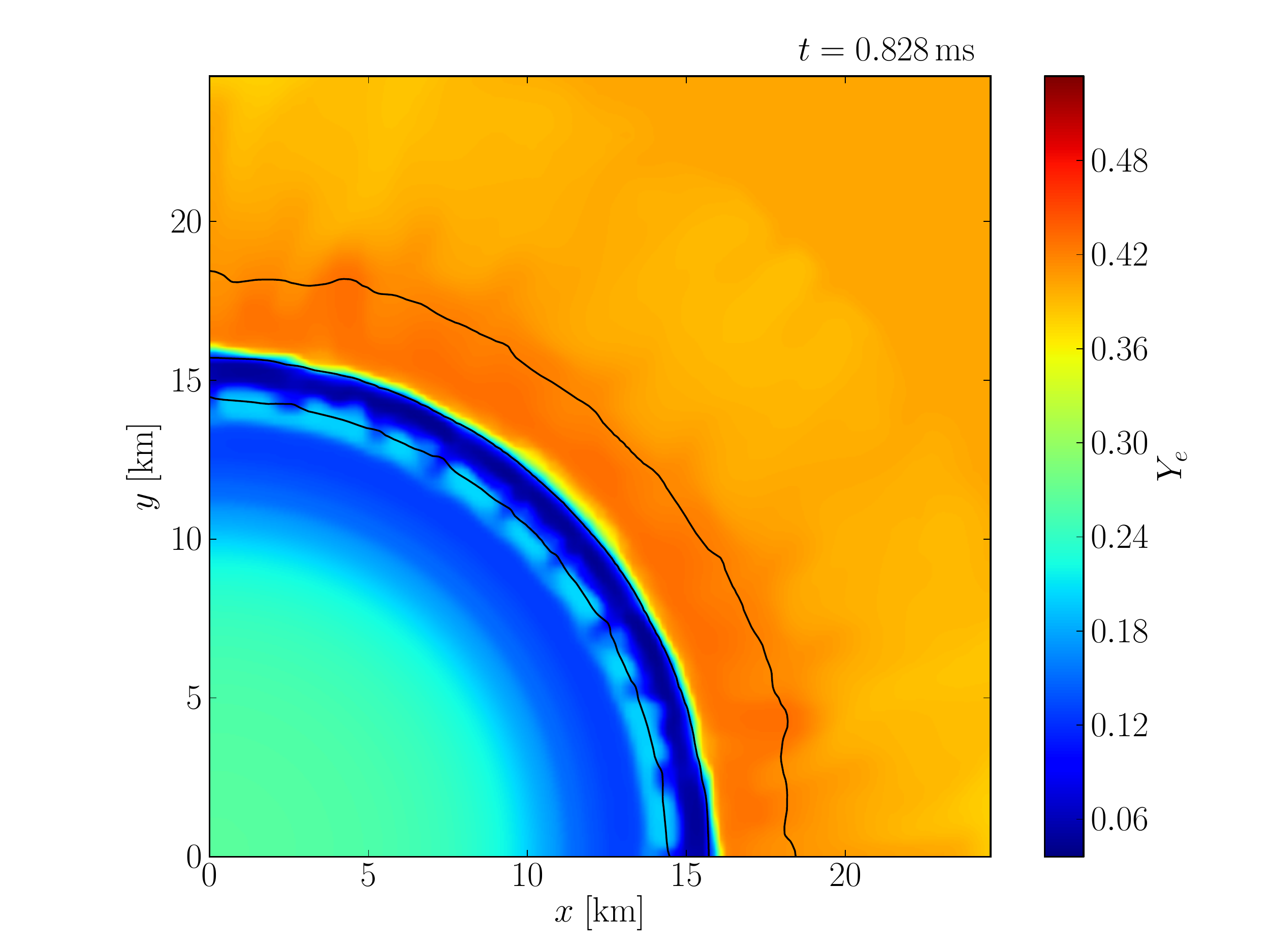}
    \hskip 1.0cm
  \includegraphics[width=0.45\textwidth]{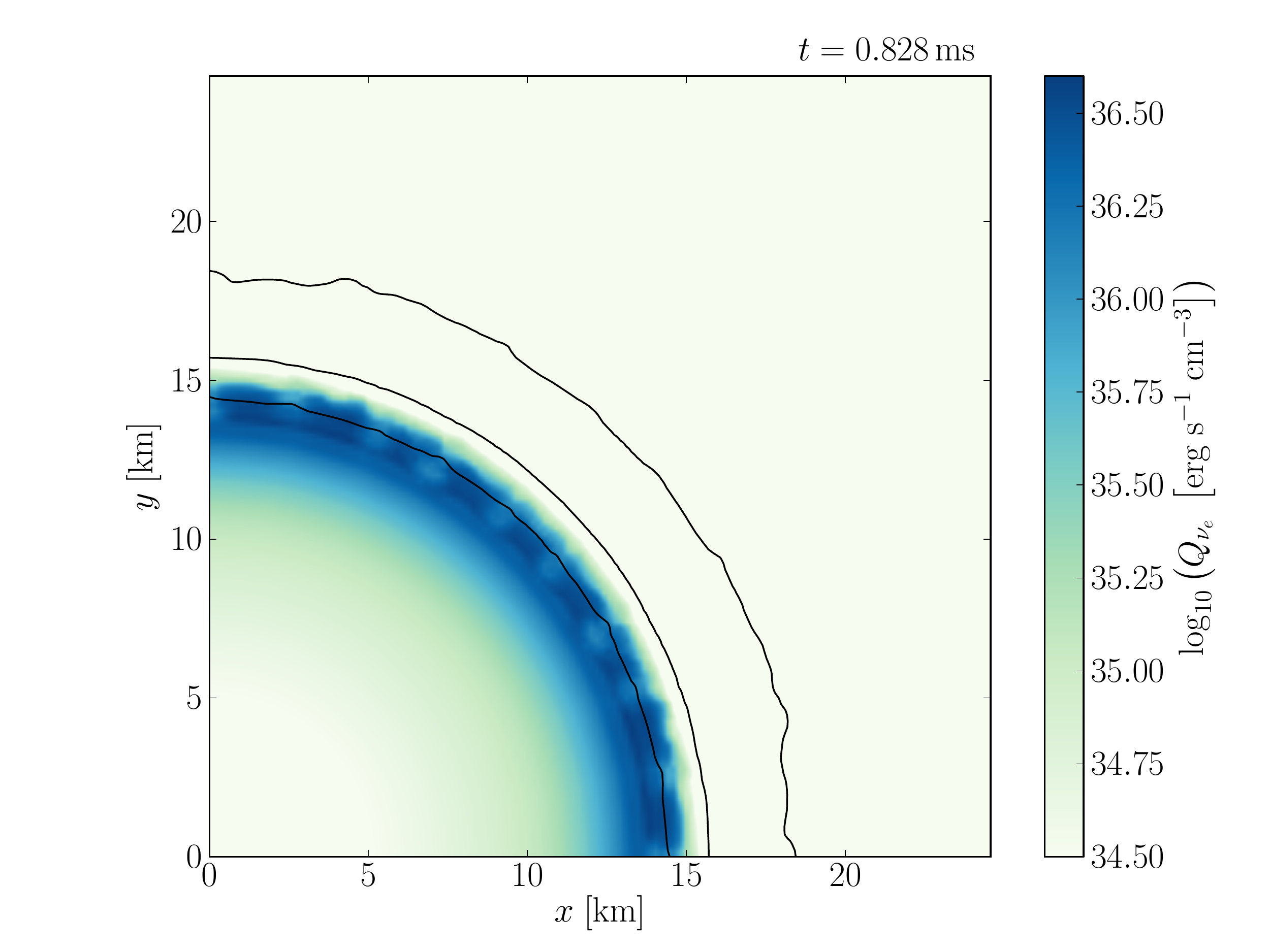}
  \includegraphics[width=0.45\textwidth]{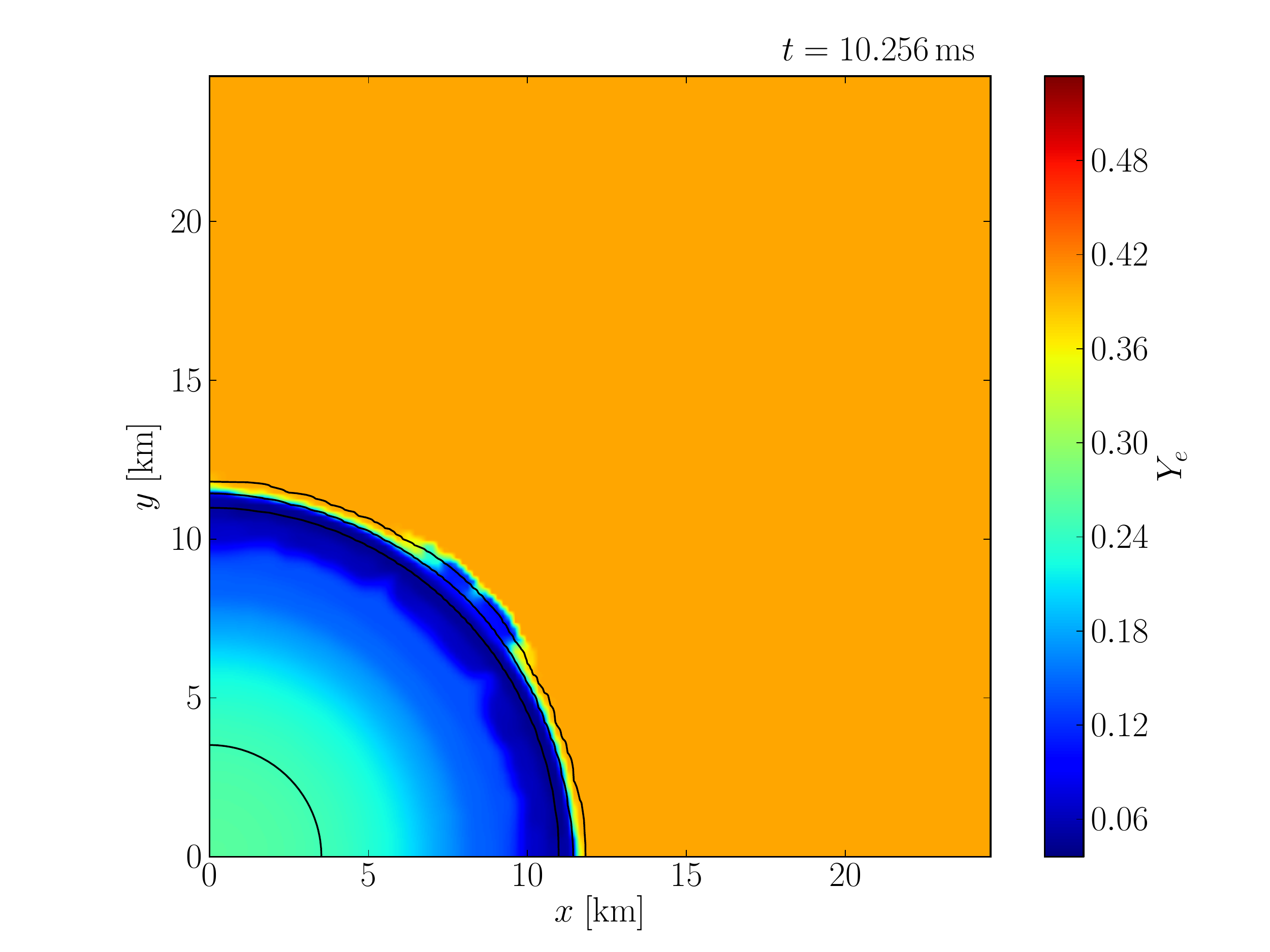}
    \hskip 1.0cm
  \includegraphics[width=0.45\textwidth]{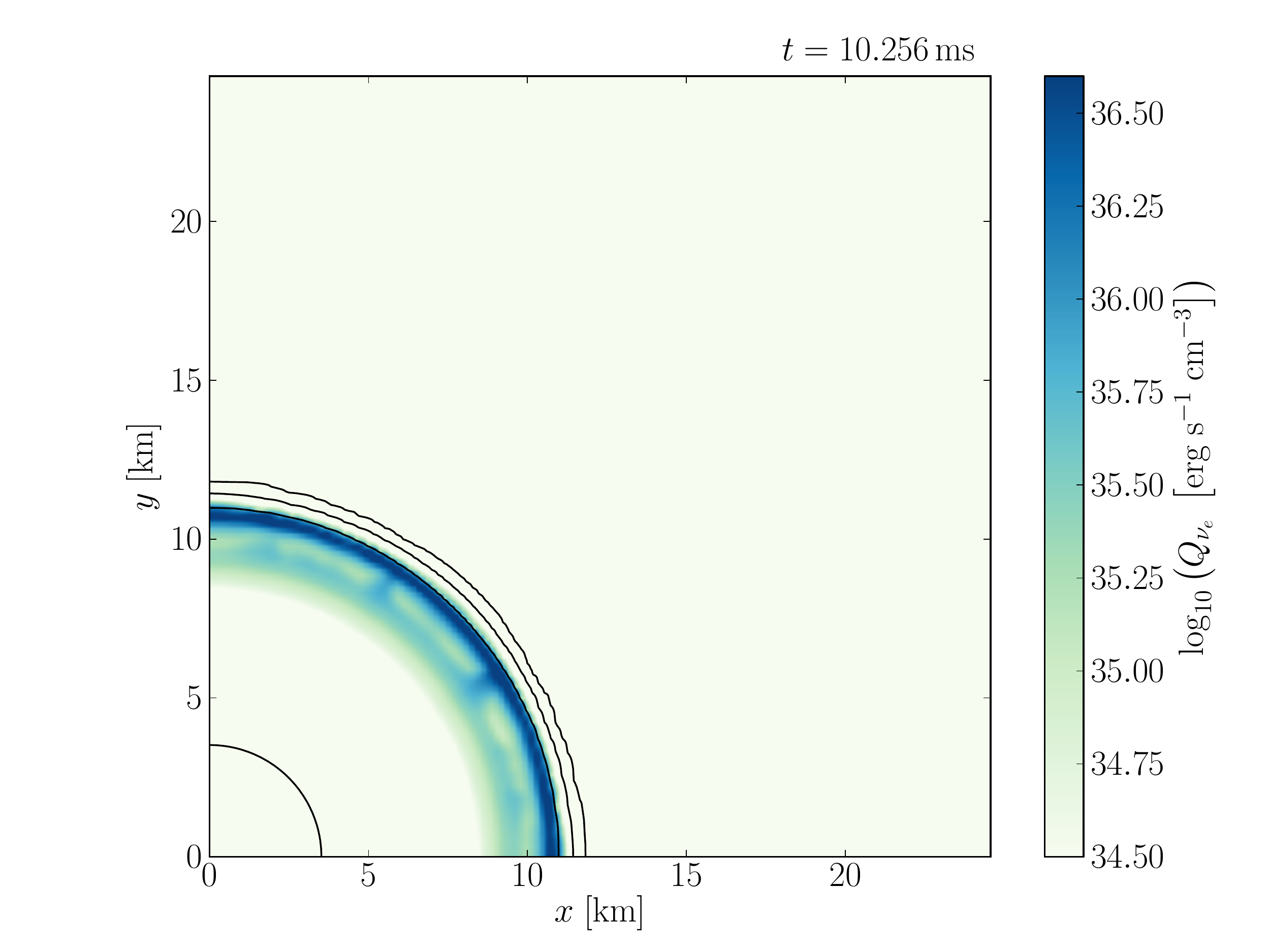}
  \caption{Neutrino-induced collapse of model \texttt{NIC-As} (see
    Table~\ref{tab:nic_models}). Left panels: Electron fraction on
    the $(x,y)$ plane at different times of the evolution. In the panel
    in the middle is visible the convective instability developing near
    the star surface. The last panel shows the onset of the gravitational
    collapse. Right panels: The local electron-neutrino
    emissivity also on the $(x,y)$ plane.}
 \label{fig:NIC_2D} 
\end{figure*}

Note that the oscillations of all models in the right panel of
Fig.~\ref{fig:nic_rhoc_ev} show the presence of the fundamental mode with
frequencies of the order of $\sim  0.7 \usk\kilo\hertz$, but also of the
first overtone with frequencies around $\sim  4.5 \usk\kilo\hertz$.
This mode is excited mostly in the outer layers of the star by the rapid
expansion and rarefaction driven by the neutrino emission and is
therefore strongly imprinted in the neutrino luminosities (\cf
Fig.~\ref{fig:nic_lum_evol}). A similar behavior has been reported
already in the calculations discussed in Ref.~\cite{Gourgoulhon1995}.

On a much longer time scale of $\sim 9.8 \usk\milli\second$, but following
a similar dynamics, also the stable model \texttt{NIC-As} encounters the
gravitational instability, and the dynamical collapse to a BH begins. In
this case, a secular increase of the central rest-mass density is clearly
visible that was of course not present in the absence of neutrino
losses. This secular increase is due to the combined effects of a slow
deleptonization and to the cooling of the outer layers of the star close
to the neutrinosphere, which lead the star to cross the stability limit
in rest-mass density, $\rho_{c,{\rm max}}$ (shown as red dashed line).

\begin{figure}
  \begin{center}
    \includegraphics[width=0.45\textwidth]{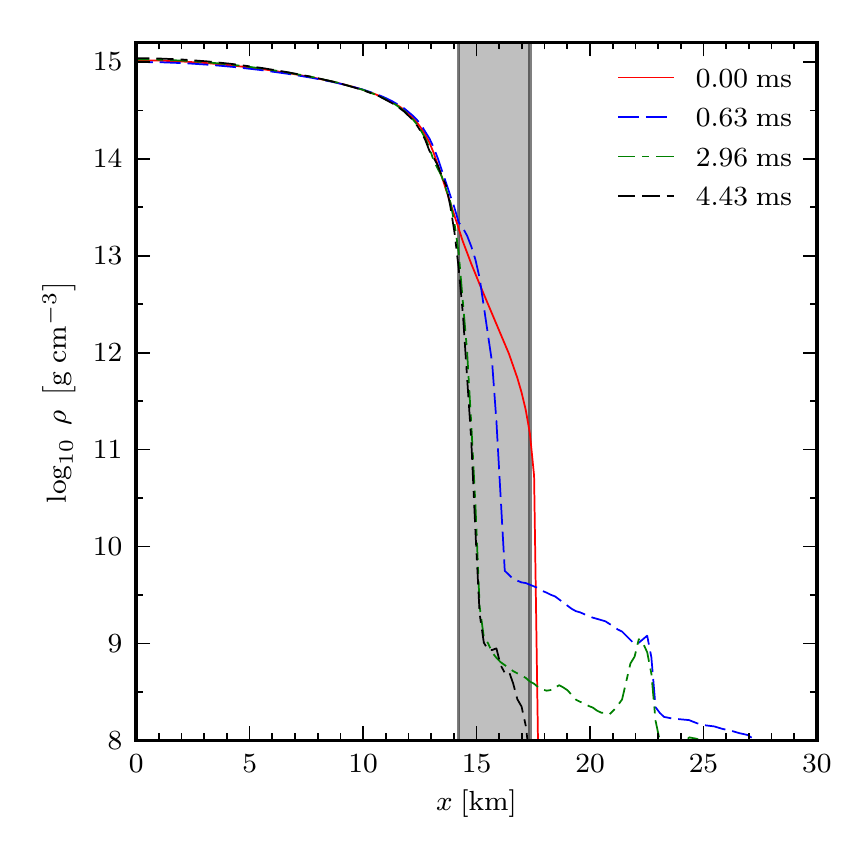}
    \caption{Evolution of the rest-mass density profile of the model
      \texttt{NIC-As}. In red the initial profile of the density solution
      of the TOV equation for a constant temperature of $30~\textrm{MeV}$
      and $\beta$-equilibrium. The position of the \referee{neutrinosphere}
      for the electron-neutrinos sweeps the shaded region
      during the evolution.}
    \label{fig:profile}
  \end{center}
\end{figure}

Additional information on the dynamics of the stable model
\texttt{NIC-As} is presented in Fig.~\ref{fig:NIC_2D}. The left panels
show the electron fraction at representative times on the $(x,y)$ plane,
while the right panels show the electron-neutrino emissivity. In all
panels we also plot isodensity contours, with the outermost one marking
the position of the stellar surface (\ie the position of the first fluid
elements at density equal to the that of the atmosphere). Note the very
rapid deleptonization caused by the neutrino losses, which reduce the
electron fraction in the outer layers of the star to values $\Efr
\lesssim 0.06$. Note also that the rapid cooling of the outer envelope of
the star leads to a net reduction of the stellar radius, which decreases
from $\sim 17.5 \usk\kilo\meter$ down to $\sim 14 \usk\kilo\meter$ (\cf
the second outermost contour line marking $1/2$ of the initial central
rest-mass density). This effective radius also coincides with the
location of the maximum electron-neutrino emission at all times during
the simulation (\cf right panels).

Another interesting aspect that emerges from Fig.~\ref{fig:NIC_2D} is the
appearance of a convective layer at a radius around $15 \usk\kilo\meter$
(\cf the middle row of panels).  
We recall that a
convective instability is expected to develop if the Ledoux
criterion~\cite{Ledoux1947,Epstein1979} is satisfied, \ie if
\begin{equation}
\label{eq:convec}
  C_L(r)=\tpd{\rho}{s}{p,Y_e}
  \frac{ds}{dr}+\tpd{\rho}{Y_e}{p,s}\frac{dY_e}{dr} \ge 0 \,.
\end{equation}

Using this criterion we find that the radius at which $C_L$ becomes
positive is approximately $14.5 \usk\kilo\meter$, which is where in fact
the first eddies appear in our simulations (this result indeed applies to
all of our models in this Sec.). The unstable region is located around
the neutrinospheres and the convection actively enhances the
deleptonization, contributing to an increase in the neutrino luminosity
and consequently to the cooling of the hot outer
envelope~\cite{Keil1996}. During the evolution, the convective layer
moves inwards until it reaches the neutrinosphere. The neutrino-opaque
interior, however, is not affected by convection, at least over the
time scale of our simulations and thus no significant changes in the
composition of the core are measured (\cf the bottom row of
panels). Finally, we note that the slight asymmetry that can be detected
along the diagonal on the last panels of Fig.~\ref{fig:NIC_2D} is simply
due to the use of Cartesian coordinates and converges away with
resolution.

A more careful look at the deleptonization occurring in the stellar
envelope and the corresponding neutrino losses can be made by analyzing
the rest-mass density profile for the stable model
\texttt{NIC-As}. This is shown in Fig.~\ref{fig:profile}, where different
lines refer to different times in the evolution. Note how the outer
envelope of the star, sustained by the lepton pressure, undergoes a rapid
cooling leaving exposed the hot stellar interior. The neutrinosphere also
shrinks, following the motion of the stellar surface. Outside the
neutrinosphere the effective emission rates given by
Eqs.~(\ref{eq:effective_rates_a})--(\ref{eq:effective_rates_b}) are
dominated by the free emission, thus leading to a rapid release of
energy. This initial burst in neutrino emission can also be clearly seen
in Fig.~\ref{fig:nic_lum_evol}, which reports the evolution of the total
neutrino luminosity for the models described in
Table~\ref{tab:nic_models}, and where the total neutrino luminosity
reaches values over $10^{55}$ erg s$^{-1}$ for all of the models
considered. As the external envelope is cooled down, it contracts, thus
contributing to the secular increase in central rest-mass density
discussed in Fig.~\ref{fig:nic_rhoc_ev}. The cooling is effective at all
times and the stars continue to emit neutrinos with a luminosity that has
decreased by about 1 order of magnitude over $10 \usk\milli\second$ for
the long-lasting model \texttt{NIC-As}. Once again we remark that most of
this emission is produced in the outer layers of the star since the
interior is mostly optically thick.

In summary, the simulations reported in this Sec. provide convincing
evidence that stellar configurations near the maximum mass of nonrotating
stellar models can be induced to collapse to a BH if subject to radiative
losses. The very same models that collapse do not show any sign of
instability in the absence of radiative losses and even models that are
metastable (\ie on the unstable branch) can migrate to the stable branch,
where they oscillate stably with large amplitude. Although rather
idealized (we are considering only spherical stars under very controlled
conditions), this result has an impact also on those studies concerned
with the formation of a PNS in simulations of stellar-core collapse or of
a HMNS in the case of BNS mergers.  \referee{While in fact we provide
  evidence that neutrino-induced collapse of stars near the stability
  threshold is indeed possible, as already suggested in
  Ref.~\cite{Paschalidis2012}, our results also indicate that this
  process is more delicate than it may appear at first sight, being
  possibly restricted to a small portion of the space of parameters.}

\section{Discussion and conclusions}

We have presented the implementation in the \texttt{Whisky} code of a
new scheme to describe the evolution of nuclear matter by means of a
microphysical finite-temperature EOSs and by a simplified neutrino
leakage scheme (NLS) to account for radiative losses via neutrinos and
the corresponding changes in composition. Although this is not the first
time that a NLS is employed in numerical simulations, even in general
relativity, we have tried to provide ample and pedagogic details about
our implementation, including a detailed description of a novel and
robust technique for the transformation from conserved to primitive
variables to be used with tabulated EOSs. We expect that this information
will be of help for anyone wanting to reproduce our results or wishing to
implement a NLS in a general-relativistic-hydrodynamical code.

Using the new code we have carried out an extensive investigation of the
properties and limitations of the NLS for isolated relativistic stars to
a level of detail that, to the best of our knowledge, has not been
discussed before in a general-relativistic context. In particular, we
have presented a series of tests on linear and nonlinear oscillations of
cold and hot NSs, which have given us the possibility to test the code
against the predictions of the linear-perturbation
theory. \referee{Overall, we have demonstrated the ability of the code to
  account for the neutrino emission from an isolated neutron star within
  the approximation of the NLS and to capture, at least as a first
  approximation, the radiative losses and the rapid changes taking place
  in the matter properties.}

In addition, we have explored some novel aspects of the dynamics of hot,
isolated NSs and, in particular, the conditions under which neutrino
radiative losses can induce their collapse to a BH. These conditions
cannot be assessed conclusively using considerations on quasistationary
transitions across equilibrium models, but rather need the use of fully
nonlinear and self-consistent numerical simulations. 

More specifically, using the newly developed code we have presented the
results of simulations of hot nonrotating NS models close to the maximum
mass and shown that the combined effects of neutrino cooling and
deleptonization can drive a stable model to collapse gravitationally to a
BH, thus confirming previous results obtained with 1D calculations
\cite{Gourgoulhon1995}. On the other hand, we have also shown that
radiative losses are neglected, the evolution leads to stable stars
oscillating in their eigenmodes. Although this scenario has not been
investigated before in three-dimensional simulations, it really applies
only to NSs with very high temperatures close to those reached by PNSs or
HMNSs, with the instability being triggered only in an extremely narrow
range of masses near the maximum one.

The numerous tests considered here are admittedly idealized, but they
provide a well-controlled environment in which to understand the
relationship between the matter dynamics and the neutrino emission, which
is of course important to understand the neutrino signals from more
complicated scenarios, such as BNS mergers, and which will represent the
future application of our code.

\begin{acknowledgments}
We thank D. Radice, C. D. Ott, Y. Eriguchi, I. Hinder, M. Obergaulinger,
G. Shen, P. Cerd\'{a}-Dur\'{a}n and N. de Brye, for useful discussions
and comments. A special thank goes to Jos\'e Mar\'ia Ib\'a\~nez, for his
initial work on the stability of hot stellar models and for the numerous
discussions on this topic. 
This work has also benefited from the stimulating atmosphere of the MICRA workshop in 2011.
We also thank Aaryn Tonita for the essential
support and collaboration in the development of the code.  Finally, we
are also grateful to the anonymous referee for the thorough and helpful
report. The numerical computations were performed on the supercomputing
cluster at the AEI. This work has been partially supported by the
''Compstar'', a Research Networking Programme of the European Science
Foundation. Partial support also comes from the VESF Grant
(EGO-DIR-69-2010) and the DFG Grant SFB/Transregio 7. JAF also
acknowledges support by the Spanish MICINN (AYA 2010-21097-C03-01) and by
the {\it Generalitat Valenciana} (PROMETEO-2009-103).
\end{acknowledgments}

\appendix

\section{Neutrino free emission rates and opacity sources}
\label{app_a}

In this appendix we will review the various weak-interaction processes
with the NS matter at different densities, temperatures and
compositions. Our primary goal is to obtain the expressions for the
{neutrino emissivity}, $\Q$ (energy emitted via neutrinos per second and
baryon) and the {neutrino emission number rate}, $\R$ (number of
neutrinos emitted per second and baryon). These explicit expression are
used in Sec.~\ref{sec:leakage} to compute the effective source terms
accounting for neutrino radiation in the hydrodynamic evolution
equations. We should note that all of the material presented here can be
found, although dispersed, in several other publications, \eg
\cite{Ruffert96b, Ruffert97, Ruffert99b, Ruffert01, Rosswog02,
  Rosswog:2003b,Rosswog:2003,Wilson2003}, and that our goal here is
mostly that of providing a single-point reference for these expressions.

We start by recalling that there is a variety of neutrino emission
mechanisms that are important at different densities and temperatures
within the NS matter. The most important of those mechanisms are:
\textit{(i)} electron and positron capture on nucleons ($\beta$ process);
\textit{(ii)} electron-positron pair annihilation; \textit{(iii)}
transverse plasmon decay.

The emission rates corresponding to the three mechanisms above can be
approximated through simple analytical formulas that 
depend only on quantities provided by the EOS tables.
We start from the simplest
and most powerful neutrino emission from hot and dense nuclear matter,
the {direct Urca process}. It consists of two reactions, the
$\beta$-decay and the electron capture on free nucleons ($n$):
\begin{align}
  e^+ + n &\to p + \nua , & e^- + p &\to n + \nu_e.
\end{align}
This is the process that drives the nucleons toward the
$\beta$-equilibrium where the chemical potentials of the electron
neutrino and antineutrino are zero, $\mu_{\nue,\nua}=0$, and the rates
of both reactions are the same, leaving the composition of the matter
unchanged. The neutrino emissivity and the neutrino
emission number rates for the $\beta$-decay are given respectively by
the following expressions~\cite{Bruenn85}
\begin{align}
  Q_{pc} (\nua)=  n_b^{-1} \beta \eta_{pn}T^6F_5(-\eta_e) \left[1-f_{\nua}\right]_{pc}\,,\\
  R_{pc} (\nua)=  n_b^{-1} \beta \eta_{pn}T^5F_4(-\eta_e)
  \left[1-f_{\nua}\right]_{pc}.
\end{align}
For the electron capture they are similarly defined as
\begin{align}
  Q_{ec}(\nu_e) &=  n_b^{-1}\beta \eta_{np}T^6F_5(\eta_e) \left[1-f_{\nue}\right]_{ec}\,,\\
  R_{ec}(\nu_e) &=  n_b^{-1}\beta \eta_{np}T^5F_4(\eta_e)
  \left[1-f_{\nue}\right]_{ec}\,,
\end{align}
where $\left[1-f_{\nu}\right]$ is the blocking factor 
for a neutrino Fermi-Dirac distribution, $f_{\nu}$,
and $\beta$ is a constant defined as
\begin{equation}
  \beta =  \frac{\pi}{h^3c^2}\frac{1+3\alpha^2}{(m_ec^2)^2}\sigma_0 
\end{equation}
where 
$\sigma_0 \approx 1.705 {\times} 10^{-44} \usk\centi\meter\squared$, 
the weak axial-vector coupling constant is $\alpha \approx 1.23$, 
and $\eta_i \equiv \mu_i/T$ are the degeneracy parameters 
for the different particle species. Temperatures are always expressed
in units of energy. 

The occupation probability of electrons, positrons, neutrinos, nucleons
and nuclei follows the relativistic Fermi-Dirac distribution with a
temperature equal to that of the fluid. The relativistic Fermi
integrals of order $N$ are defined as
\begin{equation}\label{eq::fermi}
  F_N(\eta) = \int_0^{\infty}\frac{x^N dx}{e^{x-\eta}+1}\,.
\end{equation}
and they depend on the value of the relativistic chemical potential, 
$\eta$, for each different fermion species. 
The evaluation of the Fermi integrals cannot be done analytically.
Although the numerical integration is relatively straightforward, it
is convenient to use an approximate analytic solution as suggested in
\cite{Takahashi1978}. 
After the interaction with the electron (positron), the phase space of 
the degenerate proton (neutron) is reduced by the Pauli blocking factor,
$\eta_{np}$, given by 
\begin{equation}\label{eq:block_np}
  \eta_{np} = \frac{n_n - n_p}{e^{\hat\eta}-1} \,,
\end{equation}
where $n_n$ and $n_p$ are the neutron and proton number density and
$\hat{\eta}$ is the difference between neutron and proton relativistic
degeneracy parameters. In regions where $n_p > n_n$ and $\mu_n < \mu_p$,
and where the fraction of free nucleons is small but not zero, the
Eq.~\eqref{eq:block_np} can lead to nonphysical values of the blocking
factors. We simply approximate the Eq.~\eqref{eq:block_np} in the
nondegenerate regime (for densities below $\rho = 2.0 \times 10^{12}
~{\rm g~cm}^{-3}$), with $\eta_{np} = n_n$.

Finally, the lepton phasespace is reduced by the Pauli principle, which
is taken into account by the approximate expressions
\begin{eqnarray}
  \left[1-f_{\nue}\right]_{ec} &\cong& \left\{ 1+ \exp \left[ -\frac{F_5\left(\eta_e\right)}{F_4\left(\eta_e\right)}-\eta_{\nue}\right]\right\}^{-1}\,, \nonumber \\\\
  \left[1-f_{\nua}\right]_{pc} &\cong& \left\{ 1+ \exp \left[
    -\frac{F_5\left(-\eta_e\right)}
    {F_4\left(-\eta_e\right)}-\eta_{\nua}\right] \right\}^{-1} \,. \nonumber \\
\end{eqnarray} 
In low density and high temperature nondegenerate nuclear matter,
annihilation of electron-positron pairs is an extremely efficient
neutrino mechanism
\begin{align}
  &e^+ + e^-  \to \bar{\nu}_{e} + \nu_{e}\,\\
  &e^+ + e^-  \to \nuax + \nux\,.
\end{align}
This process can be accurately calculated from temperature and
electron fraction of the matter (see \citep{Itoh1996} and for a
review on the topic see \citep{Yokovlev2001}) by the following formula
for the emission number rates for electron-neutrino and antineutrino

\begin{multline}
  R_{e^+ e^-}(\nue, \nua) =  \frac{ (C_1+C_2)_{\nue \nua}}{36 \,n_b}\frac{\sigma_0 c}{(m_e c^2)^2} 
  \times \\
  \epsilon_4(e^+) \epsilon_4(e^-)   
  \left[1- f_{\nue}(\eta_e, \eta_{\nue})\right]_{e^-e^+}  
  \times \\
  \left[1- f_{\nua}(\eta_e, \eta_{\nua})\right]_{e^-e^+}\,, 
\end{multline}
where $m_e$ is the electron mass, 
$C_1$ and $C_2$ are given in terms of the normalized vector
\referee{($C_V=0.962$)} and axial constants ($C_A=\frac{1}{2}$)
\begin{equation}
  (C_1+C_2)_{\nue \nua} = (C_V - C_A)^2 + (C_V + C_A)^2\,.
\end{equation}
For convenience we define the energy moments of the Fermi-Dirac
distribution for massive particles (electrons and positrons) as
\begin{equation}
  \epsilon_N(e^{\mp}) = \frac{8\pi}{(hc)^3}T^{N} F_{N-1}(\pm \eta_e)\,,
\end{equation}
where $N$ is an integer number. The neutrino emission rate due to
pair
annihilation, for the heavy neutrinos $\nu_{\tau}$ and $\nu_{\mu}$ (and for 
the respective antiparticles), is
given by
\begin{multline}
  R_{e^+ e^-}(\nux) = \frac{ (C_1+C_2)_{\nux \nuax}}{9\, n_b}\frac{\sigma_0 c}{(m_e c^2)^2}\times
\\
  \epsilon_4(e^+) \epsilon_4(e^-)  \left[1- f_{\nuax}(\eta_e, \eta_{\nux})\right]_{e^+ e^-}\,,   
\end{multline}
with $(C_1+C_2)_{\nux} = (C_V - C_A)^2 + (C_V + C_A-2)^2$, and
with the phasespace blocking factor for the different neutrino
species approximated as in \cite{Cooperstein1986,Cooperstein1987} by
\begin{multline}
    \left[1- f_{_I}(\eta_e, \eta_{_I})\right]_{e^-e^+} \cong \\
    \left\{ 1+ \exp \left[ - \left( \frac{1}{2}
          \frac{F_3\left(\eta_e\right)}{F_4\left(\eta_e\right)} +
          \frac{1}{2}
          \frac{F_3\left(-\eta_e\right)}{F_4\left(-\eta_e\right)}
          -\eta_{_I} \right) \right] \right\}^{-1}\,.
\end{multline}
The neutrino emissivities due to electron-positron annihilation for
the different neutrino species are
\begin{align}
  Q_{e^+ e^-}(I) =& R_{e^+ e^-}(I)
  \frac{\epsilon_5(e^-)\epsilon_4(e^+)+\epsilon_4(e^-)\epsilon_5(e^+)}
  {\epsilon_4(e^-)\epsilon_4(e^+)}\,.
\end{align}

At medium densities and high temperatures, one of the major sources of
neutrino emission comes from plasmon decay. We recall that plasmons
are quanta of electromagnetic field in a plasma and they can appear
with two distinct polarizations, longitudinal and transverse. In the
regime of our simulations, the longitudinal polarization can be
neglected~\citep{Schinder87,Kantor2007}, and the plasmon decay can be
written as
\begin{align}
  &\gamma \to \nua + \nue\,\\
  &\gamma \to \nuax + \nux\,.
\end{align}
where $\gamma$ stands for a plasmon and where the decay of a plasmon
can lead to neutrinos of different flavors. In practice, we use an
approximate fitting formula for the neutrino emission given
by~\citep{Bruenn85}
\begin{multline}
    R_{\gamma}(\nue, \nua) =   
    n_b^{-1}\frac{\pi^3}{3\alpha_{\mathrm{f}}}  C_V^2 \frac{\sigma_0 c}{(m_e c^2)^2} \gamma^6\, \times \\
    \frac{T^8}{(hc)^6} \exp(-\gamma) \left(1+\gamma\right)\times\\
    \left[ 1+ f_{\nue}(\eta_e,\eta_{\nue})\right]\left[ 1+ f_{\nua}(-\eta_e,\eta_{\nua})\right],
\end{multline}
\begin{multline}
    R_{\gamma}(\nux) = 
    n_b^{-1}\frac{4\pi^3}{3\alpha_{\mathrm{f}}}  (C_V-1)^2 \frac{\sigma_0 c}{(m_e c^2)^2} \gamma^6\, \times \\
    \frac{T^8}{(hc)^6} \exp(-\gamma) \left(1+\gamma\right)
    \left[ 1+ f_{\nux}(\eta_e,\eta_{\nux})\right],
\end{multline}
where $\alpha_{\mathrm{f}}=1/137.036$ is the fine-structure constant 
and 
\begin{eqnarray}
\gamma\approx f_{\mathrm{p}}\sqrt{\frac{1}{3}(\pi^2+3\eta_e^2)}\,
\end{eqnarray}
with $f_{\mathrm{p}}=5.565 \times 10^{-2}$
being the dimensionless plasma frequency (taken from \cite{Ruffert96b}).
The relative
blocking factor in the case of plasmon decay can then be approximated
as
\begin{multline}
  \left[1- f_{_I}(\eta_e,\eta_{_I})\right]_{\gamma}\cong \\
\left\{ 1+ \exp  \left[ - \left( 1 + \frac{1}{2}
        \frac{\gamma^2}{1+\gamma}- \eta_{_I} \right) \right]
  \right\}^{-1}\,,
\end{multline}
where the index $I$ runs over the different neutrino species.
The energy losses by plasmon decay are given by
\begin{eqnarray}
  Q_{\gamma}(I) &=& R_{\gamma}(I) \frac{1}{2} T \left(2+\frac{\gamma^2}{1+\gamma}\right).
\end{eqnarray}
For a more sophisticated treatment of the plasmon decay using a fitting formula
for the neutrino emissivity in the presence of strongly degenerate relativistic electrons 
see \cite{Yokovlev2001,Haft1994,Kantor2007}.

Another important aspect of the NLS scheme is of course given by the
neutrino opacities, which in turn depend on the different scattering
and absorption processes between neutrinos and the
matter. Particularly important and thus included in our treatment are
the following scattering processes:
\begin{itemize}
\item neutrino scattering on heavy nuclei
\item neutrino scattering on free nucleons
\item electron-flavor neutrinos absorption on free nucleons
\end{itemize}

To quantify the associated opacities we start from the simple analytic
formula for the neutrino scattering on free protons, neutrons, and
heavy nuclei, \ie
\begin{eqnarray}
  \sigma_{\nu_{_I},s}(p,n) &=& \frac{1}{4}   \frac{\sigma_0 }{(m_e c^2)^2} E_{_I}^2\,,\\
  \sigma_{\nu_{_I},s}(^A_ZX) &=& \frac{1}{16} \frac{\sigma_0 }{(m_e c^2)^2} E_{_I}^2 A^2\left(1-\frac{Z}{A}\right)\,,
\end{eqnarray}
$E_{_I}$ are the neutrino energies, $Z$ the atomic number, and $A$
the mass number for the representative heavy nuclei in the EOS. The
cross section for electron-neutrino absorption on nucleons, either free or
bound inside heavy nuclei, is larger than for any other process. An
analytic expression to account for this is given
by~\citep{Tubbs1975,Bruenn85}
\begin{align}
  \label{eq:sigma_nue}
  \sigma_{\nue,a}(n) &=  \frac{1+3\alpha^2}{4}\frac{\sigma_0 }{(m_e c^2)^2}E_{\nue}^2 \left[1-f_{e^-}(\eta_{\nue})\right]\,,\\
  \label{eq:sigma_nua}
  \sigma_{\nua,a}(p) &= \frac{1+3\alpha^2}{4}\frac{\sigma_0 }{(m_e c^2)^2}E_{\nua}^2
  \left[1-f_{e^+}(\eta_{\nua})\right] \,,
\end{align}
where the blocking factors are given by the following expressions
\begin{eqnarray}
  \left[1-f_{e^-}(\eta_{\nue})\right] &\cong&  \left\{ 1+ \exp \left[
    -\frac{F_5(\eta_{\nue})}{F_4(\eta_{\nue})}-\eta_{\nue}\right]
  \right\}^{-1}\,, \nonumber \\ \\
  \left[1-f_{e^+}(\eta_{\nua})\right] &\cong&  \left\{ 1+ \exp \left[
    -\frac{F_5(\eta_{\nua})}{F_4(\eta_{\nua})}-\eta_{\nua}\right]
  \right\}^{-1}\,. \nonumber \\
\end{eqnarray}
We should point out that the neutrino scattering on nuclei can be the
dominant source of opacity when a significant fraction of heavy
nuclei is present~\cite{Rosswog:2003b}. The heavy nuclei are
particularly abundant below nuclear saturation density and
temperatures below $15 \usk\mega\electronvolt$. Above these limits 
the nucleons can be
found mostly in the form of free protons and neutrons.
  
Using expressions~\eqref{eq:sigma_nue}--\eqref{eq:sigma_nua} it is
then possible to compute the local mean free path for each neutrino
species as 
\begin{align}
  \begin{split}
    \lambda_{_I}^{-1} &= n_p\sigma_{\nu_{_I},s}(p)
    +n_n\sigma_{\nu_{_I},s}(n)
    +n_{h}\sigma_{\nu_{_I},s}(^A_ZX)
    \\
    &\qquad +n_n\sigma_{\nu_{_I},a}(n) +
    n_p\sigma_{\nu_{_I},a}(p) \,,
  \end{split}
\end{align}
where $\sigma_{\nu_{_I},a}(n)$ is zero for all but the electron-neutrino and
$\sigma_{\nu_{_I},a}(p)$ for all but the electron-antineutrino, 
and $n_{h}$ stands for the number density of heavy
nuclei. From each of the above cross sections it is possible to
factor out the square of the neutrino energy and define a new
quantity as the {energy-independent mean free path} for the
different neutrino species
\begin{align}\label{eq:ene_free_mean_path}
  \zeta_{_I}^{-1} =
  \lambda_{_I} E_{_I}^2 \,.
\end{align}
Note this quantity depends only on the local thermodynamic state but
not on the energy of the neutrinos,
and is used to compute the energy-independent part of the optical depth, 
as explained in Sec.~\ref{sec:optical_depth}, and the diffusion rates in 
Sec.~\ref{sec:leakage}

Evaluating the integrals in the Eqs.~(\ref{eq:RD}) and (\ref{eq:QD}) 
for the diffusive emission rates, we finally obtain
\begin{align}
  R^{^D}_{_I}(\eta_{_I}^{eq})  
  &= n_b^{-1}\frac{4\pi c g_{_I}}{(hc)^3}\frac{\zeta_{_I}}{\mathcal{D}\chi ^2_{_I}}T F_0(\eta_{_I}^{eq})\\
  Q^{^D}_{_I}(\eta_{_I}^{eq}) 
  &= n_b^{-1}\frac{4\pi c g_{_I}}{(hc)^3}\frac{\zeta_{_I}}{\mathcal{D}\chi ^2_{_I}}T^2
  F_1(\eta_{_I}^{eq})
\end{align}
where the degeneracy factors are $g_{_I}=1$ for the light neutrinos
and $g_{_I}=4$ for the heavy ones.

\section{Convergence Tests}
\label{sec::convergence}
In order to verify the convergence of our code, we performed a series 
of tests using the full code, including the SHT-EOS 
and the NLS.
For our future simulations, the accuracy with which the NLS will 
describe the evolution of the matter during the collapse to a BH is of 
crucial importance. For this test, we therefore choose a TOV star with
initial constant temperature of $30 \usk\mega\electronvolt$ in 
$\beta$-equilibrium, listed in Table~\ref{tab:freq} as \texttt{uTOVh-SHT}.
The model is located on the unstable branch, and we add a small inward
pointing velocity perturbation to trigger the collapse to a BH.
The grid structure for this test is similar to the one used in the
Sec.~\ref{sec:nic}, using three fixed refinement levels. The 
finest one extends to  $22 \usk\kilo\meter$, covering the star
completely.

\begin{figure*}[t]
  \begin{center}
    \includegraphics[width=0.45\textwidth]{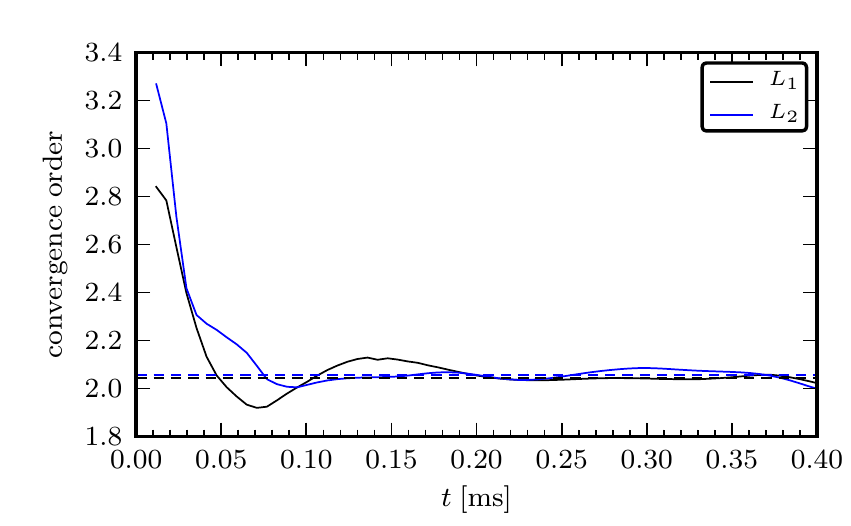}
    \hskip 1.0cm
    \includegraphics[width=0.45\textwidth]{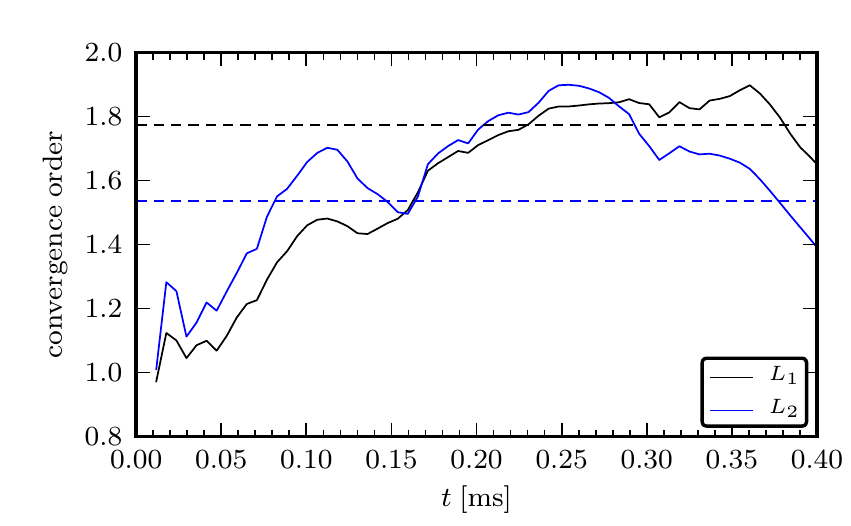}
    \includegraphics[width=0.45\textwidth]{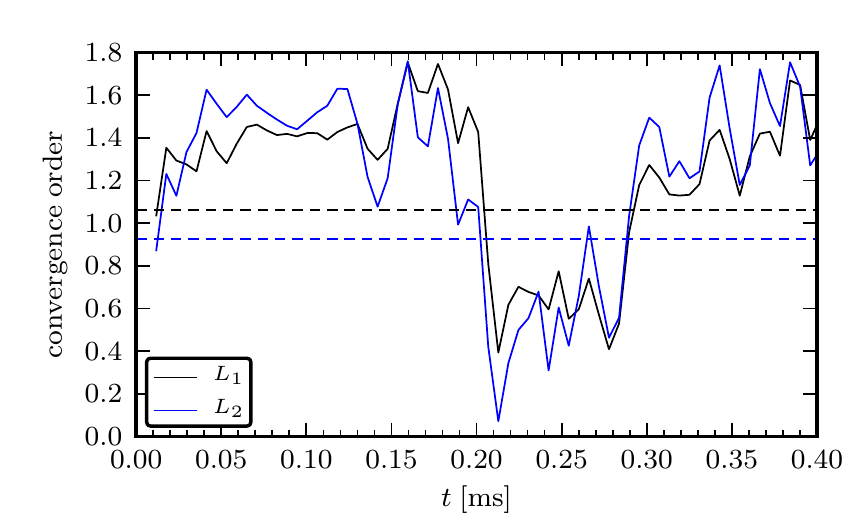}
    \hskip 1.0cm
    \includegraphics[width=0.45\textwidth]{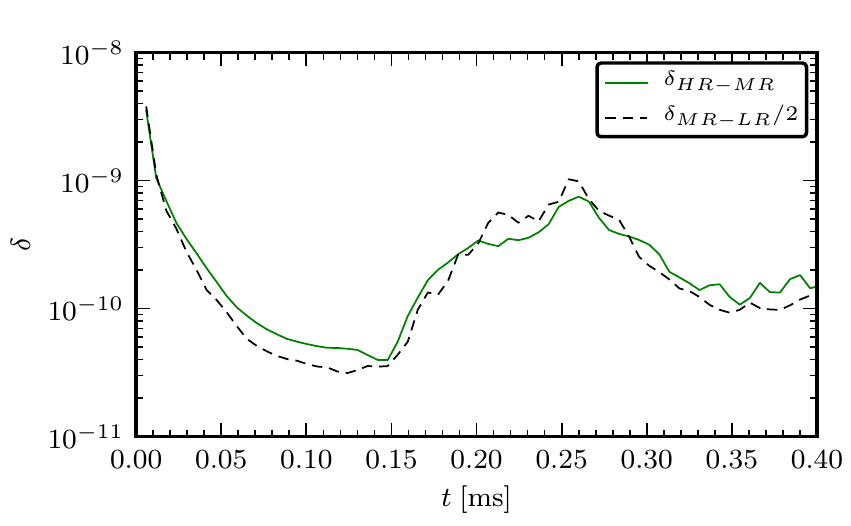}
    \caption{ \label{fig:conv_test} Time dependent convergence order 
      and residuals measured using the $L_1$- and $L_2$-norm over the 
      $(x,y)$ plane. \emph{Top left:} convergence order of the lapse 
      function. The dashed horizontal lines indicate the convergence 
      order obtained from the time-averaged residuals (see main text). 
      \emph{Top right:} convergence of energy density $\Ced$. 
      \emph{Bottom left:} convergence order of the electron-antineutrino 
      emissivity $Q_{\nua}$. \emph{Bottom right:}  the residuals for the
      same quantity, scaled assuming first order convergence.}
  \end{center}
\end{figure*}

We evolved the system
with three different grid spacings $h_\text{lo}$, $h_\text{me}$, and 
$h_\text{hi}$, where $h_\text{hi} {=} 90\usk\meter$, and 
$h_\text{hi} / h_\text{me} = h_\text{me} / h_\text{lo} = f = 2$.
For the spherical grid, on which the optical depth is computed, we use a
fixed radial resolution of $90 \usk\meter$. The reason is that we want 
to measure the convergence order of the evolution scheme, not the one of 
the interpolation to and from the spherical grid. We assume that for the 
chosen resolution, the interpolation error is not the 
dominant one.

Assuming a convergence behavior $e(h) \propto h^l$ for the error $e$
of any evolved quantity $X$, we estimate the convergence order $l$ 
of our simulations from the expression
\begin{align}\label{eq:conv_order_est}
l &\approx \frac{1}{\ln(f)} \ln\left(
     \frac{\Vert X(h_\text{lo}) - X(h_\text{me}) \Vert}
          {\Vert X(h_\text{me}) - X(h_\text{hi}) \Vert}
     \right)\,,
\end{align}
which becomes exact in the limit of infinite resolution.
We consider different norms to use in the formula above. In detail, 
we compute the 
discrete $L_1$- and $L_2$-norms over all grid points of the finest 
level that are present for each grid spacing and located on the 
equatorial plane.
By doing this at each time step (common to all resolutions), we obtain
a time dependent convergence order.
Experience tells us that such measures can be quite noisy at practical
resolutions. Therefore, we also use a global norm in 
\Eref{eq:conv_order_est} defined by the time average of the spatial 
norms. 
From this, we obtain a convergence order for the 
whole simulation.
We evaluate those measures for three representative variables, 
namely the lapse $\alpha$, the energy density $\Ced$, and the emissivity $Q$
of the electron-antineutrino. 

In Fig.~\ref{fig:conv_test}, the top left panel shows the time 
evolution of the convergence order for the lapse function. As described 
in Sec.~\ref{sec:einstein_eq}, we use a fourth order finite difference
scheme for the evolution of the spacetime. However, after few crossing 
times of the finest grid the error of the lapse becomes dominated 
by the error of the fluid in the bulk of the star. The reconstruction
scheme used for the hydrodynamical variables is expected to be at most
second order and it is not surprising that this is also the order found 
for the evolution of the spacetime quantities.

As can be seen in the top right panel, the evolution of the energy 
density $\Ced$ converges on average with order 1.7 with respect to the 
$L_1$-norm, and with order 1.5 for the $L_2$-norm. Those are typical 
values that can be expected from the PPM reconstruction procedure. 
Initially, however the error is dominated by the treatment of the
surface and only during the rapid infall of the matter we recover the
expected order of convergence.

The lower panels show the error of the effective electron-neutrino 
luminosity. 
In the right 
panel we show the residuals between high and medium resolution, 
as well as medium and low resolution, the latter rescaled assuming
a convergence order of 1.
The left panel depicts the convergence order. We find that the neutrino
sources are at most first order accurate, as expected from the linear
spatial interpolation from the spherical to the Cartesian grid and
because of the first order accurate calculation of the optical depth.
After the initial $0.15 \usk\milli\second$, which are dominated by the 
rapid cooling of the surface region, the model develops a convective 
layer in the outer envelope as described in Sec.~\ref{sec:nic}.
This causes the temporary dip seen in the convergence order, since it
is notoriously difficult to resolve convective movements numerically. 
As we could see from this analysis, the formal convergence order of our 
evolution is one, limited by the lowest order part the scheme which is 
the computation of the neutrino sources. However in most astrophysical
scenarios described with the NLS, the neutrinos are only of secondary
importance in the dynamics of the system and for this reason a higher
convergence order can be expected for the fluid quantities.

\section{Transformation from conserved to primitive variables}
\label{apdx:con2prim}

One of the most error prone parts of relativistic-hydrodynamic codes is
the conversion from the evolved conserved variables to the primitive
variables, which cannot be expressed in closed analytic form and requires
some sort of root finding procedure. Moreover, there are corner cases for
which numerical errors in the evolution lead to conserved variables which
violate physical bounds, in some cases such that no solution exists at
all. Those corner cases appear frequently at fluid-vacuum boundaries, \eg
NS surfaces, or close to the center of black holes. In both cases it is
however possible to apply corrections such that the bulk behavior of the
system is not influenced too strongly by the few affected grid
points. Using tabulated EOSs further complicates the problem because the
validity range of such EOS is limited, and because their derivatives tend
to be noisy. It is crucial to have a scheme that takes all those corner
cases into account, and that is able to apply corrections in a clearly
defined way if necessary. In the following, we discuss the corner cases
and present a scheme that is able to recognize them and take actions
based on an error policy that can be chosen based on the problem at
hand. Further, we describe the error policy used for the simulations in
this article.

For convenience, we introduce the following definitions
\begin{align}
  a &= \frac{p}{\Rmd\left(1+\Sed\right)}, &
  z &= \Lf v, \\  
  q &= \frac{\Ced}{\Crmd}, &
  r &= \frac{\sqrt{\Cmom^i \Cmom_i}}{\Crmd}, &
  k &= \frac{r}{1+q}.
\end{align}
It is easy to show the following relations
\begin{align}
  z    &= \frac{r}{h}, &
  \Rmd &= \frac{\Crmd}{\Lf}, &
  \Lf &= \sqrt{1 + z^2}, \label{eq:c2p_z_rmd_lf}
\end{align}
and further
\begin{align}
  \Sed &= \Lf q - z r + \Lf - 1,  \label{eq:c2p_sed}\\
  h &= (1+\Sed)(1 + a) = \left( \Lf - z k \right) \left(1 +
    q\right)\left( 1 + a \right).
  \label{eq:c2p_enthalpy}
\end{align}

\subsubsection*{Equation of state requirements}
For the recovery procedure, the only function needed to describe the EOS
is $a(\Rmd,\Sed,\Efr)$. Derivatives are not required. It is crucial for 
our scheme that the EOS satisfies a few reasonable constraints. First, 
the matter should satisfy the dominant energy condition, \ie the pressure 
should be smaller than the total energy density. We also exclude negative
pressure. Second, the sound speed computed under the assumption of
constant $\Efr$ has to be real valued and smaller than the speed of light, 
\ie the EOS should respect causality. We thus strictly require that
\begin{align}
  0 &\le a \le 1, \label{eq:eos_bound_a}\\
  0 &\le c_s^2 \le 1. \label{eq:eos_bound_csnd}
\end{align}
where $c_s$ is the sound speed at constant composition.

Not all values of $(\Rmd, \Sed, \Efr)$ are physically
meaningful. Moreover, the range in which the EOS is known or valid
might be limited. This information is crucial in order to deal with
corner cases correctly. We therefore define a validity region for
each EOS by
\begin{align}
  0 &\le \Rmd^\text{min} \le \Rmd \le \Rmd^\text{max},
  \label{eq:c2p_range_rmd}\\
  0 &\le \Sed^\text{min}(\Rmd,\Efr) \le \Sed \le \Sed^\text{max}(\Rmd,
  \Efr), 
  \label{eq:c2p_range_sed}\\
  0 &\le \Efr^\text{min} \le \Efr \le \Efr^\text{max} \le 1.
  \label{eq:c2p_range_efr}
\end{align}
Note the validity range for the internal energy depends on density and 
electron fraction. Usually, the lower bound is the zero-temperature limit. Note
also we require positive internal energy. This is not a restriction on
the EOS, but a restriction on the formal baryon mass constant used to
define rest-mass density $\Rmd$ in terms of the baryon number density.

\subsubsection*{Bounds for the conserved variables}
The conserved variables are subject to some physical and technical
constraints, which we will derive in the following.

Since the baryon number density is always positive, we obviously
require $\Crmd \ge 0$. For the benefit of the evolution scheme, we also
require that $\Rmd \ge \Rmd_\text{atmo}$, which implies $\Crmd \ge
\Rmd_\text{atmo}$. This is not essential to the recovery scheme,
however. Requiring the total energy density to be nonnegative
implies $q\ge -1$. Moreover, we require $\Sed \ge 0$, which leads to
\begin{align}
    q &= \Lf h - 1 - \frac{p}{\Lf\Rmd}
    \ge \Lf - 1 + \Lf\Sed \ge 0.
\end{align}
The conserved tracer for the electron fraction is trivially limited by 
$\Crmd \Efr^\text{min} \le \hat{Y}_e \le \Crmd \Efr^\text{max}$,
since $\Efr = \hat{Y}_e / \Crmd$.

Next, we compute bounds on the total momentum density, expressed in
terms of the variable $k$. We start from
\begin{align}
  k(v,a) &= v \frac{1 + a}{1+v^2 a},&
    \frac{\partial}{\partial a} k(v,a) \ge 0.
\end{align}
Together with \Eref{eq:eos_bound_a}, we obtain the following order
\begin{align}
  0 &\le \frac{1}{2} k \le v \le k \le \frac{2v}{1+v^2} <  1. 
  \label{eq_c2p_bound_vk}
\end{align}
The inequality $k < 1$ is a direct consequence of the dominant
energy condition. In terms of the evolved variables, it is equivalent
to $S \le \Crmd + \Ced$. If we decide to limit the velocity admitted
during an evolution to a certain value $v_\text{max}$, we obtain a
necessary (but not sufficient) condition for $k$
\begin{align}
  k &< k_\text{max} 
    = \frac{2 v_\text{max}}{1 + v_\text{max}^2} < 1.
\end{align}

If any of those bounds is violated, it
depends on the error handling policy whether the run is aborted or the
values are adjusted such that the bounds are satisfied. If the policy
allows it, we make the following adjustments. First, if
$\Crmd<\Rmd_\text{atmo}^\text{cut}$, we set all variables to artificial
atmosphere with $v=0$, and $\Rmd=\Rmd_\text{atmo}$, with $\Rmd_0 <
\Rmd_\text{atmo} \le \Rmd_\text{atmo}^\text{cut}$. If $q<0$ we set
$q=0$ while keeping $k$ and $\Crmd$ constant. If $k > k_\text{max}$, we
rescale the conserved momentum density such that $r=k_\text{max}
\left(1+q\right)$. Finally, the electron fraction is limited to the 
allowed range.

Note that satisfying the above bounds for the conserved variables does 
not guarantee a valid solution yet. It however allows to safely compute
values of the primitive variables using the method described in the 
following. The result will then be subject to further checks.

\subsubsection*{Solving for the primitive variables}
We formulate the problem of computing the primitive variables as a
root finding problem for some scalar function $f$. There are many
choices, not all of which are well behaved and efficient. Since we
already have a bracketing of the velocity, it seems natural to choose
the velocity as the independent variable. However, it has proven
advantageous to use the equivalent quantity $z$ to eliminate all
problems for strongly relativistic cases. For given conserved
variables, we can use
Eqs.~\eqref{eq:c2p_z_rmd_lf}--\eqref{eq:c2p_enthalpy} to express
primitive variables as functions of $z$, defining
\begin{align}
  \tilde{\Lf}(z) &= \sqrt{1+z^2}, \qquad
  \tilde{\Rmd}(z) = \frac{\Crmd}{\tilde{\Lf}(z)},
  \label{eq:c2p_lf_from_z} \\
  \tilde{\Sed}(z) &= \tilde{\Lf}(z) q - z r +
  \frac{z^2}{1+\tilde{\Lf}(z)}.
  \label{eq:c2p_sed_from_z}
\end{align}
Note we replaced $\Lf - 1$ in \Eref{eq:c2p_sed} by an equivalent
expression that is numerically more accurate for small velocities.
The electron fraction, which can be computed directly from the evolved
tracer variable, plays the role of a fixed parameter in the following 
and will be omitted to shorten notation.

The functions defined by
Eqs.~\eqref{eq:c2p_lf_from_z}--\eqref{eq:c2p_sed_from_z} are
smooth and free of singularities for all values of $z$. However, they
can produce values for $(\tilde{\Sed}, \tilde{\Rmd})$ outside the
validity range of the EOS. In order to avoid having to take this into
account during the root finding procedure, we simply extend the EOS
function $a(\Rmd, \Sed)$ to $\mathbb{R}^2$ by the prescription
\begin{align}
  \hat{\Rmd}(\Rmd)
  &\equiv \max\left(\min\left(\Rmd^\text{max}\,, \Rmd\right)\,,\Rmd^\text{min}\right), \\
  \hat{\Sed}(\Sed, \Rmd) &\equiv
  \max\left(\min\left(\Sed^\text{max}\left(\hat{\Rmd}\left(\Rmd\right)\right)\,,
      \Sed\right)\,,
    \Sed^\text{min}\left(\hat{\Rmd}\left(\Rmd\right)\right)\right), \\
  \hat{a}(\Sed, \Rmd) &\equiv a\left(\hat{\Rmd}\left(\Rmd\right)\,,
    \hat{\Sed}\left(\Sed, \Rmd \right ) \right).
\end{align}
Next, we define
\begin{align}
  \tilde{a}(z)  &= \hat{a}(\tilde{\Rmd}(z)\,, \tilde{\Sed}(z)), \\
  \tilde{h}(z) &= (1+\tilde{\Sed}(z))(1 + \tilde{a}(z)).
\end{align}
Finally, we obtain the desired function from \Eref{eq:c2p_z_rmd_lf}
\begin{align}
  f(z) = z - \frac{r}{\tilde{h}(z)}.
\end{align}
This function is well behaved for all values of $z$. In the Newtonian
limit, it is even linear. Due to \Eref{eq_c2p_bound_vk} any root
$z_0$ is confined to the interval $[z_-, z_+]$ given by
\begin{align}
  z_{-} &= \frac{k/2}{\sqrt{1 - k^2/4}}\,, & z_{+} &= \frac{k}{\sqrt{1 -
      k^2}}.
\end{align}
Note this bracketing is quite tight. In the Newtonian limit the upper
bound becomes an equality, \ie $v=k$. We now show that $f(z)$ always
has a root in the above interval. First, we express $f$ as
\begin{align}
  f(z)
  &=z -\frac{k}{ \left( \tilde{\Lf} - z k \right) \left( 1 + \tilde{a}
    \right) }.
\end{align}
A straightforward computation then yields
\begin{align}
  f(z_-) &\le -\frac{\tilde{\Lf}_- \tilde{v}_-^3}{1-2\tilde{v}_-^2}
  \le 0, & f(z_+) &\ge 0.
\end{align}
Since $f$ is continuous, we have shown the existence of a root. Once
it is found, the velocity is computed from
\begin{align}
  v^i &= \frac{\Cmom^i}{\Crmd \tilde{\Lf} \tilde{h}}.
\end{align}
The root can still correspond to an invalid solution, in the sense
that $(\tilde{\Rmd},\tilde{\Sed})$ lies outside the validity
range of the EOS, or that $v>v_\text{max}$. If the error policy
allows it, we make the following adjustments. First, if $\Rmd <
\Rmd_\text{atmo}^\text{cut}$, we set all variables to the artificial
atmosphere values. Second, if the specific energy density is not in
the EOS validity range \Eref{eq:c2p_range_sed}, we replace $\Sed \to
\hat{\Sed}\left(\tilde{\Sed}, \tilde{\Rmd}\right)$, while keeping
density and velocity constant. In terms of the conserved variables,
this means keeping $\Crmd$ and $k$ constant while changing $q$ according
to
\begin{align}\label{eq:adj_q}
  (1+q) &\to (1+q) \frac{1+\hat{\Sed}(\tilde{\Sed}\,,
    \tilde{\Rmd})}{1+\tilde{\Sed}}.
\end{align}
Finally, if $v>v_\text{max}$, we rescale the velocity such that
$v=v_\text{max}$, keeping $\Crmd$ constant. Since this slightly increases
$\Rmd$ (due to the Lorentz factor), we also limit $\Sed$ again to the
range allowed at the adjusted density. In case of adjustments,
pressure and conserved variables are also recomputed consistently.

In order to determine the root of $f(z)$ numerically, we use the 
Illinois algorithm, a variant of the regula falsi method with 
superlinear convergence. This method is very robust since it keeps
the root bracketed, never evaluates the function outside the initial 
interval, and most importantly does not require derivatives. Given 
the current status of tabulated EOSs, one cannot assume that the EOS 
is reasonable smooth or that the derivatives are meaningful everywhere. 
This discourages the use of root finding methods which require a 
derivative, \eg Newton-Raphson. 

We tested our method over a large parameter range. The  number of 
required iterations to determine $z$ with an accuracy better than 
$10^{-7}$ depends mainly on the Lorentz factor and only weakly on 
density and temperature. For the case of the LS-EOS, only $3$ 
calls to the EOS are required if $z<0.1$, while $10$ calls are still 
sufficient for Lorentz factors up to $100$.

\subsubsection*{Uniqueness of the solution}
Now that we have shown that a solution exists, we need to prove that
it is also unique. Since $f$ is continuous, we only have to show that 
$f'(z)>0$ at any root. For arbitrary $z$, we find
\begin{align}
  \begin{split}
    \frac{{d} f}{{d} z} =& \frac{r}{\tilde{h}}
    \left[\left(q+1\right)\left(\tilde{v}-k\right)
      \left(\frac{1}{1+\tilde{\Sed}} +\frac{1}{1+\tilde{a}}
        \frac{\partial \hat{a}}{\partial\Sed} \right) \right. \\ &
    \left. -\frac{1}{1+\tilde{a}}
      \frac{\tilde{\Rmd}\tilde{v}}{\tilde{\Lf}} \frac{\partial
        \hat{a}}{\partial\Rmd} \right] + 1.
  \end{split}
\end{align}
For a solution $f(z)=0$, we can rewrite this as
\begin{align}\label{eq:c2p_dfdz}
  \begin{split}
    \frac{{d} f}{{d} z} &= 1 - v^2 B, \\
    B &= 
    \hat{a} \left[
        1 + \frac{\partial \ln \left(1 + \hat{a} \right)}{\partial \ln
          \left(1+\Sed \right)} \right]
         + \frac{\partial \ln\left(1+\hat{a}\right)}{\partial
        \ln\left(\Rmd\right)} .
  \end{split}
\end{align}
If the solution is valid, $\hat{a}=a$. We then find that 
$B=c_s^2 \le 1$, where $c_s$ is the sound speed at constant electron 
fraction, and hence $f^\prime(z) > 0$.
With that we have proven that to each physically meaningful set of
conserved variables, there exists exactly one set of primitives. In
order to show that there is no additional invalid solution either, one 
can simply verify numerically that $B<1$ along the boundaries of the 
validity region of a given EOS.

\subsubsection*{Error handling policy}
In the following, we present the error policy that determines under
what conditions unphysical values of the evolved variables are
corrected, as described previously, and when to abort the run. Most
errors occur at stellar surfaces, where the densities are low and
corrections therefore have a minor (but not negligible) impact on the
global dynamics. We therefore consider some errors tolerable for low
densities $\Rmd < \Rmd_\text{strict}$. Further, we are more lenient
in a region near the center of black holes, which we define in terms
of the lapse function $\alpha < \alpha_\text{BH}$, where the constant
$\alpha_\text{BH}$ is chosen on a case by case basis such that it
occurs only well inside the apparent horizon. One harmless error that
occurs frequently when evolving zero-temperature initial data is that
$\Sed < \Sed^\text{min}$, simply because $\Sed^\text{min}$ corresponds
to the zero (or low) temperature limit and even small numerical errors
can push $\Sed$ below that boundary. In detail, we distinguish the
following cases:
\begin{enumerate}
\item $\Sed<\Sed^\text{min}$. Never an error, adjust $\Sed$.
\item $\Sed>\Sed^\text{max}$. Adjust for low density or at BH center,
  else fatal error.
\item $\Rmd < \Rmd_\text{cut}$. Never an error, set everything to
  atmosphere.
\item $\Rmd > \Rmd_1$. Always a fatal error.
\item $v>v_\text{max}$. Adjust velocity for low density or at BH
  center, else fatal error.
\item $Y_e$ out of range. Adjust electron fraction for low density or
  at BH center, else fatal error.
\end{enumerate}


\bibliography{leakage}


\end{document}